\documentclass[fleqn,12pt]{article}
\usepackage{amsfonts,epsfig}

\newcommand{\Ga}{\alpha}
\newcommand{\Gb}{\beta}

\newcommand{\Gd}{\delta}

\newcommand{\Gg}{\gamma}
\newcommand{\GG}{\Gamma}

\newcommand{\GL}{\Lambda}

\newcommand{\GS}{\Sigma}

\newcommand{\Gth}{\theta}
\newcommand{\GTh}{\Theta}

\newcommand{\cA}{{\scriptscriptstyle\cal A}}
\newcommand{\cB}{{\scriptscriptstyle\cal B}}

\newcommand{\cM}{{\scriptscriptstyle\cal M}}
\newcommand{\cN}{{\scriptscriptstyle\cal N}}
\newcommand{\cK}{{\scriptscriptstyle\cal K}}
\newcommand{\cL}{{\scriptscriptstyle\cal L}}

\newcommand{\CB}{{\cal B}}

\newcommand{\CM}{{\cal M}}

\newcommand{\CL}{{\cal L}}
\newcommand{\CP}{{\cal P}}

\newcommand{\cV}{{\cal V}}

\newcommand{\dA}{{\dot{A}}}
\newcommand{\dB}{{\dot{B}}}
\newcommand{\dC}{{\dot{C}}}

\newcommand{\dGa}{{\dot{\alpha}}}
\newcommand{\dGb}{{\dot{\beta}}}
\newcommand{\dGg}{{\dot{\gamma}}}
\newcommand{\dGd}{{\dot{\delta}}}

\newcommand{\Hi}{{\hat{i}}}
\newcommand{\Hj}{{\hat{j}}}
\newcommand{\Hk}{{\hat{k}}}
\newcommand{\Hl}{{\hat{l}}}

\newcommand{\Bara}{{\bar{a}}}
\newcommand{\bb}{{\bar{b}}}
\newcommand{\bc}{{\bar{c}}}
\newcommand{\bd}{{\bar{d}}}

\newcommand{\bbP}{\mathbb P}
\newcommand{\bbR}{\mathbb R}

\newcommand{\ft}[2]{{\textstyle {\frac{#1}{#2}} }}

\newcommand{\be}{\begin{equation}}
\newcommand{\ee}{\end{equation}}
\newcommand{\ben}{\begin{displaymath}}
\newcommand{\een}{\end{displaymath}}
\newcommand{\ba}{\begin{eqnarray}}
\newcommand{\ea}{\end{eqnarray}}
\newcommand{\nn}{\nonumber}
\newcommand{\non}{\nonumber\\}

\newcommand{\mathon}{\mathversion{bold}}
\newcommand{\mathoff}{\mathversion{normal}}

\makeatletter
\@addtoreset{equation}{section}
\makeatother
\renewcommand{\theequation}{\thesection.\arabic{equation}}

\newcommand{\la}{\label}

\newcommand{\Ref}[1]{(\ref{#1})}

\def\moth{\mathsurround=0pt}
\newdimen\zo \zo=0pt

\def\tick{\leaders\hrule height 0.5ex depth 0pt \hskip 0.5pt}
\def\upboxfill{$\moth \setbox\zo\hbox{\tick}%
  \hskip 2pt\hbox to 0pt{$\tick$\hss}\hrulefill \hbox to 6pt{$\tick$\hss}$}
\def\underbox#1{\offinterlineskip{\mathord{\mathop{\vtop{\moth\ialign{##\crcr
      $\hfil\displaystyle{#1}\hfil$\crcr\noalign{}
      {\upboxfill}\crcr\noalign{}}}}\limits}}}
\def\dtick{\leaders\hrule height .34pt depth .5ex \hskip 0.5pt}
\def\downboxfill{$\moth \setbox\zo\hbox{\dtick}%
  \hskip 2pt\hbox to 0pt{$\dtick$\hss}\hrulefill \hbox to 6pt{$\dtick$\hss}$}

\def\undersym#1{\underbox{{}#1}}

\newcommand{\vl}{{\vphantom{[}}}
\newcommand{\VV}[2]{{\cV^{\,#1}{}\!^\vl_{#2}}}
\newcommand{\TT}[2]{{T^\vl_{#1|#2}}}

\newcommand{\cro}{\!\times\!}
\newcommand{\equ}{\!=\!}
\newcommand{\mis}{\!-\!}
\newcommand{\pls}{\!+\!}

\newcommand{\arccosh}{\rm arccosh}

\newcommand{\ignore}[2]{}

\newcommand{\NP}{P}
\newcommand{\NQ}{Q}

\newcommand{\BGI}{\mathfrak{G}}

\begin{document}

\thispagestyle{empty}

\begin{flushright}
AEI-2002-052 \\
ITF-2002/38\\
SPIN-2002/21\\
{\tt hep-th/0207206}
\end{flushright}

\begin{center}

\mathon
{\bf\Large Vacua of Maximal Gauged $D=3$ Supergravities}
\mathoff

\bigskip\bigskip

{\bf T. Fischbacher, H. Nicolai}
\vspace{.1cm}  

{\em Max-Planck-Institut f{\"u}r Gravitationsphysik,\\
  Albert-Einstein-Institut\\
  M\"uhlenberg 1, D-14476 Potsdam, Germany\\}
{\small tf@aei-potsdam.mpg.de, 
nicolai@aei-potsdam.mpg.de}

\vspace{.5cm}
{\bf H. Samtleben}
\vspace{.1cm}  

{\em Institute for Theoretical Physics} \,and\, 
{\em Spinoza Institute},\\ 
{\em Utrecht University, Postbus 80.195, 3508 TD Utrecht, 
The Netherlands\\}
{\small h.samtleben@phys.uu.nl} 

\end{center}

\renewcommand{\thefootnote}{\arabic{footnote}}
\setcounter{footnote}{0}
\bigskip
\bigskip
\medskip

\begin{abstract}
We analyze the scalar potentials of maximal gauged three-dimensional
supergravities which reveal a surprisingly rich structure. In contrast
to maximal supergravities in dimensions $D\geq 4$, all these theories
admit a maximally supersymmetric ($N\equ16$) ground state with
negative cosmological constant $\Lambda<0$, except for the gauge group
$SO(4,4)^2$, for which $\Lambda=0$.  We compute the mass spectra of
bosonic and fermionic fluctuations around these vacua and identify the
unitary irreducible representations of the relevant background
(super)isometry groups to which they belong.

In addition, we find several stationary points which are not maximally
supersymmetric, and determine their complete mass spectra as well. In
particular, we show that there are analogs of all stationary points
found in higher dimensions, among them de Sitter vacua in the theories
with noncompact gauge groups $SO(5,3)^2$ and $SO(4,4)^2$, as well as
anti-de Sitter vacua in the compact gauged theory preserving 1/4 and
1/8 of the supersymmetries. All the dS vacua have tachyonic instabilities, 
whereas there do exist non-supersymmetric AdS vacua which are stable,
again in contrast to the $D\geq 4$ theories.
\end{abstract}

\renewcommand{\thefootnote}{\arabic{footnote}}
\setcounter{footnote}{0}

\newpage

\section{Introduction}

Maximal ($N\!=\!16$) gauged supergravities \cite{NicSam01,NicSam01a}
are the most symmetric of all known field theories in three space time
dimensions. Their unique position is not least a consequence of the
presence of the ``maximally extended'' Lie algebra $E_{8(8)}$ which
plays a very special role in their construction. In contrast to gauged
supergravities in higher dimensions the vector fields appear via a
non-abelian Chern Simons term rather than the usual Yang Mills term,
implying a non-abelian duality between scalars and vectors which has
no analog in dimensions $D>3$ because suitable non-abelian extensions
of higher rank tensor gauge theories do not appear to exist. As
required by supersymmetry and the matching of bosonic and fermionic
degrees of freedom on-shell, these vectors do not introduce new
propagating degrees of freedom over and above the scalar fields
already present in these theories (in the ungauged version of the
theory obtained by torus reduction from eleven dimensions, eight
Kaluza Klein vectors and 28 vectors coming from the rank three
antisymmetric tensor are dualized to scalar fields \cite{Juli83}). The
fact that the number of gauge fields is not a priori fixed entails a
much greater variety of gaugings with both compact and non-compact
gauge groups $G_0\subset E_{8(8)}$ than in higher dimensions.

In this paper we will focus on the semi-simple gaugings obtained in
\cite{NicSam01,NicSam01a} and investigate the associated scalar field 
potentials, which arise through the gauging. The existence of maximal
gauged supergravities with non-semi-simple gauge groups will be
demonstrated in a separate publication; again, there are more
possibilities than in higher dimensions as well as new phenomena
without higher-dimensional analogs. The potentials of gauged
$N\!=\!16$ supergravity are substantially more complicated than the
potentials of maximal gauged supergravities in dimensions
$D\!\geq\!4$, and arguably the most intricate potentials ever
encountered in the context of supergravity (and perhaps beyond). A
glimpse of their structural wealth is already offered by their
maximally supersymmetric stationary points (at the origin $\cV=I$)
which exist for all the semisimple gauge groups $G_0$, and which we
study in some detail here. Our analysis nicely exemplifies the
representation theory of supergroups $\BGI$ containing the $D\!=\!3$
AdS group $SO(2,2)$~\cite{GuSiTo86}. In fact, the model contains
representative examples of almost all such supergroups, including the
exceptional ones $G(3)$ and $F(4)$.

Although a general study of the extremal properties of the potentials
appears to be beyond reach with present techniques, considerable
progress can be made by adapting a technique first introduced by
N.~Warner~\cite{Warn83}, which consists in studying the potential on a
restricted subspace of scalar fields which are singlets under some
fixed subgroup of the gauge group. In a previous paper by one of the
authors \cite{Fisc02}, this technique was already employed to identify
a number of non-trivial stationary points for the $SO(8)\times SO(8)$
gauged theory. Here, we continue this analysis by working out the
potentials for various other gauge groups and singlet sectors, and
exhibit several new non-trivial stationary points. In addition we give
general mass formulas which allow us to compute the full mass spectra
at each of these stationary points. A good part of our analysis relies
on the computational methods developed in \cite{Fisc02}, which will be
described in greater detail in a forthcoming publication \cite{tf1}.

Let us briefly summarize the most interesting facets of our findings.
The compact gauge group $G_0=SO(8)^2$ admits AdS vacua preserving 1/4
and 1/8, respectively, of the supersymmetries. In addition, for gauge
groups $G_0=SO(8)^2$ and $SO(7,1)^2$ we identify non-supersymmetric
AdS vacua which unlike their known higher-dimensional analogs are
stable in the sense that all scalar fields satisfy the
Breitenlohner-Freedman bound~\cite{BreFre82}. For the noncompact gauge
group $G_0=SO(5,3)^2$ we find the first example of a maximal
supersymmetric model with both AdS and dS stationary points. The
potential corresponding to the gauge group $G_0=SO(4,4)^2$ even
interpolates between a dS stationary point and a maximally
supersymmetric vacuum with vanishing cosmological constant.  As a more
exotic example, we investigate the potential of the theory with
exceptional gauge group $G_2\times F_{4(-20)}$, and find a non-trivial
supersymmetric AdS stationary point, which breaks the maximal
$N=(7,9)$ supersymmetry in an asymmetric way to a residual $N=(0,1)$
supersymmetry and an unbroken $SU(3) \times SO(7)^-$ symmetry.

Further motivation for our studies comes from the appearance of gauged
supergravities in the AdS/CFT correspondence~\cite{AGMOO00}. In
particular, their scalar potentials turn out to carry the information
about holographic renormalization group flows in the boundary quantum
field theories, see e.g.~\cite{FGPW99,GPPZ00,BiFrSk01} for work in
higher dimensions. Flows in three-dimensional gauged supergravities
have recently been studied in the $N=8$ theories related to the D1-D5
system~\cite{BerSam02}. The maximal ($N=16$) theories remain to be
fully exploited in this context; in particular, they may have a role
to play in the supergravity description of matrix string
theories~\cite{IMSY98,MorSam02}. We should like to emphasize that not
much is known about the boundary (super)conformal field theories
related to the maximal AdS supergravities in three dimensions. What is
known is that, in the absence of propagating degrees of freedom in the
bulk, the pure CS theories reduce to Liouville and WZNW theories on
the boundary (or their supersymmetric extensions)
\cite{Verl90,Carl91,CoHevD95}.  An important question concerns the
extendibility of the background superisometries with more the $N=4$
supersymmetry on the boundary worldsheet to infinite dimensional
superalgebras containing the Witt-Virasoro algebra.

Finally, pure CS theories in three dimensions are known to occupy a
central place in the classification of knot invariants {\it \`a la}
Jones Witten. While the significance of the gauge groups found here,
in particular the non-compact ones, in that context is not clear (most
of the previous work \cite{Witt89,Tura94,KasRes02} is based on the
compact gauge groups $SU(2)$), one could hope that gauged supergravity
might also provide a much wider framework for investigations in
$D\!=\!3$ differential geometry and topology.

This paper is organized as follows. In section~2, we give a brief
review of the three-dimensional maximal gauged supergravities, in
particular their scalar potentials, stationarity conditions, and the
computation of the mass matrices around a given stationary point. In
section~3, we analyze in some detail the maximally $N=(8,8)$
supersymmetric vacua of the theories with gauge group $SO(p,8\mis
p)\times SO(p,8\mis p)$. The spectra of physical fields are organized
by the corresponding superextensions of the $AdS_3$ group $SO(2,2)$,
except for $p=4$ for which the ground state is Minkowskian. In
section~4, we extend this analysis to the exceptional gauge groups
which all admit a maximally supersymmetric AdS vacua. Section~5
finally is devoted to further extremal points in the scalar potentials
which do not preserve the full supersymmetry.

\mathon
\section{Potential and mass matrices}
\mathoff

Let us briefly recall the pertinent facts about gauged maximal
($N\!=\!16$) supergravity in three dimensions, which we will need
here, especially those concerning the scalar potential. For further
information we refer readers to \cite{NicSam01,NicSam01a} where the
construction of the gauged theories has been explained in great
detail. In addition we here present some new formulas which will
enable us to calculate the various mass matrices at the stationary
points under consideration.

The gauging of $N\!=\!16$ supergravity was achieved in
\cite{NicSam01,NicSam01a} by minimally coupling the scalar fields
to their dual vector fields. This induces a nonabelian duality between
vectors and scalars, which has no analog in higher dimensions.  Due to
the fact that the number of gauge fields is not determined {\it a
priori} (unlike in dimensions $D\geq 4$), there is a richer variety of
possible gauge groups, all of which are subgroups of the rigid
$E_{8(8)}$ symmetry of ungauged $N\!=\!16$ supergravity.  As in the
ungauged theory \cite{Juli83,MarSch83} the 128 propagating scalar
field of the theory are conveniently described as elements of the
coset space $E_{8(8)}/SO(16)$. Hence, the scalar potential of the
gauged theory is also a function on this coset space. It is given by
\ba\la{potential}
V &=&
-\ft18 \,g^2\,\Big(
A_{1}^{IJ}A_{1}^{IJ}-\ft12\,A_{2}^{I\dA}A_{2}^{I\dA} \Big) \;,
\ea
where the tensors $A_1$ and $A_2$, respectively, transform in the
${\bf 1}+{\bf 135}$ and ${\bf\overline{1920}}$ representations of
$SO(16)$. A third tensor, $A_3^{\dA\dB}$, transforming in the 
${\bf 1} + {\bf 1820}$, governs the Yukawa couplings of the matter
fermions to the scalars; unlike its analogs in dimensions $D\geq 4$,
it is algebraically independent of $A_1$ and $A_2$. These tensors
are defined as
\ba
A_{1}^{IJ} &=&
\ft87\,\Gth\,\Gd_{IJ}
+\ft1{7}\,\TT{IK}{JK}
\;,
\non[1ex]
A_{2}^{I\dA}&=&
-\ft17\,\GG^J_{A\dA}\,\TT{IJ}{A}
\;,
\non[1ex]
A_{3}^{\dA\dB}&=&
2\Gth\,\Gd_{\dA\dB}
+\ft1{48}\,\GG^{IJKL}_{\dA\dB}\,
\TT{IJ}{KL}
\;,
\la{AinT}
\ea
in terms of the so-called $T$-tensor
\ba\la{defT}
\TT{\cA}{\cB} &=& \VV{\cM}{\cA} \VV{\cN}{\cB} \,\Theta_{\cM\cN} \;,
\qquad \Gth ~=~ \ft1{248}\,\eta^{\cM\cN}\,\Theta_{\cM\cN} \;.
\ea
where the $E_{8(8)}$-valued matrix $\VV{\cM}{\cA}$ contains the 128
physical scalar fields of the theory. The numerical tensor
$\Theta_{\cM\cN}$ is the embedding tensor of the gauge group
$G_0\subset E_{8(8)}$. As shown in \cite{NicSam01,NicSam01a} all
consistency conditions, and in particular the maximal supersymmetry
of the gauged theory, are satisfied as a consequence of a single
algebraic condition on the embedding tensor, namely
\ba\la{27000a}
\bbP_{{\bf 27000}}\,\GTh &=& 0 \;.
\la{27000}
\ea
where $\bbP_{{\bf 27000}}$ is the projector onto the $\bf{27000}$
representation in the decomposition
\be
\big( \bf{248} \times \bf{248} \big)_{sym} =
 \bf{1} \oplus \bf{3875} \oplus \bf{27000} \;.
\ee
As also explained in \cite{NicSam01,NicSam01a} the condition
\Ref{27000a} entails that only the $SO(16)$ representations $\bf{1}$,
$\bf{135}$, $\bf{1820}$ and $\bf\overline{1920}$ can appear in $\GTh$.
More specifically, we have
\ba\la{Theta}
\GTh_{IJ|KL} &=& -2 \Gth \Gd^{IJ}_{KL} +
  2 \Gd\undersym{_{I[K} \, \Xi_{L]J} \, } + \Xi_{IJKL}  \non
\GTh_{IJ|A}&=& - \ft17 \GG^{[I}_{A\dA} \Xi^{J]\dA}   \non
\GTh_{A|B} &=& \Gth \Gd_{AB} + \ft1{96} \Xi_{IJKL} \GG^{IJKL}_{AB}
\;,
\ea
where $\Xi_{IJ}$, $\Xi_{IJKL}$ and $\Xi^{I\dA}$ denote the $\bf{135}$, 
$\bf{1820}$ and $\bf\overline{1920}$ representations of $SO(16)$, 
respectively (hence $\Xi_{II} = 0$ and $\GG^I_{A\dA} \Xi^{I\dA} = 0$,
and $\Xi_{IJKL}$ is completely antisymmetric in its four indices).
For the semisimple gauge groups identified in~\cite{NicSam01a} the 
embedding tensor $\GTh$ has no component transforming as the 
$\bf\overline{1920}$ representation, and we will therefore set
\be
\Xi^{I\dA} = 0 \;,
\ee
in the remainder of this paper. As we will explain elsewhere, however,
this component is needed for the non-semisimple gaugings.

Stationary points of the scalar potential \Ref{potential} are
characterized by
\ba\la{extremum}
\frac{\Gd V}{\Gd \GS^A}=0
\qquad&\Longleftrightarrow&\qquad
3\,A_1^{IM}A_2^{M\dA} = A_3^{\dA\dB}A_2^{I\dB} \;,
\la{stationary}
\ea
where the derivative is taken w.r.t.\ to a left invariant vector field
$\GS^A$ along the coset manifold $E_{8(8)}/SO(16)$. By an adaptation
of the arguments of \cite{GuRoWa86}, it has been shown in
\cite{NicSam01a} that the number of unbroken supersymmetries at a
stationary point is determined by the number of eigenvalues $\Ga_i$ of
$A_1^{IJ}$ satisfying
\ba\la{SUSY}
16\,\alpha_i^2&=&
A_{1}^{IJ}A_{1}^{IJ}-\ft12\,A_{2}^{I\dA}A_{2}^{I\dA} =
\frac{4}{g^2L^2} \;,
\ea
Here $L$ denotes the AdS scale, which is set by the value $V_0$ of the
potential at the stationary point, viz.
\be
\frac4{g^2 L^2} \equiv -\frac8{g^2} V_0 \;,
\la{AdSlength}
\ee
(as is well known, unbroken supersymmetry requires the value of $V_0$
to be non-positive). Maximal supersymmetry is then equivalent to
$A_{2}^{I\dA}=0$.

By use of the formulas given in section~4.3 of \cite{NicSam01a}
it is straightforward to compute the scalar mass matrix at
any given stationary point, which is given by the matrix of
second derivatives.
\ba\label{massmatrix1}
-4g^{-2}\CM_{AB} &\equiv&
-8g^{-2}\,\frac{\Gd^2}{\Gd \Sigma^A\Gd \Sigma^B}\; V ~=~ \non[1ex]
&=& \ft34 \left( \GG^I_{A\dA} A_2^{J\dA} A_2^{J\dB} \GG^I_{\dB B} +
          \GG^I_{A\dA} A_2^{J\dA} A_2^{I\dB} \GG^J_{\dB B} \right) 
\non[1ex]
&& + \ft34 A_1^{IJ} A_1^{IJ} \Gd_{AB} - \ft34 A_1^{II} T_{A|B} +
     \ft12 \GG^I_{A\dA} A_1^{IJ} A_3^{\dA\dB} \GG^J_{\dB B} \non[1ex]
&& - \ft14 \GG^I_{A\dA} A_3^{\dA\dC} A_3^{\dC\dB} \GG^I_{\dB B}
   + \ft14 \GG^I_{A\dA} A_3^{\dA\dC} \GG^I_{\dC C} T_{C|B} \;.
\la{massmatrix}
\ea
Because the derivatives have been taken w.r.t.\ to a left invariant
vector field, the scalar kinetic term is uniformly normalized
\be
\CL_{\rm kin} = \ft14 e \partial^\mu \GS^A \partial_\mu \GS^A + \dots
\;, 
\ee
independently of which stationary point of the potential one is
expanding around.

A substantial part of this paper will be devoted to studying the mass
matrices at the origin $\cV = I$. By \Ref{defT} the $T$-tensor then
coincides with the embedding tensor, i.e.\ $T=\GTh$.  Since
$\GTh_{IJ|A} = 0$ for all the semisimple gaugings considered in this
paper, we have
\be
A_2^{I\dA} \Big|_{\cV=I} = 0 \;,
\ee
and the stationarity condition \Ref{extremum} is trivially satisfied.
By the same token, all these stationary points preserve maximal
supersymmetry. Observe that this is not true in dimensions $D\geq 4$
where the origin $\cV=I$ is not a stationary point of the
potential, unless the gauge group is compact.  Not unexpectedly, the
scalar mass matrix \Ref{massmatrix} simplifies considerably when
$\cV=I$: with
\ba
A_1^{IJ} &=& - \Gth \Gd^{IJ} + \Xi_{IJ} \non
A_3^{\dA\dB} &=& 2 \Gth \Gd^{\dA\dB} +
                 \ft1{48} \Xi_{IJKL} \GG^{IJKL}_{\dA\dB} \;,
\la{A1A3}
\ea
we obtain 
\ba\label{massmatrix2}
-4g^{-2}\CM_{AB} \Big|_{\cV=I} &=& \left(\ft34 \Xi^{IJ}\Xi^{IJ}
    - \ft1{32} \Xi_{KLMN} \Xi_{KLMN} \right)\Gd_{AB} \non[.5ex]
&& + \left(-\ft1{12} \Xi_{IM} \Xi_{MJKL} +
     \ft1{32} \Xi_{IJMN} \Xi_{MNKL} \right) \GG^{IJKL}_{AB}
\non[1.5ex] 
&& + \ft1{2304} \Xi_{IJKL} \Xi_{MNPQ} \GG^{IJKLMNPQ}_{AB} \;,
\ea
for the semi-simple gaugings with $\GTh_{IJ|A}=0$. This expression is
independent of $\Gth$ (that this should be so is obvious for $G_0 =
E_{8(8)}$, where $\GTh_{\cM\cN} = \Gth\eta_{\cM\cN}$ and the potential
is constant).  Note, however, that $\Gth$ is not a free parameter, but
fixed by group theory in relation to the other components of the
embedding tensor. The only tunable free parameter is the overall gauge
coupling constant $g$.

Let us next turn to the vector bosons. In the absence of mass terms
the vectors do not represent propagating degrees of freedom. However,
at the stationary points the $G_0$ symmetry is spontaneously broken to
its maximal compact subgroup $H_0$. Consequently, the vector fields
associated with the non-compact generators of $G_0$ will absorb the
corresponding Goldstone bosons and thereby acquire a mass in a
$D\!=\!3$ (topological) variant of the Brout-Englert-Higgs
effect. This can be directly seen from the vector field equation of
motion, namely eq.~(3.32) of \cite{NicSam01a}, which we quote here for
the reader's convenience
\ba\la{duality}
\varepsilon^{\mu\nu\rho} \GTh_{\cM\cN} {\CB_{\nu\rho}}^\cN &=&
    2e\, \GTh_{\cM\cN} {\cV^\cN}_A \CP^{\mu A} + \dots
\;,
\ea
(the dots stand for fermionic terms not relevant for our
argument). At a given stationary point, this equation reduces to
\ba
\varepsilon^{\mu\nu\rho} \GTh_{\cM\cN}\,{\CB_{\nu\rho}}^\cN &=& 
  2eg\,\GTh_{\cM\cK} {\cV^\cK}_A {\cV^\cL}_A
\GTh_{\cN\cL}\,  {\CB_{\mu}}^\cN \;,
\ea
forming a set of massive self-duality equations \cite{ToPivN84}. The
vector masses may hence be obtained as eigenvalues of the matrix
\ba
g^{-1}\CM^{\rm vec}_{AB} &=& 
{\cV^\cM}_A\, \GTh_{\cM|\cN} {\cV^\cN}_B ~=~ T_{A|B} \;,
\la{vectormass1}
\ea
which at the symmetric vacuum $\cV = I$, simplifies to the noncompact
part of the embedding tensor
\ba
g^{-1}\CM^{\rm vec}_{AB}\Big|_{\cV=I}  &=& \GTh_{A|B} ~=~
\Gth \Gd_{AB} + \ft1{96} \Xi_{IJKL} \GG^{IJKL}_{AB}\;.
\la{vectormass2}
\ea
It is then evident that only those vector fields corresponding to
non-compact generators in the gauge group acquire a mass at $\cV =
I$. In this way, some propagating bosonic degrees of freedom are
shifted from the scalar sector to the vector fields. As we will see
very explicitly, this effect is beautifully realized for all gaugings.

For the maximally supersymmetric stationary points, for which
$A_2^{I\dA} = 0$, the fermion masses are simply given by the
eigenvalues of the matrix $A_3^{\dA\dB}$. All gravitinos remain
massless (with a formal mass term dictated by the ambient AdS
geometry) and will pair up with the dreibein and the massless vector
fields transforming under the unbroken compact subgroup $H_0 \subset
G_0$ according to the sign of the associated eigenvalues $\Ga_i$ in
\Ref{SUSY} into (nonpropagating) supermultiplets so as to reproduce
the purely topological Lagrangians of \cite{AchTow86}. For all other
stationary points supersymmetry is partially broken, and we have
$A_2^{I\dA}\neq 0$. In that case, some of the fermions become
Goldstinos, and the gravitinos acquire a mass by the super
Brout-Englert-Higgs effect. To find out which supersymmetries are
preserved one must look for Killing spinors satisfying $\Gd\psi^I =
\Gd\chi^\dA = 0$. Designating the variations along the direction of
broken supersymmetry by $\varphi^I$, we split the matter fermions as
\be
\chi^\dA = \eta^\dA + A_2^{I\dA} \varphi^I \; , 
\quad 
\psi^I_\mu = \tilde\psi^I_\mu + \widehat{D}_\mu \varphi^I \;,
\la{split}
\ee
where $A_2^{I\dA} \eta^\dA = 0$, thereby diagonalizing the fermionic
mass terms, such that the masses can be read off directly from the
eigenvalues of $A_1^{IJ}$ and $A_3^{\dA\dB}$ at the stationary point
in question. In summary, there is thus a fermionic analog of the
transferral of physical degrees of freedom from the matter fields to
some of the previously non-propagating gauge fields, in precise
agreement with the supermultiplet structure required by the background
superisometries.

\mathon
\section{Maximally supersymmetric vacua
for gauge groups $G_0 = SO(p,8\mis p)\cro SO(p,8\mis p)$}
\mathoff

In this section and the following one we concentrate on the
maximally supersymmetric stationary points and determine the mass
matrices for all gauge groups identified in \cite{NicSam01a}.
In particular, we will demonstrate that the mass spectra for the various
gauge groups are indeed consistent with the representation theory
of the corresponding supergroups as far as it has been developed
\cite{GuSiTo86,GunSca91}. In addition, we determine the
representations and spectra for the exceptional supergroups $G(3)$ and
$F(4)$, which apparently do not admit an oscillator construction of
the type considered in \cite{GuSiTo86,GunSca91}.

\subsection{Embedding of the gauge groups}

The gauge groups $SO(p,8-p)\times SO(p,8-p)$ for $p=0, 1,\dots ,4$ are
the only known solutions of \Ref{27000a} with vanishing singlet
contribution, i.e.\ $\Gth = 0$. They are embedded into $E_{8(8)}$
via its $SO(8,8)$ subgroup (with $p+q=8$)
\be
SO(p,q)\times SO(p,q)\subset SO(8,8) \subset E_{8(8)} \;.
\ee
At the supersymmetric extremum for $\cV =I$, the symmetry is broken
down to the maximally compact subgroup
\be
H_0 = SO(p)\times SO(q)\times SO(p)\times SO(q)\subset SO(8)\times
SO(8)   \;.
\la{H0}
\ee
It is useful to note here that, apart from the case $p\equ1$, the
verification that $\cV=I$ is indeed a (maximally supersymmetric)
stationary point does not even require explicit knowledge of the
embedding tensor \Ref{27000a} and the vanishing of $\GTh_{IJ|A}$ for
the gauge groups considered here, but is a direct consequence of the
fact that there is no $H_0$-invariant tensor in the decomposition of
the ${\bf \overline{ 1920}}$. Thus $A_{2}$ must vanish at the
$H_0$-invariant point, implying stationarity and maximal
supersymmetry.  We emphasize this point because in dimensions $D\geq
4$ the $A_2$-tensor does contain singlets w.r.t.\ to the unbroken
compact gauge group when $\cV=I$, violating supersymmetry and the
stationarity condition, see section~\ref{higherD}.

Let us now study the embedding in somewhat more detail: under the
$SO(8) \times SO(8)$ subgroup of $E_{8(8)}$ the relevant $SO(16)$
representations decompose as follows\footnote{Contact with the results
for $D=4, N=8$ is established by noting that w.r.t.\ the diagonal
$SO(8)$ the $E_{7(7)}$ subgroup consists of the representations ${\bf
28} + {\bf 35}_v + {\bf 35}_s + {\bf 35}_c$, while the $SL(2)$, which
commutes with it is made out of the three singlets arising in the
decomposition of the ${\bf 120}$ and the ${\bf 128}_s$.}
\ba\la{SO16decomp}
\bf{16}_v &\longrightarrow& (\bf{8}_v, \bf{1})  + (\bf{1},\bf{8}_v)  \non
\bf{120} &\longrightarrow& (\bf{28}, \bf{1})  + (\bf{1},\bf{28})
                               + (\bf{8}_v, \bf{8}_v)   \non
\bf{128}_s &\longrightarrow&  (\bf{8}_s, \bf{8}_s) + (\bf{8}_c, \bf{8}_c) \non
\bf{128}_c &\longrightarrow&  (\bf{8}_s, \bf{8}_c) + (\bf{8}_c, \bf{8}_s)
\;.
\ea
Accordingly, we split the vector indices $I$ as $a\equiv I$ for $I \in
\{1,\dots,8\}$, and $\Bara \equiv I-8$ for $I\in \{9,\dots,16\}$.  The
compact part of $SO(8,8)$ is then composed out of the two $\bf{28}$
representations occurring in the decomposition of $\bf{120}$, while
its non-compact part is identified with the $(\bf{8}_s, \bf{8}_s)$. In
terms of $SO(8)$ $\Gg$-matrices, the embedding tensor reads
\ba
\GTh_{ab|cd} &=& \frac1{4} \left( \NP_{\Ga\Gg} \NP_{\Gb\Gd} -
                 \NQ_{\Ga\Gg} \NQ_{\Gb\Gd} \right)
                 \Gg^{ab}_{\Ga\Gb} \Gg^{cd}_{\Gg\Gd} \non
\GTh_{\Bara\bb|\bc\bd} &=& 
             \frac1{4} \left(\NQ_{\Ga\Gg} \NQ_{\Gb\Gd} -
                   \NP_{\Ga\Gg} \NP_{\Gb\Gd} \right)
             \Gg^{\Bara\bb}_{\Ga\Gb} \Gg^{\bc\bd}_{\Gg\Gd} \non[1ex]
\GTh_{\Ga\Gb|\Gg\Gd} &=& 2 \left( \NP_{\Ga\Gg} \NQ_{\Gb\Gd} -
                 \NQ_{\Ga\Gg} \NP_{\Gb\Gd} \right) \;,
\label{thetatensor}
\ea
with all other components vanishing (in particular, all components
with dotted indices: $\GTh_{\dGa\dGb | \dGg\dGd} = \dots = 0$).
The symbols $\NP$ and $\NQ$ are defined by
\ba
\NP_{\Ga\Gb} &=& \left\{
\begin{array}{rl} \Gd_{\Ga\Gb} & \quad\mbox{for}\;\;
\Ga,\Gb \in \{1,\dots,p\} \\
 0 & \quad\mbox{otherwise}
\end{array}\right. \;, \non
\NQ_{\Ga\Gb} &=& \left\{
\begin{array}{rl} \Gd_{\Ga\Gb} & \quad\mbox{for}\;\;
\Ga,\Gb \in \{p+1,\dots,8\} \\
 0 & \quad\mbox{otherwise}
\end{array}\right. \;,
\la{PQ}
\ea
such that $\NP_{\Ga\Gb} + \NQ_{\Ga\Gb} = \Gd_{\Ga\Gb}$, which shows
explicitly the embedding~\Ref{H0}, cf.~\cite{NicSam01a}. From the form
of $\GTh$ it is evident that the ratio of the two coupling constants
is $-1$. Furthermore, it is straightforward to verify that this tensor
indeed satisfies the projection condition \Ref{27000}. To this end, we
rewrite the compact part $\GTh_{IJ|KL}$ as
\ba
\GTh_{ab|cd} &=& -\ft1{2}\, \gamma^{abcd}_{\Ga\Gb} \NP_{\Ga\Gb}
+\ft3{8}(p-q) \, \delta_{ab}^{cd} \;,\non
\GTh_{\Bara\bb|\bc\bd} &=& 
\ft1{2}\,  \gamma^{\Bara\bb\bc\bd}_{\Ga\Gb} \NP_{\Ga\Gb}
-\ft3{8}(p-q) \, \delta_{\Bara\bb}^{\bc\bd} \;,\nn
\label{thetatensor2}
\ea
which shows that they are indeed of the form \Ref{Theta} with
$\Gth=0$. Using a triality rotated version of the decomposition of
$SO(16)$ $\GG$-matrices given in \cite{Nico87} (see appendix~A), one
likewise verifies the last equation in \Ref{Theta} for $\GTh_{A|B}$.

For later use, we also record the results for the tensors $A_1$ and
$A_3$ for $\cV=I$, which are
\be
A_1^{ab} = \ft14 (p-q) \Gd^{ab} \; , \quad
A_1^{\Bara\bb} = - \ft14 (p-q) \Gd^{\Bara\bb} \;,
\la{A1Ipq}
\ee
and (using the decomposition of the ${\bf 128}_c$ in \Ref{SO16decomp})
\be
A_3^{\Ga\dGg \, \Gb\dGd} = \ft12\,  \Gd_{\dGg\dGd}
    \left( q \NP_{\Ga\Gb} +  p \NQ_{\Ga\Gb} \right) \; , \quad  
A_3^{\dGa\Gg \, \dGb \Gd} =   - \ft12\,  \Gd_{\dGa\dGb}
    \left( q \NP_{\Gg\Gd} + p \NQ_{\Gg\Gd} \right) \;.
\la{A3Ipq}
\ee
The value of the cosmological constant at the maximally supersymmetric
vacuum is hence given by
\be
\la{LSOpq}
\Lambda = -\frac{2}{L^2} = -2g^2\,(p-q)^2 \;.
\ee
In particular, $\Lambda$ vanishes for $p=q$, i.e.\ for gauge group $G_0
=SO(4,4) \times SO(4,4)$.

\mathon
\subsection{$SO(8)\times SO(8)$}
\mathoff

For the compact gauging, the gauge group remains unbroken at the
origin. The background isometry group around this point is
$\BGI=O\!Sp(8|2,\bbR)\times O\!Sp(8|2,\bbR)$ which has the bosonic
part $\left( SL(2,\bbR)\times SO(8)\right)^2$. The physical spectrum
is given by the tensor product of two (left and right) singleton
supermultiplets according to
\ba
( 8_v + 8_s ,\, 8_v + 8_s ) \;,
\ea
under $SO(8)_L\cro SO(8)_R$. Representation theory of
$O\!Sp(8|2,\bbR)$ gives the conformal weights of the states in this
multiplet \cite{GuSiTo86}
\ba
\begin{tabular}{|c||c|c|} \hline
$SO(8)$  &  $8_v$ & $8_s$ \\
\hline
$\ell_0$ & $\ft14$ & $\ft34 $
\\ \hline
\end{tabular}
\label{mult8}
\ea
{}From this one reads off that the physical spectrum around the origin
consists of 128 scalars and 128 spin-$\ft12$ fields with
\ba
\begin{tabular}{|c||c|c|c|} \hline
fields & $SO(8)_L\times SO(8)_R$ & $(\ell_0,\bar{\ell}_0)$ & $m^2L^2$
\\
\hline\hline
scalars & $(8_v,8_v)$ & $(\ft14,\ft14)$ & $-\ft3{4}$
\\ \cline{2-4}
& $(8_s,8_s)$ & $(\ft34,\ft34)$ & $-\ft3{4}$
\\ \hline
fermions & $(8_v,8_s)$ & $(\ft14,\ft34)$ & 0
\\ \cline{2-4}
& $(8_s,8_v)$ & $(\ft34,\ft14)$ & 0
\\ \hline
\end{tabular}
\label{spec8}
\ea
The relation between mass and conformal dimension
$\Delta=\ell_0+\bar{\ell_0}$ in three dimensions is given
by~\cite{Witt98,HenSfe98} 
\ba
\Delta(\Delta-2) = m^2L^2 &&  \mbox{for scalars} \non
(\Delta - 1)^2 =  m^2L^2 &&  \mbox{for fermions} \non
(\Delta - 1)^2 =  m^2L^2 &&  \mbox{for massive selfdual vectors} \;,
\la{MDelta}
\ea
which gives the mass values in the last column of \Ref{spec8} in units
of the inverse AdS length~\Ref{AdSlength}. They indeed agree with the
spectrum computed from \Ref{A1A3}, \Ref{massmatrix2}. All mass
eigenvalues satisfy the Breitenlohner-Freedman
bound~\cite{BreFre82,MezTow85}
\ba
m^2 L^2 \ge -1 \;,
\la{BF}
\ea
the stationary point hence is stable as is implied by supersymmetry.
The metric and the massless gravitino fields form a separate
(unphysical) ``multiplet'' together with the massless selfdual vector
fields. They transform in the adjoint representation of
$\BGI$.

\mathon
\subsection{$SO(7,1)\times SO(7,1)$}
\mathoff

The background isometry group at the origin is $\BGI=F(4)\times F(4)$,
whose bosonic part is $\left(SL(2,\bbR)\times SO(7)\right)^2$. The
physical spectrum around this point is given by the tensor product of
two (left and right) massless unitary supermultiplets according to
\ba
( 1+8+7 ,\, 1+8+7\,) \;,
\la{SM7}
\ea
under $SO(7)_L\cro SO(7)_R$. To the best of our knowledge, the
representation theory of $F(4)$ has not been worked out so far. We can
however invert the reasoning which for the compact gauge group led to
\Ref{spec8}, and derive the conformal weights from the masses of the
supergravity fields. Computing the masses from \Ref{A1A3},
\Ref{massmatrix2}, \Ref{vectormass2} gives rise to
\ba
\begin{tabular}{|c||c|c|c|} \hline
fields & $H_0$ &
$(\ell_0,\bar{\ell}_0)$ & $m^2L^2$ \\
\hline\hline
scalars & (1,1) & $(\ft43,\ft43)$ & $ \ft{16}{9} $
\\ \cline{2-4}
& $(8,8)$ & $(\ft56,\ft56)$ & $  -\ft{5}{9} $
\\ \cline{2-4}
& $(7,7)$ & $(\ft13,\ft13)$ & $ -\ft{8}{9} $
\\ \hline
fermions & $ (1,8) $ & $(\ft43,\ft56)$ & $\ft{49}{36} $
\\ \cline{2-4}
& $ (7,8)$ & $  (\ft13 , \ft56) $ & $ \ft1{36}$
\\ \hline
vectors & $ (1,7)$ & $ (\ft43,\ft13) $ & $\ft49$
\\ \cline{2-4}
& $ (7,1)$ & $ (\ft13, \ft43) $ & $\ft49$
\\ \hline
\end{tabular}
\label{spec7}
\ea
For simplicity, we have omitted half of the fermionic fields which
arise with opposite chirality. Note that fourteen vector fields have
become massive due to the Brout-Englert-Higgs like effect,
corresponding to the noncompact directions in the gauge group. The
corresponding massless scalar (Goldstone) fields have not been
included in the table. The conformal dimensions in \Ref{spec7} have
been computed via \Ref{MDelta}. This confirms the structure of the
spectrum as a tensor product of $F(4)$ supermultiplets~\Ref{SM7} whose
conformal dimensions are given by
\ba
\begin{tabular}{|c||c|c|c|} \hline
$SO(7)$  &  $7$ & $8$ & $1$ \\
\hline
$\ell_0$ & $\ft13 $ & $\ft56 $ & $\ft43$
\\ \hline
\end{tabular}
\label{mult7}
\ea
Note, that this poses a highly nontrivial consistency check on the
masses obtained in our supergravity computation. Furthermore, it is
obvious from these values that the oscillator construction developed
in \cite{GuSiTo86} does not apply here because it can only produce
values $\ell_0$ which are multiples of $\ft14$.

\mathon
\subsection{$SO(6,2)\times SO(6,2)$}
\mathoff

The background isometry group at the origin is $\BGI=SU(4|1,1)\times
SU(4|1,1)$. The physical spectrum around this point is given by the
tensor product of two (left and right) massless unitary
supermultiplets according to
\ba
\left( 1^{+2} + 4^+ + 6^0 + \overline{4}{}^- + 1^{-2} , \,
1^{+2} + \overline{4}{}^+ + 6^0 + 4^- + 1^{-2} \right) \;.
\la{SM6}
\ea
Representation theory of $SU(4|1,1)$ gives the conformal weights of
the states in this multiplet~(cf.~\cite{GuSiTo86}, table 3, $n=4$)
\ba
\begin{tabular}{|c||c|c|c|c|c|} \hline
$SO(6)\cro U(1)$  &  $1^{+2}$ & $4^+$ & $6^0$ &
$\overline{4}{}^-$ & $1^{-2}$ \\
\hline
$\ell_0$ & $\ft32$ & $1 $ & $\ft12$ & $1$ & $\ft32$
\\ \hline
\end{tabular} \;.
\label{mult6}
\ea
Note that this is the unique supermultiplet of $SU(4|1,1)$ which upon
tensoring a left with a right copy reproduces the correct spins for
the supergravity fields, including the 24 massive (selfdual) vector
fields which correspond to the noncompact directions of the gauge
group. In particular, this rules out the similar multiplet
of~\cite{GuSiTo86} (table 2), whose states combine the same $SU(4)$
quantum numbers~\Ref{SM6} with different values of $\ell_0$, giving
rise to massive spin-2 states which do not occur in the supergravity.

{}From \Ref{mult6} one may read off the physical spectrum around
the origin
\ba
\begin{tabular}{|c||c|c|c|} \hline
fields & $H_0$ &
$(\ell_0,\bar{\ell}_0)$ & $m^2L^2$ \\
\hline\hline
scalars & $(1^{+2},1^{+2})$ & $(\ft32,\ft32)$ & $ 3 $
\\ \cline{2-4}
& $(4^+,\overline{4}{}^+)$ & $(1,1)$ & $ 0 $
\\ \cline{2-4}
& $(6^0,6^0)$ & $(\ft12,\ft12)$ & $ -1 $
\\ \cline{2-4}
& $(\overline{4}{}^-,4^-)$ & $(1,1)$ & $ 0 $
\\ \cline{2-4}
& $(1^{-2},1^{-2})$ & $(\ft32,\ft32)$ & $ 3 $
\\ \cline{2-4}
& $(1^{+2},1^{-2})$ & $(\ft32,\ft32)$ & $ 3 $
\\ \cline{2-4}
& $(1^{-2},1^{+2})$ & $(\ft32,\ft32)$ & $ 3 $
\\ \cline{2-4}
& $(\overline{4}^-,\overline{4}{}^+)$ & $(1,1)$ & $ 0 $
\\ \cline{2-4}
& $4^+,4^-)$ & $(1,1)$ & $ 0 $
\\ \hline
fermions & $ (1^{+2},\overline{4}{}^+)$ & $ (\ft32,1) $ & $\ft94$
\\ \cline{2-4}
& $ (4^{+},6^0)$ & $ (1,\ft12) $ & $\ft14$
\\ \cline{2-4}
& $ (6^0,4^-)$ & $ (\ft12,1) $ & $\ft14$
\\ \cline{2-4}
& $ (\overline{4}{}^-,1^{2-})$ & $ (1,\ft32) $ & $\ft94$
\\ \cline{2-4}
& $ (1^{2+},4^-)$ & $ (\ft32,1) $ & $\ft94$
\\ \cline{2-4}
& $ (4^{+},1^{-2})$ & $ (1,\ft32) $ & $\ft94$
\\ \hline
vectors & $ (1^{+2},6^0)$ & $ (\ft32,\ft12) $ & $1$
\\ \cline{2-4}
& $ (1^{-2},6^0)$ & $ (\ft12, \ft32) $ & $1$
\\ \hline
\end{tabular}
\label{spec6}
\ea
which again gives complete agreement with the masses computed in
supergravity from \Ref{A1A3}, \Ref{massmatrix2}, \Ref{vectormass2}. As
above, we have omitted half of the vector and half of the fermion
fields which arise with opposite chirality.

\mathon
\subsection{$SO(5,3)\times SO(5,3)$}
\mathoff

The background isometry group at the origin is
$\BGI=O\!Sp(4^*|4)\times O\!Sp(4^*|4)$. The physical spectrum around
this point is given by the tensor product of two (left and right)
massless unitary supermultiplets according to
\ba
\Big( (1,3)+(4,2)+(5,1) ,\, (1,3)+(4,2)+(5,1) \Big) \;.
\ea
Representation theory of $O\!Sp(4^*|4)$ gives the conformal weights of
the states in this multiplet~\cite{GunSca91}
\ba
\begin{tabular}{|c||c|c|c|} \hline
$SO(5)\cro SO(3)$  & (5,1) & (4,2) & (1,3) \\
\hline
$\ell_0$ & $1$ & $\ft32$ & $2$
\\ \hline
\end{tabular}
\label{mult5}
\ea
{}From this one may read off the physical spectrum around the origin
\ba
\begin{tabular}{|c||c|c|c|} \hline
fields & $H_0$ &
$(\ell_0,\bar{\ell}_0)$ & $m^2L^2$ \\
\hline\hline
scalars & $(5,1,5,1)$ & $(1,1)$ & $ 0 $
\\ \cline{2-4}
& $(4,2,4,2)$ & $(\ft32,\ft32)$ & $ 3 $
\\ \cline{2-4}
& $(1,3,1,3)$ & $(2,2)$ & $ 8 $
\\ \hline
fermions & $ (5,1,4,2) $ & $(1,\ft32)$ & $\ft94$
\\ \cline{2-4}
& $ (4,2,1,3)$ & $  (\ft32,2) $ & $\ft{25}{4}$
\\ \hline
vectors & $ (5,1,1,3)$ & $ (1,2) $ & $4$
\\ \cline{2-4}
& $ (1,3,5,1)$ & $ (2, 1) $ & $4$
\\ \hline
\end{tabular}
\label{spec5}
\ea
This again agrees with the spectrum computed from \Ref{A1A3},
\Ref{massmatrix2}, \Ref{vectormass2}.

\mathon
\subsection{$SO(4,4)\times SO(4,4)$}
\mathoff

{}From \Ref{A1Ipq} it follows that for this gauge group both tensors
$A_1$ and $A_2$ vanish at the origin. The theory hence possesses a
maximally supersymmetric Minkowski
vacuum.\footnote{In~\cite{NicSam01a}, this vacuum had been
misidentified as an AdS stationary point.} From \Ref{A1A3},
\Ref{massmatrix2}, \Ref{vectormass2}, we find the spectrum 
\ba
\begin{tabular}{|c||c|c|} \hline
fields & $\#$ & $m^2$ \\
\hline\hline
scalars & $96$ & $ 4g^2 $
\\ \hline
fermions & $128$ & $ 4g^2 $
\\ \hline
vectors & $32$ & $4g^2$
\\ \hline
\end{tabular}
\label{spec4}
\ea

\section{Maximally supersymmetric vacua for exceptional gaugings}

One notable peculiarity of three-dimensional maximal gauged
supergravity is the possibility to have exceptional gauge
groups. Thus, the question naturally arises whether these gaugings
which do not have any higher-dimensional counterparts also possess
nontrivial extremal structures. Again, all of these theories admit a
maximally supersymmetric AdS vacuum at the scalar origin $\cV=I$. If
we gauge the full $E_{8(+8)}$, the scalar potential reduces to a
cosmological constant. The smaller we choose the gauge group the
richer we expect the structure of the potential to become, and
therefore, it is particularly interesting to study the second-simplest
case, $E_{7(+7)}\times SL(2)$, as well as the most compact form of the
smallest exceptional gauge group, $G_2\times F_{4(-20)}$. We will
discuss the embedding of these two gauge groups and their maximally
supersymmetric vacua in some detail. For all the other possible
exceptional gauge groups~\cite{NicSam01a}, we simply list the
supermultiplet structures of the physical field content around the
maximally supersymmetric AdS vacuum.

\mathon
\subsection{$G_{2}\times F_{4(-20)}$}
\mathoff

For this gauge group, the embedding tensor assumes a rather simple
form. The relevant $SO(16)$ representations decompose as follows under
the subgroup $G_2\times SO(9)$
\ba
{\bf 16}_v &\longrightarrow& ({\bf 7}, {\bf 1}) + ({\bf 1}, {\bf 9})
\non 
{\bf 120} &\longrightarrow& ({\bf 14}, {\bf 1}) + ({\bf 1}, {\bf 36}) 
          + ({\bf 7}, {\bf 1}) +   ({\bf 7}, {\bf 9}) \non
{\bf 128}_s &\longrightarrow& ({\bf 1}, {\bf 16}) + ({\bf 7}, {\bf 16})
\;.
\ea
Accordingly, we split the $SO(16)$ vector indices as $I=(i,\Hj)$. The
embedding tensor then reads
\ba
\GTh_{ij|kl} &=& 12 {P_{ij}}^{kl} \equiv 8 \Gd_{ij}^{kl} + 2 C_{ijkl}
\;,
\non 
\GTh_{\Hi\Hj |\Hk\Hl} &=& - 8 \Gd_{\Hi\Hj}^{\Hk\Hl}  \;, \non
\GTh_{\Ga\Gb} &=& - \Gd_{\Ga\Gb}
\;,
\ea
with all other components zero. Here $P$ is the projector onto the $G_2$ 
subgroup of $SO(7)$, with $C_{ijkl}$ the $G_2$ invariant tensor made
out of the octonionic structure constants \cite{GunNic95}, obeying
\be
C_{ijmn} C_{mnkl} = 8 \Gd_{ij}^{kl} - 2 C_{ijkl}
\;.
\ee
We see that the ratio of coupling constants is indeed
$(-3/2)$. Furthermore, $\GTh$ can be brought into the form \Ref{Theta}
with
\be
\Gth = -1 \; , \quad
\Xi_{IJ} = {\rm diag} \, ( 9 \Gd_{ij}, -7 \Gd_{\Hi\Hj}) \; , \quad
\GTh_{AB} = - \Gd_{AB} + C_{AB} \;,
\ee
with the $G_2$ invariant tensor
\be
C_{AB} =  \Gd_{\Ga\Gb}\, {\rm diag} \, ( -7,\Gd_{ij}) \;,
\ee 
showing that this indeed gives a solution of the projection
condition~\Ref{27000}. At the origin $\cV=I$, the gauge group is
broken down to its maximally compact subgroup $G_2\times
SO(9)$. Together with the 16 supercharges and the $AdS_3$ group
$SO(2,2)$ this combines into the background isometry group
$\BGI=G(3)_L\cro O\!Sp(9|2,{\mathbb R})_R$, i.e.\ the supersymmetries
split as $N=(7,9)$. The spectrum around this point is given by the
tensor product of two (left and right) supermultiplets of $G(3)$ and
$O\!Sp(9|2,{\mathbb R})$, respectively. Comparing to the masses
computed from \Ref{A1A3}, \Ref{massmatrix2}, \Ref{vectormass2}, we
identify their conformal weights as
\ba
\begin{tabular}{|c||c|c|c|} \hline
$(G_2)_L$  & & $7$ & $1$ \\
\hline
$\ell_0$ && $\ft34$ & $\ft54$ \\
\hline
{\large $\times$}\\
\hline
$SO(9)_R$ & $16$ & $16$ & \\
\hline
$\bar{\ell}_0$ &$\ft14$ & $\ft34$ &
\\ \hline
\end{tabular}
\label{multgf2}
\ea
With \Ref{MDelta}, this gives the correct supergravity masses, which
we do not write out explicitly here. Note that \Ref{multgf2}
consistently describes 16 massive selfdual vector fields,
corresponding to the noncompact directions in the gauge group.

\mathon
\subsection{$E_{7(7)}\times SL(2)$}
\mathoff

In terms of the $SO(8)\times SO(8)$ decomposition of section~2.1,
the embedding tensor of this gauge group is given by
\ba
\GTh_{ab|cd} &=& {(P_1)_{ab}}^{cd} \equiv \Gd_{ab}^{cd} \; , \non
\GTh_{ab|\bc\bd} &=& {(P_1)_{ab}}^{\bc\bd} \equiv \Gd_{ab}^{\bc\bd} 
\; , \non
\GTh_{\Bara\bb|cd} &=& {(P_1)_{\Bara\bb}}^{cd} \equiv 
\Gd_{\Bara\bb}^{cd} \; , \non
\GTh_{\Bara\bb|\bc\bd} &=& {(P_1)_{\Bara\bb}}^{\bc\bd} \equiv
       \Gd_{\Bara\bb}^{\bc\bd}   \non
\GTh_{a\bb|c\bd} &=& - \GTh_{\bb a|c\bd} ~=~
- \GTh_{a\bb|\bd c} ~=~
\GTh_{\bb a|\bd c} ~=~
{(P_1)_{a\bb}}^{c\bd} -  3{(P_2)_{a\bb}}^{c\bd} \non
\GTh_{\Ga\Gb|\Gg\Gd} &=& \Gd_{(\Ga}^\Gg \Gd_{\Gb)}^\Gd -
                         \ft12 \Gd_{\Ga\Gb} \Gd^{\Gg\Gd}   \non
\GTh_{\dGa\dGb|\dGg\dGd} &=& \Gd_{(\dGa}^\dGg \Gd_{\dGb)}^\dGd -
                         \ft12 \Gd_{\dGa\dGb} \Gd^{\dGg\dGd}
\ea
where $P_1$ and $P_2$ are projectors onto the $SU(8)$ and $U(1)$
subgroups of $SO(16)$, respectively, with
\ba
{(P_1)_{a\bb}}^{c\bd}  &:=& \Gd_{(a}^c \Gd_{b)}^d -
                         \ft18 \Gd_{ab} \Gd^{cd}     \non
{(P_2)_{a\bb}}^{c\bd} &:=& \ft18 \Gd_{ab} \Gd^{cd} \;.
\ea
We see that the relative coupling strength is indeed $(-3)$.
Furthermore the embedding tensor is invariant under triality
rotations interchanging ${\bf 35}_v \rightarrow {\bf 35}_s
\rightarrow {\bf 35}_c \rightarrow {\bf 35}_v$. The background
isometry group at $\cV=I$ is given by the $N=(16,0)$ supergroup
$\BGI=SU(8|1,1)_L\cro SU(1,1)_R$. The physical spectrum is described
by tensoring a (left) supermultiplet of $SU(8|1,1)$~\cite{GuSiTo86},
with a singlet on the right
\ba
\begin{tabular}{|c||c|c|c|c|c|} \hline
$SU(8)_L\cro U(1)_L$  & $70^0$ & $56^{+}+\overline{56}^{-}$ &
$28^{+2} + \overline{28}^{-2}$ & $8^{+3}+\overline{8}^{-3}$ &
$1^{+4}+1^{-4}$ \\
\hline
$\ell_0$ & $\ft12$ & $1$& $\ft32$ & $2$ & $\ft52$
\\ \hline
{\large $\times$}\\ \hline
$I_R$ & && $1$ &&  \\
\hline
$\bar{\ell}_0$ & && $\ft32$ &&
\\ \hline
\end{tabular}
\label{multe71}
\ea
The supergravity masses are again obtained from \Ref{MDelta}. The
tensor product~\Ref{multe71} consistently includes 70 massive selfdual
vector fields ($\Delta=(\ft12,\ft32)$) corresponding to the noncompact
directions of $E_{7(7)}$ and 2 massive selfdual vector fields of
opposite spin ($\Delta=(\ft52,\ft32)$) associated with the noncompact
directions of $SL(2)$.

\subsection{Spectra of the other exceptional gaugings}

Here we list the physical mass spectra around the maximally
supersymmetric vacuum for the remaining exceptional gaugings. They
again factor into tensor products under the two factors of the
background isometry group $\BGI=\BGI_L\times\BGI_R$. For simplicity,
we restrict to giving the conformal dimensions $\ell_0$,
$\bar{\ell}_0$ for the states in these factors, from which
three-dimensional spins and masses may be extracted via
$s=|\ell_0-\bar{\ell}_0|$, $\Delta=\ell_0+\bar{\ell}_0$, and
\Ref{MDelta}.

\begin{itemize}

\item
\mathon 
$G_{2(2)}\times F_{4(4)}$ :\qquad
\mathoff
$\BGI=D^1(2,1;-\ft23)_L \times O\!Sp(4^*|6)_R$
\ba
\begin{tabular}{|c||c|c|c|c|} \hline
$SU(2)_L\cro SU(2)_L$  & & $(1,2)$ & $(2,1)$ & \\
\hline
$\ell_0$ && $\ft32$ & $2$ & \\
\hline
{\large $\times$}\\
\hline
$SU(2)_R\cro U\!Sp(6)_R$ & $(1,14)$ & $(2,14')$ & $(3,6)$ & $(4,1)$ \\
\hline
$\bar{\ell}_0$ &$1$ & $\ft32$ & $2$ & $\ft52$
\\ \hline
\end{tabular}
\label{multgf1}
\ea

\item
\mathon 
$E_{6(6)}\times SL(3)$ :\qquad
\mathoff
$\BGI=O\!Sp(4^*|8)_L\times SU(1,1)_R$
\ba
\begin{tabular}{|c||c|c|c|c|c|} \hline
$SU(2)_L\cro U\!Sp(8)_L$  & $(1,42)$& $(2,48)$ & $(3,27)$ & $(4,8)$ &
$(5,1)$ \\
\hline
$\ell_0$ &$1$& $\ft32$ & $2$ & $\ft52$ & $3$
\\ \hline
{\large $\times$}\\ \hline
$I_R$ & && $1$ &&  \\
\hline
$\bar{\ell}_0$ & && $2$ &&
\\ \hline
\end{tabular}
\label{multe61}
\ea

\item
\mathon 
$E_{6(2)}\times SU(2,1)$ :\qquad
\mathoff
$\BGI=SU(6|1,1)_L \times D^1(2,1;-\ft12)_R $
\ba
\begin{tabular}{|c||c|c|c|c|} \hline
$SU(6)_L\cro U(1)_L$  & $20^0$ & $15^+ + \overline{15}^-$ &
$6^{+2}+\overline{6}^{-2}$ & $1^{+3}+1^{-3}$ \\
\hline
$\ell_0$ & $\ft12$ & $1$ & $\ft32$ & $2$  \\
\hline
{\large $\times$}\\
\hline
$SU(2)_R\cro SU(2)_R$ &  & $(2,1)$ & $(1,2)$ &  \\
\hline
$\bar{\ell}_0$ & & $1$ & $\ft32$ &
\\ \hline
\end{tabular}
\label{multe62}
\ea

\item
\mathon 
$E_{6(-14)}\times SU(3)$ :\qquad
\mathoff
$\BGI=O\!Sp(10|2,{\mathbb R})_L\times SU(3|1,1)_R$
\ba
\begin{tabular}{|c||c|c|c|} \hline
$SO(10)_L $  & $16$ & $\overline{16}$ &
\\
\hline
$\ell_0$   & $\ft14$ & $\ft34$ &  \\
\hline
{\large $\times$}\\
\hline
$SU(3)_R\cro U(1)_R$ &  & $3^+ +\overline{3}{}^-$ & $1^{+2}+1^{-2}$ \\
\hline
$\bar{\ell}_0$ & & $\ft34$ & $\ft54$
\\ \hline
\end{tabular}
\label{multe63}
\ea

\item
\mathon 
$E_{7(-5)}\times SU(2)$ :\qquad
\mathoff
$\BGI=O\!Sp(12|2,{\mathbb R})_L \times D^1(2,1;-\ft13)_R$
\ba
\begin{tabular}{|c||c|c|c|} \hline
$SO(12)_L$  & $32$ & $\overline{32}$ & \\
\hline
$\ell_0$ & $\ft14$ & $\ft34$ & \\
\hline
{\large $\times$}\\
\hline
$SU(2)_R\cro SU(2)_R$ &  & $(2,1)$ & $(1,2)$   \\
\hline
$\bar{\ell}_0$ & & $\ft34$ & $\ft54$
\\ \hline
\end{tabular}
\label{multe72}
\ea

\item
\mathon 
$E_{8(8)}$ :\qquad
\mathoff
$\BGI=O\!Sp(16|2,{\mathbb R})_L\times SU(1,1)_R$

This gauging is special in that the scalar potential becomes trivial
(a negative cosmological constant), and all scalar fields are absorbed
into vector fields. The theory can thus be considered as a maximally
supersymmetric $SO(16)$ CS theory coupled to 128 massive selfdual
vectors and 128 spin$-\ft12$ fields. The spectrum is simply
obtained from tensoring the singleton multiplet of $O\!Sp(16,2|\bbR)$
with a singlet on the right:
\ba
\begin{tabular}{|c||c|c|c|} \hline
$SO(16)_L$  & $128$ & $\overline{128}$ & \\
\hline
$\ell_0$ & $\ft14$ & $\ft34$  &
\\ \hline
{\large $\times$}\\ \hline
$I_R$ & && $1$  \\
\hline
$\bar{\ell}_0$ & && $\ft54$
\\ \hline
\end{tabular}
\label{multe8}
\ea

\end{itemize}

\section{Vacua without maximal supersymmetry}

We here do not aim at an exhaustive classification of non-trivial
stationary points, but rather would like to discuss and to illustrate
some salient features, in particular those that have no analogs in
higher-dimensional gauged supergravities. For more details, especially
concerning the computational aspects, readers are referred to a
forthcoming article by one of the authors \cite{tf1}.

\subsection{Vacua in higher dimensional noncompact gaugings}
\la{higherD}

Let us first briefly recall the known structure of extrema in higher
dimensional maximal noncompact gaugings. In $D\equ4$, the noncompact
$SO(p,q)$ gaugings ($p\pls q=8$) were originally obtained in
\cite{Hull84a,Hull84b} by analytic continuation of the compact gauged
theory~\cite{deWNic82}. The structure of the scalar potential is
similar to \Ref{potential}:
\ba
V&=& -g^2 \left(
\ft34 A_1^{ij}A_1^{ij} - \ft1{24} A_2^i{}_{jkl}A_2^i{}_{jkl}\right)
\;,
\ea
where the indices $i, j, \dots$ denote $SU(8)$ indices. The tensors
$A_1$ and $A_2$ are functions on the coset manifold $E_{7(7)}/SU(8)$
and transform in the ${\bf 36}$ and ${\bf 420}$ of $SU(8)$,
respectively. Together, they form the ${\bf 912}$ of $E_{7(7)}$.

At the origin $\cV=I$, the gauge group is broken down to its maximally
compact subgroup $H_0=SO(p)\times SO(q)$. But unlike in three
dimensions, this point is {\it not} a stationary point in the
noncompact gaugings. This is because, except for the compact gauged
theory, the tensor $A_2^i{}_{jkl}$ does not vanish at $\cV=I$, as may
be anticipated from the fact that the ${\bf 420}$ contains singlets
under $H_0=SO(p)\times SO(q)$ unless $p=0$. Vanishing $A_2^i{}_{jkl}$
would imply stationarity and maximal supersymmetry, but this would be
incompatible with the non-existence of proper superextensions of the
four-dimensional AdS group, i.e.\ simple supergroups containing
$SO(3,2) \times \big(SO(p) \times SO(8-p)\big)$ as maximal bosonic
subgroup for $p\neq 0$ \cite{Kac77,Nahm78}. Recall that, by contrast,
in three dimensions no singlets appear in the decomposition of $A_2$
under $H_0$, and this was sufficient to imply the existence of
maximally supersymmetric stationary points.

The search for stationary points of the noncompact potentials has been
pursued in~\cite{HulWar85b} by restricting the potential to singlets
under certain subgroups of the gauge group. Summarizing their results,
there is no stationary point of the $SO(7,1)$ gauged theory which
leaves the $G_2\subset SO(7)$ invariant, and no stationary point with
at least $SU(3)$ invariance in the $SO(6,2)$ gauged theory. The
potential of the $SO(5,3)$ gauged theory on the other hand does
exhibit a stationary point away from the origin with $SO(5)\times
SO(3)$ residual symmetry \cite{Hull85}. It is found by computing the
potential in the truncation to the only singlet under $SO(5)\times
SO(3)$ and has a positive cosmological constant. Similarly, it has
been found, that the $SO(4,4)$ gauged theory admits a dS vacuum with
remaining $SO(4)\times SO(4)$ symmetry. Both these dS points have been
shown to be unstable in the sense that they admit tachyonic scalar
fluctuations with $V''=-2V$~\cite{AhnWoo01,KLPS02}.

In five dimensions the potential for the $SO(p,q)$ ($p\pls q=6$)
gauged theory is given by~\cite{GuRoWa86}
\ba
V&=& -g^2 \left(
\ft6{45^2} A_1^{ab}A_1^{ab} - \ft1{96} A_{2}^{abcd}
A_{2}^{abcd}\right) \;,
\ea
where the indices $a, b, \dots$ now denote $USp(8)$ indices. The
tensors $A_1$ and $A_2$ transform in the ${\bf 36}$ and ${\bf 315}$ of
$USp(8)$, respectively, and together combine into the ${\bf 351}$ of
$E_{6(6)}$.

Again, the scalar origin $\cV=I$ is not a stationary point in the
noncompact gaugings --- the ${\bf 315}$ under $H_0=SO(p)\times SO(q)$
contains singlets unless $p=0$. The existence of critical points in
the $SO(5,1)$ and $SO(4,2)$ gauged theories which preserve at least an
$SO(5)$ and $SO(4)\times SO(2)$ subgroup, respectively, has been
excluded in \cite{GuRoWa86}. A stationary point with positive
cosmological constant, again for $\cV\neq I$, has been identified 
in the $SO(3,3)$ gauged theory. Presumably it, too, is unstable.

\mathon
\subsection{$SO(8)\times SO(8)$}
\mathoff

Several stationary points breaking the diagonal of this group down to a
group containing $SU(3)$ have been presented in \cite{Fisc02}. All of
these correspond to known stationary points of $D=4, N=8$
supergravity. Here, we want to complete this list by also giving
analogs of $D=4$ stationary points breaking $SO(8)$ down to $SO(7)^-$
and $G_2$.

In what follows, we will designate by $SO(7)^+$ the subgroup of
$SO(8)$ stabilizing the spinor $\psi_\alpha=\delta_{\alpha8}$, and by
$SO(7)^-$ the subgroup stabilizing the co-spinor
$\phi_{\dot\alpha}=\delta_{\dot\alpha\dot8}$. With the conventions of
appendix A, their intersection $G_2$ will also stabilize the vector
$v^i=\delta^{i8}$.  Those generators of $E_8$ which are invariant
under $(G_2)_{\rm diag}$ form an $SL(2)\times SL(2)$ subalgebra (where
one of these $SL(2)$'s is just the $SL(2)$ from $E_{7(+7)}\times
SL(2)\subset E_{8(+8)}$ which will show up whenever we form a diagonal
$SO(p,8-p)$). Hence we parametrize the four-dimensional manifold of
$(G_2)_{\rm diag}$ singlets in the coset $E_8/SO(16)$ by
\be
\mathcal{V}=\exp\left(v V\right)\exp\left(s S\right)
\exp\left(-v V\right)\;\exp\left(w W\right)\exp\left(z Z\right)
\exp\left(-w W\right) \;,
\ee
where the generators $V$, $S$ respectively $W$, $Z$ corresponding to
one compact and one noncompact generator of each $SL(2)$. Using the
same decomposition as in \Ref{SO16decomp}, these generators read
explicitly
\be\label{vswz}
\begin{array}{lclclcl}
V^{\mathcal C}{}_{\mathcal B}&=&
\left(2\delta^{a8}\delta^{\bar b8}-\frac{1}{4}\delta^{a\bar b}\right)
\;f_{\underline{[a\bar b]}\mathcal{B}}{}^{\mathcal C}&\quad&
S^{\mathcal C}{}_{\mathcal B}&=&\left(2\delta^{\alpha8}
\delta^{\beta8}-\frac{1}{4}\delta^{\alpha\beta}\right)\;
f_{\alpha\beta\mathcal{B}}{}^{\mathcal C}\\
W^{\mathcal C}{}_{\mathcal B}&=&\frac{1}{4}\delta^{a\bar b}\;
f_{\underline{[a\bar b]}\mathcal{B}}{}^{\mathcal C}&\quad&
Z^{\mathcal C}{}_{\mathcal B}&=&\frac{1}{4}\delta^{\alpha\beta}\;
f_{\alpha\beta\mathcal{B}}{}^{\mathcal C}\;.
\end{array}
\ee
Using the $SO(8)\times SO(8)$ embedding tensor $(\ref{thetatensor})$,
the corresponding potential reads
{\small
\be
\begin{array}{lcl}
-8g^{-2}V&=&\frac{243}{8}+\frac{7}{2}\,\cosh(2\,s)+\frac{49}{8}\,\cosh(4\,s)+\frac{1141}{64}\,\cosh(s)\,\cosh(z)\\
&&+\frac{427}{64}\,\cosh(3\,s)\,\cosh(z)-\frac{7}{64}\,\cosh(5\,s)\,\cosh(z)-\frac{25}{64}\,\cosh(7\,s)\,\cosh(z)\\
&&+\frac{21}{8}\,\cos(4\,v)-\frac{7}{2}\,\cos(4\,v)\,\cosh(2\,s)+\frac{7}{8}\,\cos(4\,v)\,\cosh(4\,s)\\
&&-\frac{21}{64}\,\cos(4\,v)\,\cosh(s)\,\cosh(z)+\frac{21}{64}\,\cos(4\,v)\,\cosh(3\,s)\,\cosh(z)\\
&&+\frac{7}{64}\,\cos(4\,v)\,\cosh(5\,s)\,\cosh(z)-\frac{7}{64}\,\cos(4\,v)\,\cosh(7\,s)\,\cosh(z)\\
&&-\frac{1645}{128}\,\cos(v-w)\,\sinh(z)\,\sinh(s)+\frac{651}{128}\,\cos(v-w)\,\sinh(z)\,\sinh(3\,s)\\
&&+\frac{7}{128}\,\cos(v-w)\,\sinh(z)\,\sinh(5\,s)-\frac{49}{128}\,\cos(v-w)\,\sinh(z)\,\sinh(7\,s)\\
&&-\frac{315}{64}\,\cos(3\,v+w)\,\sinh(z)\,\sinh(s)+\frac{133}{64}\,\cos(3\,v+w)\,\sinh(z)\,\sinh(3\,s)\\
&&-\frac{7}{64}\,\cos(3\,v+w)\,\sinh(z)\,\sinh(5\,s)-\frac{7}{64}\,\cos(3\,v+w)\,\sinh(z)\,\sinh(7\,s)\\
&&+\frac{35}{128}\,\cos(7\,v+w)\,\sinh(z)\,\sinh(s)-\frac{21}{128}\,\cos(7\,v+w)\,\sinh(z)\,\sinh(3\,s)\\
&&+\frac{7}{128}\,\cos(7\,v+w)\,\sinh(z)\,\sinh(5\,s)-\frac{1}{128}\,\cos(7\,v+w)\,\sinh(z)\,\sinh(7\,s)
\end{array}
\label{g2so8}
\ee
}

Setting $v=0$, this reduces to the potential on the three-dimensional
manifold of $SO(7)^+_{\rm diag}$ singlets 
{\small
\be\label{so7pso8}
\begin{array}{lcl}
-8g^{-2}V&=&33+7\,\cosh(4\,s)+\frac{35}{2}\,\cosh(s)\,\cosh(z)+7\,\cosh(3\,s)\,\cosh(z)\\
&&-\frac{1}{2}\,\cosh(7\,s)\,\cosh(z)-\frac{35}{2}\,\cos(w)\,\sinh(z)\,\sinh(s)\\
&&+7\,\cos(w)\,\sinh(z)\,\sinh(3\,s)-\frac{1}{2}\,\cos(w)\,\sinh(z)\,\sinh(7\,s)
\end{array}
\ee } 
\noindent
whose only nontrivial stationary point is located at $w=\pi$,
$s=-z=\ft12\,\arccosh\,2$, with remaining $SO(7)^+\times SO(7)^+$
invariance and completely broken
supersymmetry~\cite{Fisc02}.\footnote{Whenever we give coordinates for
stationary points, we list only one representative and do not consider
trivial sign flips.} The value of the cosmological constant at this
vacuum is $\Lambda=-50g^2$. Recalling that the central charge of the
associated conformal algebra on the boundary goes proportional in
$\sqrt{1/\Lambda}$, \cite{BroHen86,HenSke98}, we find from \Ref{LSOpq}
that
\ba
\frac{c_{SO(7)}}{c_{SO(8)}} &=& 
\sqrt{\frac{\GL_{SO(8)}}{\GL_{SO(7)}}} ~=~ \frac45 \;,
\ea
i.e.\ a {\em rational} value for the ratio of central charges of the
boundary theories associated with the different vacua. The scalar
masses at this extremum are computed with \Ref{massmatrix1} and give
\ba
\begin{tabular}{|c||c|c|c|}\hline
$SO(7)^+\cro SO(7)^+$ & $(1,1)$ & $(8,8)$ & $(7,7)$ \\ \hline
$m^2L^2$ & $\ft{96}{25}$ & $-\ft{9}{25}$ & $-\ft{24}{25}$ \\ \hline
\end{tabular}
\la{specN0}
\ea
in units of the inverse AdS length $L$, together with $14$ Goldstone
scalars. The full mass spectrum is collected in \Ref{E-p8-so7}. In
particular, this vacuum despite being non-supersymmetric and in
contrast to its higher-dimensional analoga is {\em stable} in the
sense that all scalar fields satisfy the Breitenlohner-Freedman
bound~\Ref{BF}.\footnote{Several nonsupersymmetric stable AdS vacua in
the three-dimensional half-maximal ($N=8$) gauged supergravities have
been found in~\cite{BerSam02}.} Moreover, their associated conformal
dimensions computed from
\Ref{MDelta}, \Ref{specN0} are all rational. 

The corresponding potential on the manifold of $SO(7)^-_{\rm diag}$
singlets is obtained by replacing the generator $S$ by $\tilde
S^{\mathcal C}{}_{\mathcal
B}=\left(2\delta^{\dot\alpha\dot8}\delta^{\dot\beta\dot8}-
\frac{1}{4}\delta^{\dot\alpha\dot\beta}\right)\;
f_{\dot\alpha\dot\beta\mathcal{B}}{}^{\mathcal C}$ and reads {\small
\be\label{so7mso8}
\begin{array}{lcl}
-8g^{-2}V&=&33+7\cosh(4\,s)+\frac{35}{2}\,\cosh(s)\,\cosh(z)+7\,\cosh(3\,s)\,\cosh(z)\\
&&-\frac{1}{2}\,\cosh(7\,s)\,\cosh(z)+\frac{35}{2}\,\sinh(z)\,\sin(w)\,\sinh(s)\\
&&-7\,\sinh(z)\,\sin(w)\,\sinh(3\,s)+\frac{1}{2}\,\sinh(z)\,\sin(w)\,\sinh(7\,s)
\end{array}
\ee
} 
\noindent
which differs from the previous one only by a rotation in $w$. This
means that in contrast to $N=8$, $D=4$, the other $SO(7)$ stationary
point, here at $w=-\pi/2$, $s=z=\ft12\,\arccosh\,2$, with
$SO(7)^-\times SO(7)^-$ symmetry, has the same value of the
cosmological constant, and the same mass spectrum.

Although $(\ref{g2so8})$ is too complicated for a detailed analytic
treatment, it is nevertheless possible to extract information about
the location of further extrema either numerically\footnote{Note that
$\partial_v V=0$ can readily be solved for $v$ in terms of $w, z, s$,
thereby considerably reducing the complexity of this problem.} or by
educated inspection. This allows to identify a further stationary
point at $v=w=\frac{1}{4}\,\pi$,
$s=z=\frac{1}{2}\,\arccosh\,\ft{7}{3}$ which breaks $SO(8)\times
SO(8)$ down to $G_2\times G_2$, preserving $N=(1,1)$ supersymmetry.
Again, the ratio of central charges associated with this vacuum and
the origin comes out to be rational
\ba
\frac{c_{G_2}}{c_{SO(8)}} &=& 
\sqrt{\frac{\GL_{SO(8)}}{\GL_{G_2}}} ~=~ \frac34 \;.
\ea
The scalar mass spectrum is given by 
\ba
\begin{tabular}{|c||c|c|c|c|}\hline
$G_2\cro G_2$ & $(1,1)$ & $(1,1)$ & $(7,7)$ & $(7,7)$ \\ \hline
$m^2L^2$ & $\ft{65}{16}$ & $\ft9{16}$ & $-\ft{7}{16}$ & 
$-\ft{15}{16}$ \\ \hline
\end{tabular}
\ea
together with $28$ Goldstone scalars. By further computation, one may
verify that the physical spectrum around this vacuum is organized in
terms of $N=(1,1)$ supermultiplets and the external $G_2\cro
G_2$. More precisely, the original chiral multiplet \Ref{mult8} breaks
into $N=1$ multiplets according to
\ba
\begin{tabular}{|c||c|c|}\hline
$G_2$ & $1$ & $7$  \\ \hline
$\ell_0$ & $\ft{13}{8} , \ft98$ & $\ft58 , \ft18$ \\ \hline
\end{tabular}
\la{specN1}
\ea
Tensoring a left and a right copy of \Ref{specN1} reproduces the full
physical spectrum at this vacuum given in \Ref{E-p8-g2}, in particular
the $14$ massive spin$-\ft32$ fields and $28$ massive vector fields
corresponding to broken super- and gauge symmetries.

Although numerical evidence suggests that there is no further
stationary point of $(\ref{g2so8})$ which is not equivalent to one of
those given here, a proof is still lacking. For completeness, we
include here the physical spectrum around the $N=(2,2)$ supersymmetric
vacuum with remaining $SU(3)\cro SU(3)\cro U(1)\cro U(1)$ symmetry,
found in~\cite{Fisc02}. This vacuum is located at
\be
\begin{array}{lcll}
\mathcal{V}^{\mathcal{C}}{}_{\mathcal{B}}&=&\exp\Biggl(&\frac{1}{8}\,\arccosh(3)\,\Bigl(f_{\alpha\beta\mathcal{B}}{}^{\mathcal{C}}\bigl(\Gamma_{\alpha\beta}^{1234}+\Gamma_{\alpha\beta}^{1256}+\Gamma_{\alpha\beta}^{1278}-\delta_{\alpha\beta}\bigr)\\
&&&-f_{\dot\alpha\dot\beta\mathcal{B}}{}^{\mathcal{C}}\bigl(\Gamma_{\dot\alpha\dot\beta}^{1357}-\Gamma_{\dot\alpha\dot\beta}^{1467}+\Gamma_{\dot\alpha\dot\beta}^{1458}+\Gamma_{\dot\alpha\dot\beta}^{1368}\bigr)\Bigr)\Biggr)\;.\\
\end{array}
\ee

The spectrum is organized in terms of
$N=(2,2)$ supermultiplets and the external $SU(3)\cro SU(3)$ (note
that $U(1)$ is the R-symmetry of the associated $N=2$ superconformal
algebra). The original chiral multiplet \Ref{mult8} breaks into $N=2$
multiplets according to
\ba
\begin{tabular}{|c||c|c|}\hline
$SU(3)$ & $1$ & $6$  \\ \hline
$\ell_0$ & $\ft{10}{6} , \ft76, \ft76, \ft46$ & $\ft46 , \ft16$ 
\\ \hline
\end{tabular}
\la{specN2}
\ea
from which one again reproduces the physical mass spectrum
\Ref{E-p8-su3} via \Ref{MDelta}.

\mathon
\subsection{$SO(7,1)\times SO(7,1)$}
\mathoff

In four dimensions, there is no stationary point with $G_2$ symmetry
in the theory with noncompact gauge group $SO(7,1)$. For comparison,
we will again compute the potential restricted to the four-dimensional
manifold of $(G_2)_{\rm diag}$ singlets. Since our conventions are
just such that $G_2$ leaves the last vector, spinor, and co-spinor
index invariant, this calculation parallels the $SO(8)\times SO(8)$
case, only with a different embedding tensor $\Theta$.

Here, the potential on the manifold of $G_{2,\rm diag}$ singlets is
{\small
\be
\begin{array}{lcl}
-8g^{-2}V&=&\frac{909}{32}-\frac{7}{8}\,\cosh(2\,s)-\frac{49}{32}\,\cosh(4\,s)+\frac{6461}{512}\,\cosh(s)\,\cosh(z)\\
&&-\frac{1001}{512}\,\cosh(3\,s)\,\cosh(z)-\frac{203}{512}\,\cosh(5\,s)\,\cosh(z)-\frac{137}{512}\,\cosh(7\,s)\,\cosh(z)\\
&&+\frac{21}{8}\,\cos(2\,v)+\frac{63}{32}\,\cos(4\,v)+\frac{7}{2}\,\cos(2\,v)\,\cosh(2\,s)\\&&
-\frac{49}{8}\,\cos(2\,v)\,\cosh(4\,s)-\frac{21}{8}\,\cos(4\,v)\,\cosh(2\,s)+\frac{21}{32}\,\cos(4\,v)\,\cosh(4\,s)\\
&&+\frac{5145}{1024}\,\cos(2\,v)\,\cosh(s)\,\cosh(z)-\frac{5229}{1024}\,\cos(2\,v)\,\cosh(3\,s)\,\cosh(z)\\
&&+\frac{217}{1024}\,\cos(2\,v)\,\cosh(5\,s)\,\cosh(z)-\frac{133}{1024}\,\cos(2\,v)\,\cosh(7\,s)\,\cosh(z)\\
&&-\frac{21}{512}\,\cos(4\,v)\,\cosh(s)\,\cosh(z)-\frac{63}{512}\,\cos(4\,v)\,\cosh(3\,s)\,\cosh(z)\\
&&+\frac{147}{512}\,\cos(4\,v)\,\cosh(5\,s)\,\cosh(z)-\frac{63}{512}\,\cos(4\,v)\,\cosh(7\,s)\,\cosh(z)\\
&&-\frac{105}{1024}\,\cos(6\,v)\,\cosh(s)\,\cosh(z)+\frac{189}{1024}\,\cos(6\,v)\,\cosh(3\,s)\,\cosh(z)\\
&&-\frac{105}{1024}\,\cos(6\,v)\,\cosh(5\,s)\,\cosh(z)+\frac{21}{1024}\,\cos(6\,v)\,\cosh(7\,s)\,\cosh(z)\\
&&-\frac{28987}{2048}\,\cos(v-w)\,\sinh(z)\,\sinh(s)-\frac{651}{512}\,\cos(v+w)\,\sinh(z)\,\sinh(s)\\
&&-\frac{2835}{2048}\,\cos(v-w)\,\sinh(z)\,\sinh(3\,s)-\frac{3255}{512}\,\cos(v+w)\,\sinh(z)\,\sinh(3\,s)\\
&&-\frac{623}{2048}\,\cos(v-w)\,\sinh(z)\,\sinh(5\,s)+\frac{21}{512}\,\cos(v+w)\,\sinh(z)\,\sinh(5\,s)\\
&&-\frac{343}{2048}\,\cos(v-w)\,\sinh(z)\,\sinh(7\,s)-\frac{63}{512}\,\cos(v+w)\,\sinh(z)\,\sinh(7\,s)\\
&&+\frac{1323}{1024}\,\cos(3\,v-w)\,\sinh(z)\,\sinh(s)-\frac{7875}{2048}\,\cos(3\,v+w)\,\sinh(z)\,\sinh(s)\\
&&-\frac{385}{1024}\,\cos(3\,v-w)\,\sinh(z)\,\sinh(3\,s)+\frac{2541}{2048}\,\cos(3\,v+w)\,\sinh(z)\,\sinh(3\,s)\\
&&+\frac{35}{1024}\,\cos(3\,v-w)\,\sinh(z)\,\sinh(5\,s)+\frac{609}{2048}\,\cos(3\,v+w)\,\sinh(z)\,\sinh(5\,s)\\
&&-\frac{49}{1024}\,\cos(3\,v-w)\,\sinh(z)\,\sinh(7\,s)-\frac{399}{2048}\,\cos(3\,v+w)\,\sinh(z)\,\sinh(7\,s)\\
&&+\frac{35}{2048}\,\cos(5\,v-w)\,\sinh(z)\,\sinh(s)+\frac{315}{1024}\,\cos(5\,v+w)\,\sinh(z)\,\sinh(s)\\
&&-\frac{189}{2048}\,\cos(5\,v-w)\,\sinh(z)\,\sinh(3\,s)+\frac{63}{1024}\,\cos(5\,v+w)\,\sinh(z)\,\sinh(3\,s)\\
&&+\frac{175}{2048}\,\cos(5\,v-w)\,\sinh(z)\,\sinh(5\,s)-\frac{189}{1024}\,\cos(5\,v+w)\,\sinh(z)\,\sinh(5\,s)\\
&&-\frac{49}{2048}\,\cos(5\,v-w)\,\sinh(z)\,\sinh(7\,s)+\frac{63}{1024}\,\cos(5\,v+w)\,\sinh(z)\,\sinh(7\,s)\\
&&+\frac{315}{2048}\,\cos(7\,v+w)\,\sinh(z)\,\sinh(s)-\frac{189}{2048}\,\cos(7\,v+w)\,\sinh(z)\,\sinh(3\,s)\\
&&+\frac{63}{2048}\,\cos(7\,v+w)\,\sinh(z)\,\sinh(5\,s)-\frac{9}{2048}\,\cos(7\,v+w)\,\sinh(z)\,\sinh(7\,s)
\end{array}
\ee
}
\noindent
which unfortunately is again too complicated for a detailed analytic
treatment. Again, by using numerical guidance we find a nontrivial AdS
extremum located at $v=w=-\frac{\pi}{2}$,
$s=z=\frac{1}{2}\,\arccosh\,2$, with a remaining symmetry of
$G_2\times G_2$ and mass spectrum collected in~\Ref{E-p7-g2}. The
cosmological constant takes the value $\Lambda=-211g^2/8$. Note
that also this vacuum is {\em stable} although it does not preserve
any supersymmetry. Further stationary points of this potential might
exist. Upon restriction to $SO(7)_{\rm diag}$ singlets, we obtain
\be
\begin{array}{lcl}
-8g^{-2}V&=&33-7\cosh(4\,s)+\frac{35}{2}\,\cosh(s)\,\cosh(z)-7\,\cosh(3\,s)\,\cosh(z)\\
&&-\frac{1}{2}\,\cosh(7\,s)\,\cosh(z)-\frac{35}{2}\,\cos(w)\,\sinh(z)\,\sinh(s)\\
&&-7\,\cos(w)\,\sinh(z)\,\sinh(3\,s)-\frac{1}{2}\,\cos(w)\,\sinh(z)\,\sinh(7\,s)
\end{array}
\ee
which does not possess a nontrivial stationary point.

\mathon
\subsection{$SO(6,2)\times SO(6,2)$}
\mathoff

The four-dimensional theory with gauge group $SO(6,2)$ has no
stationary point with remaining $SU(3)$ symmetry. In three dimensions,
there are twelve $SU(3)_{\rm diag}$ singlets among the 128 scalars,
seven $SO(6)_{\rm diag}$ singlets, and five singlets under
$(SO(6)\times SO(2))_{\rm diag}$, so we consider breaking the gauge
group down to $SO(6)_{\rm diag}$ here. The seven singlets come from
the noncompact directions of $SL(3)\times SL(2)$ commuting with
$SO(6)$ whose generators $p_{1\ldots8},q_{1\ldots3}$ are given by
\ben
\begin{array}{lclclcl}
p_1{}^{\mathcal{C}}{}_{\mathcal{B}}&=&\multicolumn{5}{l}{\frac{1}{2}\left(\delta^i_1\delta^j_7+\delta^i_2\delta^j_8-\delta^i_3\delta^j_6+\delta^i_4\delta^j_5\right)\left(\delta_i^I\delta_j^{J-8}-\delta_i^{J-8}\delta_j^{I}+\delta_i^{I-8}\delta_j^J-\delta_i^{J}\delta_j^{I-8}\right)f_{\underline{[IJ]}\mathcal{B}}{}^{\mathcal{C}}}\\
p_2{}^{\mathcal{C}}{}_{\mathcal{B}}&=&\multicolumn{5}{l}{\frac{1}{2}\delta^{ij}\left(\delta_i^{I-8}\delta_j^J-\delta_i^I\delta_j^{J-8}\right)f_{\underline{[IJ]}\mathcal{B}}{}^{\mathcal{C}}}\\
p_3{}^{\mathcal{C}}{}_{\mathcal{B}}&=&\multicolumn{5}{l}{-\frac{1}{2}\left(\delta^i_1\delta^j_7+\delta^i_2\delta^j_8-\delta^i_3\delta^j_6+\delta^i_4\delta^j_5\right)\left(\delta_i^I\delta_j^J-\delta_i^J\delta_j^I-\delta_i^{I-8}\delta_j^{J-8}+\delta_i^{J-8}\delta_j^{I-8}\right)f_{\underline{[IJ]}\mathcal{B}}{}^{\mathcal{C}}}\\
p_4{}^{\mathcal{C}}{}_{\mathcal{B}}&=&\multicolumn{5}{l}{\frac{1}{2}\left(\delta^{\dot\gamma}_1\delta^{\dot\delta}_2-\delta^{\dot\gamma}_3\delta^{\dot\delta}_4+\delta^{\dot\gamma}_5\delta^{\dot\delta}_6+\delta^{\dot\gamma}_7\delta^{\dot\delta}_8\right)\left(\delta^{\dot\alpha}_{\dot\gamma}\delta^{\dot\beta}_{\dot\delta}-\delta^{\dot\beta}_{\dot\gamma}\delta^{\dot\alpha}_{\dot\delta}\right)f_{\dot\alpha\dot\beta\mathcal{B}}{}^{\mathcal{C}}}\\
p_5{}^{\mathcal{C}}{}_{\mathcal{B}}&=&\frac{1}{2}\,\delta^{\dot\alpha\dot\beta}\,f_{\dot\alpha\dot\beta\mathcal{B}}{}^{\mathcal{C}}&\qquad&
p_6{}^{\mathcal{C}}{}_{\mathcal{B}}&=&\left(\delta^\alpha_7\delta^\beta_8-\delta^\alpha_8\delta^\beta_7\right)f_{\alpha\beta\mathcal{B}}{}^{\mathcal{C}}\\
p_7{}^{\mathcal{C}}{}_{\mathcal{B}}&=&\frac{1}{2}\,\delta^{\alpha\beta}\,f_{\alpha\beta\mathcal{B}}{}^{\mathcal{C}}&\qquad&
p_8{}^{\mathcal{C}}{}_{\mathcal{B}}&=&p_7{}^{\mathcal{C}}{}_{\mathcal{B}}-2\left(\delta^\alpha_7\delta^\beta_7+\delta^\alpha_8\delta^\beta_8\right)f_{\alpha\beta\mathcal{B}}{}^{\mathcal{C}}\\
q_1{}^{\mathcal{C}}{}_{\mathcal{B}}&=&\multicolumn{5}{l}{\frac{1}{4}\left(\delta^i_1\delta^j_7+\delta^i_2\delta^j_8-\delta^i_3\delta^j_6+\delta^i_4\delta^j_5\right)\left(\delta_i^I\delta_j^J-\delta_i^J\delta_j^I+\delta_i^{I-8}\delta_j^{J-8}-\delta_i^{J-8}\delta_j^{I-8}\right)f_{\underline{[IJ]}\mathcal{B}}{}^{\mathcal{C}}}\\
q_2{}^{\mathcal{C}}{}_{\mathcal{B}}&=&\frac{1}{2}\left(\delta^\alpha_7\delta^\beta_7-\delta^\alpha_8\delta^\beta_8\right)f_{\alpha\beta\mathcal{B}}{}^{\mathcal{C}}&\qquad&
q_3{}^{\mathcal{C}}{}_{\mathcal{B}}&=&\frac{1}{2}\left(\delta^\alpha_7\delta^\beta_8+\delta^\alpha_8\delta^\beta_7\right)f_{\alpha\beta\mathcal{B}}{}^{\mathcal{C}}.
\end{array}
\een
The $p$-, resp. $q$-generators stand in one-to-one correspondence to
the following matrices that satisfy the same commutation relations:
{
\ben
\begin{array}{lclclclclcl}
\tilde p_1&=&\left(\begin{array}{ccc}0&0&0\\0&0&1\\0&-1&0\end{array}\right)&\quad&\tilde p_2&=&\left(\begin{array}{ccc}0&1&0\\-1&0&0\\0&0&0\end{array}\right)&\quad&\tilde p_3&=&\left(\begin{array}{ccc}0&0&-1\\0&0&0\\1&0&0\end{array}\right)\\
\tilde p_4&=&\left(\begin{array}{ccc}0&0&0\\0&0&1\\0&1&0\end{array}\right)&\quad&\tilde p_5&=&\left(\begin{array}{ccc}0&1&0\\1&0&0\\0&0&0\end{array}\right)&\quad&\tilde p_6&=&\left(\begin{array}{ccc}0&0&-1\\0&0&0\\-1&0&0\end{array}\right)\\
\tilde p_7&=&\left(\begin{array}{ccc}-1&0&0\\0&1&0\\0&0&0\end{array}\right)&\quad&\tilde p_8&=&\left(\begin{array}{ccc}1&0&0\\0&1&0\\0&0&-2\end{array}\right)&&&&\\
\tilde q_1&=&\left(\begin{array}{cc}0&1/2\\-1/2&0\end{array}\right)&\quad&
\tilde q_2&=&\left(\begin{array}{cc}1/2&0\\0&-1/2\end{array}\right)&\quad&
\tilde q_3&=&\left(\begin{array}{cc}0&1/2\\1/2&0\end{array}\right)
\end{array}
\een
}

Since acting with $SO(3)$ on traceless diagonal matrices gives all
traceless symmetric matrices, we parametrize $SL(3)\times SL(2)$ by
\be\label{sl3xsl2param}
\begin{array}{lcl}
\mathcal{V}&=&\exp\left({r_1\,p_1}\right)\,\exp\left({r_2\,p_3}\right)\,\exp\left({r_3\,p_2}\right)\\
&&\,\exp\left({zp_8-sp_7}\right)\,\exp\left({-r_3\,p_2}\right)\\
&&\exp\left({-r_2\,p_3}\right)\,\exp\left({-r_1\,p_1}\right)\\
&&\exp\left({r_5\,q_1}\right)\,\exp\left({v\,q_2}\right)\,\exp\left({-r_5\,q_1}\right)
\;,
\end{array}
\ee
and obtain the potential given in $(\ref{so62phi})$, which is
independent of $r_5$.  Since the intersection of the $SL(3)$ algebra
and the gauge group algebra is one-dimensional, this potential does
possess one trivial flat direction, aside from which we were not able
to find any further stationary points numerically.

It is tempting to reuse the parametrization $(\ref{sl3xsl2param})$ to
compute the potential of the $SO(8)\times SO(8)$ and $SO(7,1)\times
SO(7,1)$ gauged theories on the manifold of $SO(6)_{\rm diag}$
singlets, since one only has to repeat the calculation with a
different embedding tensor. The corresponding results are collected in
the appendix. For $SO(8)\times SO(8)$, we obtain $(\ref{so88sl3})$,
while for $SO(7,1)\times SO(7,1)$ the potential is given in
$(\ref{so71sl3})$.

\mathon
\subsection{$SO(5,3)\times SO(5,3)$}
\mathoff

We have seen that in four and five dimensions there is a de Sitter
vacuum in the theories with gauge groups $SO(5,3)$ and $SO(3,3)$,
respectively, which completely breaks supersymmetry, but preserves the
maximally compact $SO(5)\times SO(3)$ and $SO(3)\cro SO(3)$ subgroup
of the gauge group, respectively. Extrapolating these results, one may
expect an analogue of this point in the $SO(5,3)\cro SO(5,3)$ gauged
theory in three dimensions. We shall show now that this is indeed the
case.

{}From \Ref{spec5}, it follows that the spectrum contains no singlet
under $H_0=SO(5)\times SO(3)\times SO(5)\times SO(3)$, i.e.\ the only
point preserving the full $H_0$ is the AdS ground state described
above. Considering the diagonal $SO(5,3)$, the spectrum contains three
singlets: the obvious two from $SL(2)$ as well as a further one
\be
M^{\mathcal C}{}_{\mathcal B}=\left(\frac{3}{4}P^{(5)}_{\alpha\beta}
-\frac{5}{4}Q^{(5)}_{\alpha\beta}\right)\;f_{\alpha\beta\,
\mathcal{B}}{}^{\mathcal C}.
\ee
Parametrizing this three-dimensional manifold via
\be
\mathcal{V}=\exp(s M)\;\exp(w W)\,\exp(z Z)\,\exp(-w W) \;,
\ee
where $W, Z$ are given in $(\ref{vswz})$, we get the potential
\be
\begin{array}{lcl}
-8g^{-2}V&=&25-15\,\cosh(4\,s)-15\,\cosh(s)\,\cosh(z)+\frac{15}{2}\,\cosh(3\,s)\,\cosh(z)\\
&&+\frac{3}{2}\,\cosh(5\,s)\,\cosh(z)+15\,\cos(w)\,\sinh(z)\,\sinh(s)+\\
&&\frac{15}{2}\,\cos(w)\,\sinh(z)\,\sinh(3\,s)-\frac{3}{2}\,\cos(w)\,\sinh(z)\,\sinh(5\,s)
\end{array} 
\ee
which has a nontrivial stationary point at $w=\pi$,
$s=\frac{1}{4}\,\arccosh\,5$, $z=3\,s$. Since the corresponding
generator $\tilde M=M-3Z$ according to $(\ref{spec5})$ is precisely
the only $H_0=SO(5)\times SO(5)\times SO(3)_{\rm diag}$ singlet, the
symmetry is broken down to this group.
The cosmological constant takes the value $\Lambda=22g^2>0$. The mass
spectrum at this point is collected in \Ref{E-p5-so5}. In particular,
the scalar mass squares are given by
\ba
\begin{tabular}{|c||c|c|c|c|c|c|}\hline
$H_0$ & $(5,5,1)$ & $(4,4,3)$
& $(4,4,1)$ & $(1,1,5)$ & $(1,1,3)$ & $(1,1,1)$  \\ \hline
$m^2L_{\rm dS}^2 $ & $\ft{24}{11}$ & $\ft{45}{11}$ &
$-\ft3{11}$ &  $\ft{96}{11}$ & $0$ & $-\ft{48}{11}$ 
\\ \hline
\end{tabular}
\la{specdS1}
\ea
in units of the inverse dS length $L_{\rm dS}$, together with $33$
Goldstone bosons. Hence, this de Sitter vacuum is unstable like its
counterparts in higher dimensions. In our case, this instability is
already implied by the fact that it is smoothly connected with the
maximally supersymmetric AdS vacuum at the origin
\Ref{spec5}.

\mathon
\subsection{$SO(4,4)\times SO(4,4)$}
\mathoff

For the gauge group $SO(4,4)\times SO(4,4)$, we may find stationary
points breaking its compact subgroup down to a diagonal $SO(4)\times
SO(4)$. If we split $SO(8)_{L,R}$ via
$\alpha\rightarrow(\alpha_1,\alpha_2)$ and accordingly label the
$SO(4)$ factors, the compact subgroup of our gauge group is
$(SO(4)_{L1}\times SO(4)_{R2})\times(SO(4)_{R1}\times SO(4)_{L2})$.
In this particular case, there is more than one obvious way to form a
diagonal subgroup of the compact part of the gauge group, but we will
only consider the case corresponding to the constructions employed
above. We hence again have the two singlets $W,Z$ from $SL(2)$ as in
$(\ref{vswz})$ as well as two additional singlets from another
$SL(2)$
\be
S_1^{\mathcal C}{}_{\mathcal B}=\frac{1}{4}
\left(P^{(4)}_{\alpha\beta}-Q^{(4)}_{\alpha\beta}\right)\,
f_{\alpha\beta\mathcal B}{}^{\mathcal C},\qquad
S_2^{\mathcal C}{}_{\mathcal
B}=\frac{1}{4}\left(P^{(4)}_{\dot\alpha\dot\beta}-
Q^{(4)}_{\dot\alpha\dot\beta}\right)\,f_{\dot\alpha\dot\beta\mathcal
B}{}^{\mathcal C} \;.
\ee
We parametrize this four-dimensional manifold as
\be
\mathcal{V}=\exp(wW)\,\exp(zZ)\,\exp(-wW)\;\exp(v[S_1,S_2])\,\exp(sS_1)\,\exp(-v[S_1,S_2])
\;,
\ee
and get the potential
\be
-8g^{-2}V=24-16\,\cosh(z)-16\,\cosh(s)+8\,\cosh(s)\,\cosh(z) \;,
\ee
which does not depend on $w$ and $v$. Besides the origin, there is a
second stationary point at $|s|=|z|=z_0:=\arccosh\,2$, at which the
gauge group is broken down to $H_0=SO(4)_{L2}\times SO(4)_{R2}\times
SO(4)_{L1,R1}$. Evaluating the potential, one verifies that this
vacuum is de Sitter with $\Lambda=4g^2$. The mass spectrum is
collected in \Ref{E-p4-so4}. The scalar mass squares are given by
\ba
\begin{tabular}{|c||c|c|c|c|}\hline
$\#$ & $49$ & $32$ & $8$ & $1$  \\ \hline
$m^2L_{\rm dS}^2 $ & $12$ & $9$ & $0$ &  $-12$ 
\\ \hline
\end{tabular}
\la{specdS2}
\ea
together with $38$ Goldstone bosons. Note that there is just one
unstable direction which is required in order to run into the
maximally supersymmetric Minkowski vacuum \Ref{spec4} at the
origin. Comparing this mass with \Ref{specdS1}, we find that
there seems to be no universal value for the highest tachyonic mass 
square, contrary to the situation in higher dimensions~\cite{KLPS02}.

\subsection{Exceptional gauge groups}

Finally, we will compute some of the potentials of the exceptional
gaugings discussed in section~4. For the gauge group $E_{7(+7)}\times
SL(2)$ and using the tools we developed in previous sections, it is
natural to consider the scalar potential on the manifold of
$SO(6)\subset SO(8)$ singlets by using the parametrization
$(\ref{sl3xsl2param})$.  Since the $SL(2)$ parametrized by $v, r_5$ is
part of the gauge group, and hence corresponds to flat directions in
the potential, these two parameters drop out.  Furthermore, three of
the five noncompact directions of the $SO(6)$-invariant $SL(3)$
singlets lie in the gauge group, and the smallest group containing the
remaining two orthogonal directions is $SL(2)$, which we can
parametrize by
\be
\begin{array}{lcl}
S&=&2\delta^\alpha_{[6}\delta^\beta_{7]}\,f_{\alpha\beta\mathcal{B}}{}^{\mathcal{C}}\\
V&=&\frac{1}{2}\,\delta^{i\bar j}\,f_{\underline{i\bar j}\mathcal{B}}{}^{\mathcal{C}}\\
\mathcal{V}&=&\exp\left(v\,V\right)\,\exp\left(s\,S\right)\,\exp\left(-v\,V\right)
\;,
\end{array}
\ee
and obtain the potential
\be
-8g^{-2}V=22-6\,\cosh(4\,s) \;,
\ee
which obviously does not have any nontrivial stationary points.

A richer structure is found for the exceptional gauge group $G_2\times
F_{4(-20)}$. The main problem in this case is to find an appropriate
invariance subgroup of the gauge group small enough to show nontrivial
structure, yet big enough to produce not too many singlets. Since none
of the parametrizations given so far work well here, we choose that
particular subgroup $SU(3)\times SU(3)$ of the group $SO(8)_L\times
SO(8)_R$ which stabilizes the vectors $v_1^i=\delta^{i7}$,
$v_2^i=\delta^{i8}$, $v_3^{\bar i}=\delta^{\bar i7}$, $v_4^{\bar
i}=\delta^{\bar i8}$ as well as the spinors
$\psi^{\alpha_L}=\delta^{\alpha_L8}$,
$\psi^{\alpha_R}=\delta^{\alpha_R8}$ (and which is also a subgroup of
$G_2\times F_{4(-20)}$).

This group is stabilized by a subgroup $SU(2,1)\times SU(2,1)$ of
$E_{8(8)}$, hence we have to deal with an eight-dimensional
submanifold of the supergravity scalars here. The intersection of this
eight-dimensional manifold with the gauge group is four-dimensional,
but unfortunately, unlike for the parametrization considered in the
$E_{7(7)}\times SL(2)$ case, the smallest group containing the four
directions orthogonal to the gauge group is the full $SU(2,1)\times
SU(2,1)$, hence we parametrize the full eight-dimensional
manifold.\footnote{Admittedly, motivation to do so comes in part from
the urge to test the limits of our now improved symbolic algebra
tools. Calculation of this potential takes less than four hours on 
a decent modern x86-based Linux workstation.}
Using the generators $X_{(A,B)}$ of both $SO(3)$ subalgebras as well
as those of two noncompact directions $Y_{(A,B)}$
\be
\begin{array}{lcl}
Y_{(A)}^{\mathcal{C}}{}_{\mathcal{B}}&=&-\frac{1}{2}\left(\delta^{\dot\alpha}_{2}\delta^{\dot\beta}_{8}-\delta^{\dot\alpha}_{8}\delta^{\dot\beta}_{2}\right)f_{\dot\alpha\dot\beta\mathcal{B}}{}^{\mathcal{C}}\\
X_{(A)1}^{\mathcal{C}}{}_{\mathcal{B}}&=&2\left(\delta^{i}_{7}\delta^{\bar j}_{8}-\delta^{i}_{8}\delta^{\bar j}_{7}\right)f_{\underline{[i\bar j]}\mathcal{B}}{}^{\mathcal{C}}\\
X_{(A)2}^{\mathcal{C}}{}_{\mathcal{B}}&=&2\left(\delta^{i}_{7}\delta^{\bar j}_{7}+\delta^{i}_{8}\delta^{\bar j}_{8}\right)f_{\underline{[i\bar j]}\mathcal{B}}{}^{\mathcal{C}}\\
X_{(A)3}^{\mathcal{C}}{}_{\mathcal{B}}&=&2\left(\delta^{i}_{7}\delta^{j}_{8}\,f_{\underline{[ij]}\mathcal{B}}{}^{\mathcal{C}}-\delta^{\bar i}_{7}\delta^{\bar j}_{8}\,f_{\underline{[\bar i\bar j]}\mathcal{B}}{}^{\mathcal{C}}\right)\\
Y_{B}^{\mathcal{C}}{}_{\mathcal{B}}&=&-\frac{1}{2}\left(\delta^{\dot\alpha}_{2}\delta^{\dot\beta}_{8}+\delta^{\dot\alpha}_{8}\delta^{\dot\beta}_{2}\right)f_{\dot\alpha\dot\beta\mathcal{B}}{}^{\mathcal{C}}\\
X_{(B)1}^{\mathcal{C}}{}_{\mathcal{B}}&=&-2\left(\delta^{i}_{7}\delta^{\bar j}_{8}+\delta^{i}_{8}\delta^{\bar j}_{7}\right)f_{\underline{[i\bar j]}\mathcal{B}}{}^{\mathcal{C}}\\
X_{(B)2}^{\mathcal{C}}{}_{\mathcal{B}}&=&-2\left(\delta^{i}_{7}\delta^{\bar j}_{7}-\delta^{i}_{8}\delta^{\bar j}_{8}\right)f_{\underline{[i\bar j]}\mathcal{B}}{}^{\mathcal{C}}\\
X_{(B)3}^{\mathcal{C}}{}_{\mathcal{B}}&=&-2\left(\delta^{i}_{7}\delta^{j}_{8}\,f_{\underline{[ij]}\mathcal{B}}{}^{\mathcal{C}}+\delta^{\bar i}_{7}\delta^{\bar j}_{8}\,f_{\underline{[\bar i\bar j]}\mathcal{B}}{}^{\mathcal{C}}\right)\;,
\end{array}
\ee
we parametrize the eight-dimensional singlet manifold as
\be
\begin{array}{lcl}
\mathcal{V}&=&\exp(r_1\,X_{(A)1})\,\exp(r_2\,X_{(A)2})\,
\exp(r_3\,X_{(A)3})\\
&&\exp(r_4\,X_{(B)1})\,\exp(r_5\,X_{(B)2})\,\exp(r_6\,X_{(B)3})\\
&&\,\exp(s\,Y_{(A)})\,\exp(z\,Y_{(B)})\\
&&\exp(-r_6\,X_{(B)3})\,\exp(-r_5\,X_{(B)2})\,\exp(-r_4\,X_{(B)1})\\
&&\exp(-r_3\,X_{(A)3})\,\exp(-r_2\,X_{(A)2})\,\exp(-r_1\,X_{(A)1})
\;,
\end{array}
\ee
and obtain for the potential the somewhat lengthy expression given in
given in $(\ref{g2xf4})$. A nontrivial stationary point is located at
$r_i=0$, $z=-s=\frac{1}{2}\,\arccosh\,7$, with remaining symmetry
$SU(3)\times SO(7)^-$. The value of the cosmological constant is
$\Lambda=-25g^2/2$, i.e.\ again the ratio of associated central
charges of this vacuum and the origin \Ref{multgf2} comes out to be
rational:
\ba
\frac{c_{SU(3)\times SO(7)}}{c_{G_2\times SO(9)}} &=& 
\sqrt{\frac{\GL_{G_2\times SO(9)}}{\GL_{SU(3)\times SO(7)}}} ~=~ 
\frac45  \;.
\ea
The full mass spectrum is collected in \Ref{E-e1-so7}. In particular,
inspection of the gravitino masses shows that this vacuum preserves
$N=(0,1)$ supersymmetries.

\subsection*{Acknowledgements}

This work is partly supported by EU contract HPRN-CT-2000-00122.

\appendix
\renewcommand{\theequation}{\Alph{section}.\arabic{equation}}
\renewcommand{\thesection}{Appendix \Alph{section}:}

\section{$E_{8(+8)}$ conventions}

Since some of the results in the main text depend on our particular
choice of conventions for $E_{8(8)}$ structure constants (e.g.\ the
fact that $G_2$ can be embedded in such a way into $SO(8)$ that the
stabilized vector, spinor and co-spinor all carry the index $8$), we
state them here for reference.

Using the conventions of \cite{GrScWi87}, we define
\begin{equation}
\begin{array}{ll}
\sigma_1=\left(\begin{array}{rr}1&0\\0&1\end{array}\right)&
\sigma_x=\left(\begin{array}{rr}0&1\\1&0\end{array}\right)\\
\sigma_z=\left(\begin{array}{rr}1&0\\0&-1\end{array}\right)&
\sigma_e=\left(\begin{array}{rr}0&1\\-1&0\end{array}\right)
\end{array}
\end{equation}
from which we obtain $SO(8)$ $\gamma$-matrices using the tensor
$G_{i\lambda\mu\rho}$ implementing the $2\times2\times2\rightarrow 8$
mapping\footnote{Note that this convention, which seems to be more
widespread, accidentally is just the opposite of that implicitly used
in \cite{Fisc02}}
\begin{equation}
\begin{array}{lclclclclclclcl}
G_{1111}=1&\quad&G_{2112}=1&\quad&G_{3121}=1&\quad&G_{4122}=1\\
G_{5211}=1&\quad&G_{6212}=1&\quad&G_{7221}=1&\quad&G_{8222}=1
\end{array}
\end{equation}
as well as the abbreviation
\begin{equation}
Z(\sigma_{(A)};\sigma_{(B)};\sigma_{(C)})=\sigma_{(A)\alpha_1\dot\beta_1}\,\sigma_{(B)\alpha_2\dot\beta_2}\sigma_{(C)\alpha_3\dot\beta_3}G^{\alpha\alpha_1\alpha_2\alpha_3}G^{\beta\beta_1\beta_2\beta_3}
\end{equation}
via
\begin{equation}
\begin{array}{lcl c lcl}
\gamma^1&=&Z(\sigma_e;\sigma_e;\sigma_e)&\quad&\gamma^2&=&Z(\sigma_1;\sigma_z;\sigma_e)\\
\gamma^3&=&Z(\sigma_e;\sigma_1;\sigma_z)&\quad&\gamma^4&=&Z(\sigma_z;\sigma_e;\sigma_1)\\
\gamma^5&=&Z(\sigma_1;\sigma_x;\sigma_e)&\quad&\gamma^6&=&Z(\sigma_e;\sigma_1;\sigma_x)\\
\gamma^7&=&Z(\sigma_x;\sigma_e;\sigma_1)&\quad&\gamma^8&=&Z(\sigma_1;\sigma_1;\sigma_1)\\
\end{array}
\end{equation}
from which we form $SO(16)$ $\Gamma$-matrices using to the splitting
$J\rightarrow(j,\bar k)$ of SO(16) vector and
$A\rightarrow(\alpha\beta,\dot\gamma\dot\delta)$, $\dot
A\rightarrow(\alpha\dot\beta,\dot\gamma\delta)$ of $MW$ spinor and
co-spinor indices by
\begin{equation}
\begin{array}{lcl c lcl}
\Gamma^i_{\alpha\beta\,\gamma\dot\delta}&=&\phantom+\delta_{\alpha\gamma}\,\gamma^i_{\beta\dot\delta}
&\qquad&
\Gamma^i_{\dot\alpha\dot\beta\,\dot\gamma\delta}&=&\phantom+\delta_{\dot\alpha\dot\gamma}\,\gamma^i_{\delta\dot\beta}\\
\Gamma^{\bar i}_{\alpha\beta\,\dot\gamma\delta}&=&\phantom+\delta_{\beta\delta}\,\gamma^{\bar i}_{\alpha\dot\gamma}
&\qquad&
\Gamma^{\bar i}_{\dot\alpha\dot\beta\,\gamma\dot\delta}&=&-\delta_{\dot\beta\dot\delta}\,\gamma^{\bar i}_{\gamma\dot\alpha}.
\end{array}
\end{equation}

If we denote $SO(16)$ adjoint indices by $\underline{[IJ]}$, which
naturally decompose into $SO(16)$ vector indices\footnote{Note that a
sum over all adjoint indices has to include a double-counting
correction factor $1/2$ if it is performed as sum over antisymmetric
vector indices. {\em Whenever we implicitly sum over an adjoint index
$\underline{[IJ]}$, we include every pair of indices $I,J$ only once.}
} $I,J$ and split $E_{8(8)}$ adjoint indices
$\mathcal{A}\rightarrow(A,\underline{[IJ]})$, then $E_{8(8)}$
structure constants are given by
\begin{equation}
\begin{array}{lcl c lcl}
f_{\underline{[IJ]}\underline{[KL]}}{}^{\underline{[MN]}}&=&-8\delta_{{}^[I{}_[K}\delta^{MN}_{L{}_]J{}^]}&\quad&
f_{\underline{[IJ]}A}{}^{B}&=&\frac{1}{2}\Gamma^{IJ}_{AB}\\
f_{B\underline{[IJ]}}{}^{A}&=&\frac{1}{2}\Gamma^{IJ}_{AB}&\quad&
f_{AB}{}^{\underline{[IJ]}}&=&-\frac{1}{2}\Gamma^{IJ}_{AB}.
\end{array}
\end{equation}

\section{Vacua and mass spectra}

In this appendix, we collect the mass spectra computed around all the
stationary points identified in this paper. The tables give the
eigenvalues of $\cal{M}$, $\cal{M}^{\rm vec}$, $A_1$, and $A_3$, where
the multiplicity of each eigenvalue is given by the subscript in
parentheses. For the AdS vacua, the associated conformal dimensions
may be obtained from \Ref{MDelta}. The Goldstone modes are contained
in the $m^2=0$ eigenvalues of $\cal{M}$. The Goldstino modes among the
eigenvalues of $A_3$ are more difficult to disentangle as their
identification requires projection with the $A_2$ tensor,
cf.~\Ref{split} and the subsequent discussion. They are marked with an
asterisk and do not appear in the effective physical spectrum.

\bigskip

{$G_0 =SO(8)\times SO(8)$, remaining symmetry $SO(7)^\pm\times
SO(7)^{\pm}$}:
\begin{eqnarray}\la{E-p8-so7}
\begin{tabular}{|l|l|}\hline
$\Lambda/2g^2$&$-25$\\\hline
$\mathcal{M}/g^2$&$96_{(\times 1)},\;0_{(\times 14)},\;-9_{(\times
64)},\;-24_{(\times 49)}$\\\hline
$\mathcal{M}^{\rm vec}/g$&$6_{(\times 7)},\;0_{(\times 114)},\;-6_{(\times
7)}$\\\hline
$A_1$&$7/2_{(\times 8)},\;-7/2_{(\times 8)}$\\\hline
$A_3$&$21/2_{(\times 8)^*},\;3/2_{(\times 56)},\;-3/2_{(\times
56)},\;-21/2_{(\times 8)^*}$\\\hline
\end{tabular}
\end{eqnarray}

{$G_0 =SO(8)\times SO(8)$, remaining symmetry $G_2\times G_2$, $N=(1,1)$}:
\begin{eqnarray}\la{E-p8-g2}
\begin{tabular}{|l|l|}\hline
$\Lambda/2g^2$&$-256/9$\\\hline
$\mathcal{M}/g^2$&$1040/9_{(\times 1)},\;16_{(\times 1)},\;0_{(\times
28)},\;-112/9_{(\times 49)},\;-80/3_{(\times 49)}$\\\hline
$\mathcal{M}^{\rm vec}/g$&$20/3_{(\times 7)},\;4/3_{(\times 7)},\;0_{(\times
100)},\;-4/3_{(\times 7)},\;-20/3_{(\times 7)}$\\\hline
$A_1$&$4_{(\times 7)},\;8/3_{(\times 1)},\;-8/3_{(\times 1)},\;-4_{(\times
7)}$\\\hline
$A_3$&$12_{(\times 7)^*},\;28/3_{(\times 1)},\;4_{(\times
7)},\;4/3_{(\times 49)},\;$ 
\\&
$-4/3_{(\times 49)},\;-4_{(\times 7)},\;-28/3_{(\times
1)},\;-12_{(\times 7)^*}$\\\hline 
\end{tabular}
\end{eqnarray}

{$G_0 =SO(8)\cro SO(8)$, remaining symmetry $SU(3)\cro
SU(3)\cro U(1)\cro U(1)$, $N=(2,2)$}:
\begin{eqnarray}\la{E-p8-su3}
\begin{tabular}{|l|l|}\hline
$\Lambda/2g^2$&$-36$\\\hline
$\mathcal{M}/g^2$&$160_{(\times 1)},\;28_{(\times 4)},\;0_{(\times 38)},\;-20_{(\times 36)},\;-32_{(\times 49)}$\\\hline
$\mathcal{M}^{\rm vec}/g$&$8_{(\times 7)},\;2_{(\times 12)},\;0_{(\times 90)},\;-2_{(\times 12)},\;-8_{(\times 7)}$\\\hline
$A_1$&$5_{(\times 6)},\;3_{(\times 2)},\;-3_{(\times 2)},\;-5_{(\times 6)}$\\\hline
$A_3$&$15_{(\times 6)^*},\,11_{(\times2)},\,5_{(\times14)},\,1_{(\times42)},\,-1_{(\times42)},\,-5_{(\times14)},\,-11_{(\times2)},\,-15_{(\times 6)^*}$\\\hline
\end{tabular}
\end{eqnarray}

{$G_0 =SO(7,1)\times SO(7,1)$, remaining symmetry $G_2\times
G_2$}: 
\begin{eqnarray}\la{E-p7-g2}
\begin{tabular}{|l|l|}\hline
$\Lambda/2g^2$&$-211/16$\\\hline
$\mathcal{M}/g^2$&$195/4_{(\times 1)},\;45/2_{(\times 1)},\;0_{(\times
28)},\;-9/2_{(\times 49)},\;-33/4_{(\times 49)}$\\\hline
$\mathcal{M}^{\rm vec}/g$&$9/2_{(\times 7)},\;3_{(\times 7)},\;0_{(\times
100)},\;-3_{(\times 7)},\;-9/2_{(\times 7)}$\\\hline
$A_1$&$35/8_{(\times 1)},\;19/8_{(\times 7)},\;-19/8_{(\times
7)},\;-35/8_{(\times 1)}$\\\hline
$A_3$&$105/8_{(\times 1)^*},\;57/8_{(\times 7)^*},\;33/8_{(\times
7)},\;15/8_{(\times 49)},\;$
\\&$-15/8_{(\times 49)},\;-33/8_{(\times
7)},\;-57/8_{(\times 7)^*},\;-105/8_{(\times 1)^*}$\\\hline
\end{tabular}
\end{eqnarray}

{$G_0 =SO(5,3)\times SO(5,3)$, remaining symmetry $SO(5)\times
SO(5)\times SO(3)_{\rm diag}$}: 
\begin{eqnarray}\la{E-p5-so5}
\begin{tabular}{|l|l|}\hline
$\Lambda/2g^2$&$11$\\\hline
$\mathcal{M}/g^2$&$96_{(\times 5)},\;45_{(\times 48)},\;24_{(\times
25)},\;0_{(\times 33)},\;-3_{(\times 16)},\;-48_{(\times 1)}$\\\hline
$\mathcal{M}^{\rm vec}/g$&$6_{(\times 15)},\;0_{(\times 98)},\;-6_{(\times
15)}$\\\hline
$A_1$&$5/2_{(\times 8)},\;-5/2_{(\times 8)}$\\\hline
$A_3$&$15/2_{(\times 8^*,\;\times16)},\;9/2_{(\times 40)},\;-9/2_{(\times
40)},\;-15/2_{(\times 8^*,\;\times16)}$\\\hline
\end{tabular}
\end{eqnarray}

{$G_0 =SO(4,4)\times SO(4,4)$, remaining symmetry $SO(4)^4$, $N=16$}:
\begin{eqnarray}\la{E-p4-so44}
\begin{tabular}{|l|l|}\hline
$\Lambda/2g^2$&$0$\\\hline
$\mathcal{M}/g^2$&$4_{(\times 96)},\;0_{(\times 32)}$\\\hline
$\mathcal{M}^{\rm vec}/g$&
$2_{(\times 16)},\;0_{(\times 96)},\;-2_{(\times 16)}$ \\\hline 
$A_1$&$0_{(\times 16)}$\\\hline
$A_3$&$2_{(\times 64)},\;-2_{(\times 64)}$\\\hline
\end{tabular}
\end{eqnarray}

{$G_0 =SO(4,4)\times SO(4,4)$, remaining symmetry $SO(4)\times
SO(4)\times SO(4)_{\rm diag}$}: 
\begin{eqnarray}\la{E-p4-so4}
\begin{tabular}{|l|l|}\hline
$\Lambda/2g^2$&$2$\\\hline
$\mathcal{M}/g^2$&$12_{(\times 49)},\;9_{(\times 32)},\;0_{(\times
46)},\;-12_{(\times 1)}$\\\hline
$\mathcal{M}^{\rm vec}/g$&$3_{(\times 16)},\;0_{(\times 96)},\;-3_{(\times
16)}$\\\hline
$A_1$&$1_{(\times 8)},\;-1_{(\times 8)}$\\\hline
$A_3$&$3_{(\times 8^*,\;\times56)},\;-3_{(\times 8^*,\;\times56)}$\\\hline
\end{tabular}
\end{eqnarray}

{$G_0 =G_2\times F_{4(-20)}$, remaining symmetry $G_2\times
SO(9)$, $N=(7,9)$}: 
\begin{eqnarray}\la{E-e1-so9}
\begin{tabular}{|l|l|}\hline
$\Lambda/2g^2$&$-4$\\\hline
$\mathcal{M}/g^2$&$0_{(\times 16)},\;-3_{(\times 112)}$\\\hline
$\mathcal{M}^{\rm vec}/g$&$1_{(\times 16)},\;0_{(\times 112)}$\\\hline
$A_1$&$1_{(\times 7)},\;-1_{(\times 9)}$\\\hline
$A_3$&$2_{(\times 16)},\;0_{(\times 112)}$\\\hline
\end{tabular}
\end{eqnarray}

{$G_0 =G_2\times F_{4(-20)}$, remaining symmetry $SU(3)\times
SO(7)^-$, $N=(0,1)$}: 
\begin{eqnarray}\la{E-e1-so7}
\begin{tabular}{|l|l|}\hline
$\Lambda/2g^2$&$-25/4$\\\hline
$\mathcal{M}/g^2$&$24_{(\times 1)},\;0_{(\times 37)},\;-9/4_{(\times
48)},\;-6_{(\times 42)}$\\\hline
$\mathcal{M}^{\rm vec}/g$&$4_{(\times 1)},\;3_{(\times 6)},\;3/2_{(\times
8)},\;1_{(\times 7)},\;0_{(\times 91)},\;-1/2_{(\times 8)},\;-3_{(\times
7)}$\\\hline
$A_1$&$11/4_{(\times 1)},\;7/4_{(\times 6)},\;-5/4_{(\times 1)},\;-7/4_{(\times
8)}$\\\hline
$A_3$&$33/4_{(\times 1)^*},\;21/4_{(\times 6)^*},\;17/4_{(\times
1)},\;11/4_{(\times 8)},\;9/4_{(\times 7)},\;3/4_{(\times 48)},\;$
\\ &$-3/4_{(\times
42)},\;-7/4_{(\times 7)},\;-21/4_{(\times 8)^*}$\\\hline
\end{tabular}
\end{eqnarray}

\section{Explicit scalar potentials}

In this appendix, we collect some of the scalar potentials which are
too lengthy to be given in the main text. These are the potentials for
gauge groups $SO(8)\cro SO(8)$, $SO(7,1)\cro SO(7,1)$, and
$SO(6,2)\cro SO(6,2)$, restricted to the seven-dimensional manifold of
$SO(6)_{\rm diag}$ singlets, as well as the potential for the
exceptional gauge group $G_2\cro F_{4(-20)}$ restricted to the
manifold of $SU(3)\cro SU(3)$ singlets.

\bigskip

$\bullet\;\;$ $G_0=SO(8)\times SO(8)$, potential restricted to
singlets under $SO(6)_{\rm diag}$:
{\small 
\ben
\begin{array}{lcl}
-8g^{-2}V&=&27+3\,\cosh(4\,z)+3\,\cosh(4\,z)\,\cos(2\,r_2)-3\,\cosh(4\,z)\,\cos(2\,r_1)\\
&&-3\,\cosh(4\,z)\,\cos(2\,r_1)\,\cos(2\,r_2)+\frac{1}{4}\cosh(4\,s)+\frac{1}{4}\cosh(4\,s)\,\cos(2\,r_2)\\
&&-\frac{1}{4}\cosh(4\,s)\,\cos(2\,r_1)-\frac{1}{4}\cosh(4\,s)\,\cos(2\,r_1)\,\cos(2\,r_2)\\
&&+9\,\cosh(2\,s)\,\cosh(2\,z)+\frac{3}{4}\cosh(2\,s)\,\cosh(6\,z)\\
&&-3\,\cos(2\,r_3)\,\sinh(2\,z)\,\sinh(2\,s)+\frac{1}{4}\cos(2\,r_3)\,\sinh(6\,z)\,\sinh(2\,s)\\
&&-3\,\cosh(2\,s)\,\cosh(2\,z)\,\cos(2\,r_2)-\frac{1}{4}\cosh(2\,s)\,\cosh(6\,z)\,\cos(2\,r_2)\\
&&-3\,\cos(2\,r_2)\,\cos(2\,r_3)\,\sinh(2\,z)\,\sinh(2\,s)\\
&&+\frac{1}{4}\cos(2\,r_2)\,\cos(2\,r_3)\,\sinh(6\,z)\,\sinh(2\,s)\\
&&+3\,\cosh(2\,s)\,\cosh(2\,z)\,\cos(2\,r_1)+\frac{1}{4}\cosh(2\,s)\,\cosh(6\,z)\,\cos(2\,r_1)\\
&&-9\,\cos(2\,r_1)\,\cos(2\,r_3)\,\sinh(2\,z)\,\sinh(2\,s)\\
&&+\frac{3}{4}\cos(2\,r_1)\,\cos(2\,r_3)\,\sinh(6\,z)\,\sinh(2\,s)\\
&&+3\,\cosh(2\,s)\,\cosh(2\,z)\,\cos(2\,r_1)\,\cos(2\,r_2)\\
&&+\frac{1}{4}\cosh(2\,s)\,\cosh(6\,z)\,\cos(2\,r_1)\,\cos(2\,r_2)\\
\end{array}
\een
\be
\begin{array}{lcl}
&&-12\,\sin(2\,r_3)\,\sin(r_2)\,\sin(2\,r_1)\,\sinh(2\,z)\,\sinh(2\,s)\\
&&+\,\sin(2\,r_3)\,\sin(r_2)\,\sin(2\,r_1)\,\sinh(6\,z)\,\sinh(2\,s)\\
&&+3\,\cos(2\,r_1)\,\cos(2\,r_2)\,\cos(2\,r_3)\,\sinh(2\,z)\,\sinh(2\,s)\\
&&-\frac{1}{4}\cos(2\,r_1)\,\cos(2\,r_2)\,\cos(2\,r_3)\,\sinh(6\,z)\,\sinh(2\,s)+\,\cosh(2\,v)\\
&&+9\,\cosh(v)\,\cosh(4\,z)-3\,\cosh(v)\,\cosh(4\,z)\,\cos(2\,r_2)\\
&&+3\,\cosh(v)\,\cosh(4\,z)\,\cos(2\,r_1)+3\,\cosh(v)\,\cosh(4\,z)\,\cos(2\,r_1)\,\cos(2\,r_2)\\
&&-\frac{1}{4}\cosh(2\,v)\,\cosh(4\,s)-\frac{1}{4}\cosh(2\,v)\,\cosh(4\,s)\,\cos(2\,r_2)\\
&&+\frac{1}{4}\cosh(2\,v)\,\cosh(4\,s)\,\cos(2\,r_1)+\frac{1}{4}\cosh(2\,v)\,\cosh(4\,s)\,\cos(2\,r_1)\,\cos(2\,r_2)\\
&&+15\,\cosh(v)\,\cosh(2\,s)\,\cosh(2\,z)\\
&&-\frac{3}{4}\cosh(2\,v)\,\cosh(2\,s)\,\cosh(6\,z)+3\,\cosh(v)\,\cos(2\,r_3)\,\sinh(2\,z)\,\sinh(2\,s)\\
&&-\frac{1}{4}\cosh(2\,v)\,\cos(2\,r_3)\,\sinh(6\,z)\,\sinh(2\,s)\\
&&+3\,\cosh(v)\,\cosh(2\,s)\,\cosh(2\,z)\,\cos(2\,r_2)\\
&&+\frac{1}{4}\cosh(2\,v)\,\cosh(2\,s)\,\cosh(6\,z)\,\cos(2\,r_2)\\
&&+3\,\cosh(v)\,\cos(2\,r_2)\,\cos(2\,r_3)\,\sinh(2\,z)\,\sinh(2\,s)\\
&&-\frac{1}{4}\cosh(2\,v)\,\cos(2\,r_2)\,\cos(2\,r_3)\,\sinh(6\,z)\,\sinh(2\,s)\\
&&-3\,\cosh(v)\,\cosh(2\,s)\,\cosh(2\,z)\,\cos(2\,r_1)\\
&&-\frac{1}{4}\cosh(2\,v)\,\cosh(2\,s)\,\cosh(6\,z)\,\cos(2\,r_1)\\
&&+9\,\cosh(v)\,\cos(2\,r_1)\,\cos(2\,r_3)\,\sinh(2\,z)\,\sinh(2\,s)\\
&&-\frac{3}{4}\cosh(2\,v)\,\cos(2\,r_1)\,\cos(2\,r_3)\,\sinh(6\,z)\,\sinh(2\,s)\\
&&-3\,\cosh(v)\,\cosh(2\,s)\,\cosh(2\,z)\,\cos(2\,r_1)\,\cos(2\,r_2)\\
&&-\frac{1}{4}\cosh(2\,v)\,\cosh(2\,s)\,\cosh(6\,z)\,\cos(2\,r_1)\,\cos(2\,r_2)\\
&&+12\,\cosh(v)\,\sin(2\,r_3)\,\sin(r_2)\,\sin(2\,r_1)\,\sinh(2\,z)\,\sinh(2\,s)\\
&&-\,\cosh(2\,v)\,\sin(2\,r_3)\,\sin(r_2)\,\sin(2\,r_1)\,\sinh(6\,z)\,\sinh(2\,s)\\
&&-3\,\cosh(v)\,\cos(2\,r_1)\,\cos(2\,r_2)\,\cos(2\,r_3)\,\sinh(2\,z)\,\sinh(2\,s)\\
&&+\frac{1}{4}\cosh(2\,v)\,\cos(2\,r_1)\,\cos(2\,r_2)\,\cos(2\,r_3)\,\sinh(6\,z)\,\sinh(2\,s)
\end{array}
\label{so88sl3}
\ee
}

\bigskip
\bigskip

$\bullet\;\;$ $G_0=SO(7,1)\times SO(7,1)$, potential restricted to
singlets under $SO(6)_{\rm diag}$: 
{\small
\ben
\begin{array}{lcl}
-8g^{-2}V&=&\frac{1663}{64}+\frac{43}{64}\,\cos(2\,r_5)+\frac{9}{64}\,\cos(4\,r_3)\\
&&-\frac{3}{64}\,\cos(4\,r_3)\,\cos(2\,r_5)-\frac{3}{16}\,\cos(2\,r_2)+\frac{21}{64}\,\cos(4\,r_2)\\
&&+\frac{1}{16}\,\cos(2\,r_2)\,\cos(2\,r_5)-\frac{7}{64}\,\cos(4\,r_2)\,\cos(2\,r_5)\\
&&+\frac{3}{16}\,\cos(2\,r_2)\,\cos(4\,r_3)+\frac{3}{64}\,\cos(4\,r_2)\,\cos(4\,r_3)\\
&&-\frac{1}{16}\,\cos(2\,r_2)\,\cos(4\,r_3)\,\cos(2\,r_5)-\frac{1}{64}\,\cos(4\,r_2)\,\cos(4\,r_3)\,\cos(2\,r_5)\\
&&-\frac{9}{64}\,\cos(2\,r_1)+\frac{3}{64}\,\cos(2\,r_1)\,\cos(2\,r_5)-\frac{15}{64}\,\cos(2\,r_1)\,\cos(4\,r_3)\\
&&+\frac{5}{64}\,\cos(2\,r_1)\,\cos(4\,r_3)\,\cos(2\,r_5)+\frac{3}{16}\,\cos(2\,r_1)\,\cos(2\,r_2)\\
&&+\frac{21}{64}\,\cos(2\,r_1)\,\cos(4\,r_2)-\frac{1}{16}\,\cos(2\,r_1)\,\cos(2\,r_2)\,\cos(2\,r_5)\\
&&-\frac{7}{64}\,\cos(2\,r_1)\,\cos(4\,r_2)\,\cos(2\,r_5)-\frac{3}{16}\,\sin(4\,r_3)\,\sin(r_2)\,\sin(2\,r_1)\\
&&-\frac{3}{16}\,\sin(4\,r_3)\,\sin(3\,r_2)\,\sin(2\,r_1)-\frac{3}{16}\,\cos(2\,r_1)\,\cos(2\,r_2)\,\cos(4\,r_3)\\
&&+\frac{3}{64}\,\cos(2\,r_1)\,\cos(4\,r_2)\,\cos(4\,r_3)+\frac{1}{16}\,\cos(2\,r_5)\,\sin(4\,r_3)\,\sin(r_2)\,\sin(2\,r_1)\\
&&+\frac{1}{16}\,\cos(2\,r_5)\,\sin(4\,r_3)\,\sin(3\,r_2)\,\sin(2\,r_1)\\
&&+\frac{1}{16}\,\cos(2\,r_1)\,\cos(2\,r_2)\,\cos(4\,r_3)\,\cos(2\,r_5)\\
\end{array}
\een
\ben
\begin{array}{lcl}
&&-\frac{1}{64}\,\cos(2\,r_1)\,\cos(4\,r_2)\,\cos(4\,r_3)\,\cos(2\,r_5)+3\,\cosh(4\,z)\\
&&+3\,\cosh(4\,z)\,\cos(2\,r_2)-3\,\cosh(4\,z)\,\cos(2\,r_1)-3\,\cosh(4\,z)\,\cos(2\,r_1)\,\cos(2\,r_2)\\
&&-\frac{15}{64}\,\cosh(4\,s)+\frac{5}{64}\,\cosh(4\,s)\,\cos(2\,r_5)-\frac{9}{64}\,\cosh(4\,s)\,\cos(4\,r_3)\\
&&+\frac{3}{64}\,\cosh(4\,s)\,\cos(4\,r_3)\,\cos(2\,r_5)-\frac{3}{16}\,\cosh(4\,s)\,\cos(2\,r_2)\\
&&+\frac{3}{64}\,\cosh(4\,s)\,\cos(4\,r_2)+\frac{1}{16}\,\cosh(4\,s)\,\cos(2\,r_2)\,\cos(2\,r_5)\\
&&-\frac{1}{64}\,\cosh(4\,s)\,\cos(4\,r_2)\,\cos(2\,r_5)-\frac{3}{16}\,\cosh(4\,s)\,\cos(2\,r_2)\,\cos(4\,r_3)\\
&&-\frac{3}{64}\,\cosh(4\,s)\,\cos(4\,r_2)\,\cos(4\,r_3)+\frac{1}{16}\,\cosh(4\,s)\,\cos(2\,r_2)\,\cos(4\,r_3)\,\cos(2\,r_5)\\
&&+\frac{1}{64}\,\cosh(4\,s)\,\cos(4\,r_2)\,\cos(4\,r_3)\,\cos(2\,r_5)+\frac{9}{64}\,\cosh(4\,s)\,\cos(2\,r_1)\\
&&-\frac{3}{64}\,\cosh(4\,s)\,\cos(2\,r_1)\,\cos(2\,r_5)+\frac{15}{64}\,\cosh(4\,s)\,\cos(2\,r_1)\,\cos(4\,r_3)\\
&&-\frac{5}{64}\,\cosh(4\,s)\,\cos(2\,r_1)\,\cos(4\,r_3)\,\cos(2\,r_5)+\frac{3}{16}\,\cosh(4\,s)\,\cos(2\,r_1)\,\cos(2\,r_2)\\
&&+\frac{3}{64}\,\cosh(4\,s)\,\cos(2\,r_1)\,\cos(4\,r_2)-\frac{1}{16}\,\cosh(4\,s)\,\cos(2\,r_1)\,\cos(2\,r_2)\,\cos(2\,r_5)\\
&&-\frac{1}{64}\,\cosh(4\,s)\,\cos(2\,r_1)\,\cos(4\,r_2)\,\cos(2\,r_5)\\
&&+\frac{3}{16}\,\cosh(4\,s)\,\sin(4\,r_3)\,\sin(r_2)\,\sin(2\,r_1)\\
&&+\frac{3}{16}\,\cosh(4\,s)\,\sin(4\,r_3)\,\sin(3\,r_2)\,\sin(2\,r_1)\\
&&+\frac{3}{16}\,\cosh(4\,s)\,\cos(2\,r_1)\,\cos(2\,r_2)\,\cos(4\,r_3)\\
&&-\frac{3}{64}\,\cosh(4\,s)\,\cos(2\,r_1)\,\cos(4\,r_2)\,\cos(4\,r_3)\\
&&-\frac{1}{16}\,\cosh(4\,s)\,\cos(2\,r_5)\,\sin(4\,r_3)\,\sin(r_2)\,\sin(2\,r_1)\\
&&-\frac{1}{16}\,\cosh(4\,s)\,\cos(2\,r_5)\,\sin(4\,r_3)\,\sin(3\,r_2)\,\sin(2\,r_1)\\
&&-\frac{1}{16}\,\cosh(4\,s)\,\cos(2\,r_1)\,\cos(2\,r_2)\,\cos(4\,r_3)\,\cos(2\,r_5)\\
&&+\frac{1}{64}\,\cosh(4\,s)\,\cos(2\,r_1)\,\cos(4\,r_2)\,\cos(4\,r_3)\,\cos(2\,r_5)\\
&&+9\,\cosh(2\,s)\,\cosh(2\,z)-\frac{3}{4}\,\cosh(2\,s)\,\cosh(6\,z)+\frac{1}{4}\,\cosh(2\,s)\,\cosh(6\,z)\,\cos(2\,r_5)\\
&&-3\,\cos(2\,r_3)\,\sinh(2\,z)\,\sinh(2\,s)-3\,\cosh(2\,s)\,\cosh(2\,z)\,\cos(2\,r_2)\\
&&+\frac{3}{8}\,\cosh(2\,s)\,\cosh(6\,z)\,\cos(2\,r_2)-\frac{3}{8}\,\cosh(2\,s)\,\cosh(6\,z)\,\cos(4\,r_2)\\
&&-\frac{1}{8}\,\cosh(2\,s)\,\cosh(6\,z)\,\cos(2\,r_2)\,\cos(2\,r_5)\\
&&+\frac{1}{8}\,\cosh(2\,s)\,\cosh(6\,z)\,\cos(4\,r_2)\,\cos(2\,r_5)-3\,\cos(2\,r_2)\,\cos(2\,r_3)\,\sinh(2\,z)\,\sinh(2\,s)\\
&&-\frac{3}{8}\,\cos(2\,r_2)\,\cos(2\,r_3)\,\sinh(6\,z)\,\sinh(2\,s)\\
&&-\frac{3}{8}\,\cos(4\,r_2)\,\cos(2\,r_3)\,\sinh(6\,z)\,\sinh(2\,s)\\
&&+\frac{1}{8}\,\cos(2\,r_2)\,\cos(2\,r_3)\,\cos(2\,r_5)\,\sinh(6\,z)\,\sinh(2\,s)\\
&&+\frac{1}{8}\,\cos(4\,r_2)\,\cos(2\,r_3)\,\cos(2\,r_5)\,\sinh(6\,z)\,\sinh(2\,s)\\
&&+3\,\cosh(2\,s)\,\cosh(2\,z)\,\cos(2\,r_1)-9\,\cos(2\,r_1)\,\cos(2\,r_3)\,\sinh(2\,z)\,\sinh(2\,s)\\
&&-\frac{3}{4}\,\cos(2\,r_1)\,\cos(2\,r_3)\,\sinh(6\,z)\,\sinh(2\,s)\\
&&+\frac{1}{4}\,\cos(2\,r_1)\,\cos(2\,r_3)\,\cos(2\,r_5)\,\sinh(6\,z)\,\sinh(2\,s)\\
&&+3\,\cosh(2\,s)\,\cosh(2\,z)\,\cos(2\,r_1)\,\cos(2\,r_2)-\frac{3}{8}\,\cosh(2\,s)\,\cosh(6\,z)\,\cos(2\,r_1)\,\cos(2\,r_2)\\
&&-\frac{3}{8}\,\cosh(2\,s)\,\cosh(6\,z)\,\cos(2\,r_1)\,\cos(4\,r_2)\\
&&+\frac{1}{8}\,\cosh(2\,s)\,\cosh(6\,z)\,\cos(2\,r_1)\,\cos(2\,r_2)\,\cos(2\,r_5)\\
&&+\frac{1}{8}\,\cosh(2\,s)\,\cosh(6\,z)\,\cos(2\,r_1)\,\cos(4\,r_2)\,\cos(2\,r_5)\\
&&-12\,\sin(2\,r_3)\,\sin(r_2)\,\sin(2\,r_1)\,\sinh(2\,z)\,\sinh(2\,s)\\
&&-\frac{3}{4}\,\sin(2\,r_3)\,\sin(r_2)\,\sin(2\,r_1)\,\sinh(6\,z)\,\sinh(2\,s)\\
&&+\frac{3}{4}\,\sin(2\,r_3)\,\sin(3\,r_2)\,\sin(2\,r_1)\,\sinh(6\,z)\,\sinh(2\,s)\\
&&+3\,\cos(2\,r_1)\,\cos(2\,r_2)\,\cos(2\,r_3)\,\sinh(2\,z)\,\sinh(2\,s)\\
&&+\frac{3}{8}\,\cos(2\,r_1)\,\cos(2\,r_2)\,\cos(2\,r_3)\,\sinh(6\,z)\,\sinh(2\,s)\\
\end{array}
\een
\ben
\begin{array}{lcl}
&&-\frac{3}{8}\,\cos(2\,r_1)\,\cos(4\,r_2)\,\cos(2\,r_3)\,\sinh(6\,z)\,\sinh(2\,s)\\
&&+\frac{1}{4}\,\cos(2\,r_5)\,\sin(2\,r_3)\,\sin(r_2)\,\sin(2\,r_1)\,\sinh(6\,z)\,\sinh(2\,s)\\
&&-\frac{1}{4}\,\cos(2\,r_5)\,\sin(2\,r_3)\,\sin(3\,r_2)\,\sin(2\,r_1)\,\sinh(6\,z)\,\sinh(2\,s)\\
&&-\frac{1}{8}\,\cos(2\,r_1)\,\cos(2\,r_2)\,\cos(2\,r_3)\,\cos(2\,r_5)\,\sinh(6\,z)\,\sinh(2\,s)\\
&&+\frac{1}{8}\,\cos(2\,r_1)\,\cos(4\,r_2)\,\cos(2\,r_3)\,\cos(2\,r_5)\,\sinh(6\,z)\,\sinh(2\,s)\\
&&-\frac{43}{64}\,\cosh(2\,v)-\frac{43}{64}\,\cosh(2\,v)\,\cos(2\,r_5)+\frac{3}{64}\,\cosh(2\,v)\,\cos(4\,r_3)\\
&&+\frac{3}{64}\,\cosh(2\,v)\,\cos(4\,r_3)\,\cos(2\,r_5)-\frac{1}{16}\,\cosh(2\,v)\,\cos(2\,r_2)\\
&&+\frac{7}{64}\,\cosh(2\,v)\,\cos(4\,r_2)-\frac{1}{16}\,\cosh(2\,v)\,\cos(2\,r_2)\,\cos(2\,r_5)\\
&&+\frac{7}{64}\,\cosh(2\,v)\,\cos(4\,r_2)\,\cos(2\,r_5)+\frac{1}{16}\,\cosh(2\,v)\,\cos(2\,r_2)\,\cos(4\,r_3)\\
&&+\frac{1}{64}\,\cosh(2\,v)\,\cos(4\,r_2)\,\cos(4\,r_3)+\frac{1}{16}\,\cosh(2\,v)\,\cos(2\,r_2)\,\cos(4\,r_3)\,\cos(2\,r_5)\\
&&+\frac{1}{64}\,\cosh(2\,v)\,\cos(4\,r_2)\,\cos(4\,r_3)\,\cos(2\,r_5)-\frac{3}{64}\,\cosh(2\,v)\,\cos(2\,r_1)\\
&&-\frac{3}{64}\,\cosh(2\,v)\,\cos(2\,r_1)\,\cos(2\,r_5)-\frac{5}{64}\,\cosh(2\,v)\,\cos(2\,r_1)\,\cos(4\,r_3)\\
&&-\frac{5}{64}\,\cosh(2\,v)\,\cos(2\,r_1)\,\cos(4\,r_3)\,\cos(2\,r_5)+\frac{1}{16}\,\cosh(2\,v)\,\cos(2\,r_1)\,\cos(2\,r_2)\\
&&+\frac{7}{64}\,\cosh(2\,v)\,\cos(2\,r_1)\,\cos(4\,r_2)+\frac{1}{16}\,\cosh(2\,v)\,\cos(2\,r_1)\,\cos(2\,r_2)\,\cos(2\,r_5)\\
&&+\frac{7}{64}\,\cosh(2\,v)\,\cos(2\,r_1)\,\cos(4\,r_2)\,\cos(2\,r_5)\\
&&-\frac{1}{16}\,\cosh(2\,v)\,\sin(4\,r_3)\,\sin(r_2)\,\sin(2\,r_1)\\
&&-\frac{1}{16}\,\cosh(2\,v)\,\sin(4\,r_3)\,\sin(3\,r_2)\,\sin(2\,r_1)\\
&&-\frac{1}{16}\,\cosh(2\,v)\,\cos(2\,r_1)\,\cos(2\,r_2)\,\cos(4\,r_3)\\
&&+\frac{1}{64}\,\cosh(2\,v)\,\cos(2\,r_1)\,\cos(4\,r_2)\,\cos(4\,r_3)\\
&&-\frac{1}{16}\,\cosh(2\,v)\,\cos(2\,r_5)\,\sin(4\,r_3)\,\sin(r_2)\,\sin(2\,r_1)\\
&&-\frac{1}{16}\,\cosh(2\,v)\,\cos(2\,r_5)\,\sin(4\,r_3)\,\sin(3\,r_2)\,\sin(2\,r_1)\\
&&-\frac{1}{16}\,\cosh(2\,v)\,\cos(2\,r_1)\,\cos(2\,r_2)\,\cos(4\,r_3)\,\cos(2\,r_5)\\
&&+\frac{1}{64}\,\cosh(2\,v)\,\cos(2\,r_1)\,\cos(4\,r_2)\,\cos(4\,r_3)\,\cos(2\,r_5)\\
&&+3\,\cos(r_5)\,\sinh(4\,z)\,\sinh(v)-9\,\cos(2\,r_2)\,\cos(r_5)\,\sinh(4\,z)\,\sinh(v)\\
&&-3\,\cos(2\,r_1)\,\cos(r_5)\,\sinh(4\,z)\,\sinh(v)-3\,\cos(2\,r_1)\,\cos(2\,r_2)\,\cos(r_5)\,\sinh(4\,z)\,\sinh(v)\\
&&-\frac{5}{64}\,\cosh(2\,v)\,\cosh(4\,s)-\frac{5}{64}\,\cosh(2\,v)\,\cosh(4\,s)\,\cos(2\,r_5)\\
&&-\frac{3}{64}\,\cosh(2\,v)\,\cosh(4\,s)\,\cos(4\,r_3)-\frac{3}{64}\,\cosh(2\,v)\,\cosh(4\,s)\,\cos(4\,r_3)\,\cos(2\,r_5)\\
&&-\frac{1}{16}\,\cosh(2\,v)\,\cosh(4\,s)\,\cos(2\,r_2)+\frac{1}{64}\,\cosh(2\,v)\,\cosh(4\,s)\,\cos(4\,r_2)\\
&&-\frac{1}{16}\,\cosh(2\,v)\,\cosh(4\,s)\,\cos(2\,r_2)\,\cos(2\,r_5)\\
&&+\frac{1}{64}\,\cosh(2\,v)\,\cosh(4\,s)\,\cos(4\,r_2)\,\cos(2\,r_5)\\
&&-\frac{1}{16}\,\cosh(2\,v)\,\cosh(4\,s)\,\cos(2\,r_2)\,\cos(4\,r_3)\\
&&-\frac{1}{64}\,\cosh(2\,v)\,\cosh(4\,s)\,\cos(4\,r_2)\,\cos(4\,r_3)\\
&&-\frac{1}{16}\,\cosh(2\,v)\,\cosh(4\,s)\,\cos(2\,r_2)\,\cos(4\,r_3)\,\cos(2\,r_5)\\
&&-\frac{1}{64}\,\cosh(2\,v)\,\cosh(4\,s)\,\cos(4\,r_2)\,\cos(4\,r_3)\,\cos(2\,r_5)\\
&&+\frac{3}{64}\,\cosh(2\,v)\,\cosh(4\,s)\,\cos(2\,r_1)+\frac{3}{64}\,\cosh(2\,v)\,\cosh(4\,s)\,\cos(2\,r_1)\,\cos(2\,r_5)\\
&&+\frac{5}{64}\,\cosh(2\,v)\,\cosh(4\,s)\,\cos(2\,r_1)\,\cos(4\,r_3)\\
&&+\frac{5}{64}\,\cosh(2\,v)\,\cosh(4\,s)\,\cos(2\,r_1)\,\cos(4\,r_3)\,\cos(2\,r_5)\\
&&+\frac{1}{16}\,\cosh(2\,v)\,\cosh(4\,s)\,\cos(2\,r_1)\,\cos(2\,r_2)\\
&&+\frac{1}{64}\,\cosh(2\,v)\,\cosh(4\,s)\,\cos(2\,r_1)\,\cos(4\,r_2)\\
&&+\frac{1}{16}\,\cosh(2\,v)\,\cosh(4\,s)\,\cos(2\,r_1)\,\cos(2\,r_2)\,\cos(2\,r_5)\\
&&+\frac{1}{64}\,\cosh(2\,v)\,\cosh(4\,s)\,\cos(2\,r_1)\,\cos(4\,r_2)\,\cos(2\,r_5)\\
\end{array}
\een
\be
\begin{array}{lcl}
&&+\frac{1}{16}\,\cosh(2\,v)\,\cosh(4\,s)\,\sin(4\,r_3)\,\sin(r_2)\,\sin(2\,r_1)\\
&&+\frac{1}{16}\,\cosh(2\,v)\,\cosh(4\,s)\,\sin(4\,r_3)\,\sin(3\,r_2)\,\sin(2\,r_1)\\
&&+\frac{1}{16}\,\cosh(2\,v)\,\cosh(4\,s)\,\cos(2\,r_1)\,\cos(2\,r_2)\,\cos(4\,r_3)\\
&&-\frac{1}{64}\,\cosh(2\,v)\,\cosh(4\,s)\,\cos(2\,r_1)\,\cos(4\,r_2)\,\cos(4\,r_3)\\
&&+\frac{1}{16}\,\cosh(2\,v)\,\cosh(4\,s)\,\cos(2\,r_5)\,\sin(4\,r_3)\,\sin(r_2)\,\sin(2\,r_1)\\
&&+\frac{1}{16}\,\cosh(2\,v)\,\cosh(4\,s)\,\cos(2\,r_5)\,\sin(4\,r_3)\,\sin(3\,r_2)\,\sin(2\,r_1)\\
&&+\frac{1}{16}\,\cosh(2\,v)\,\cosh(4\,s)\,\cos(2\,r_1)\,\cos(2\,r_2)\,\cos(4\,r_3)\,\cos(2\,r_5)\\
&&-\frac{1}{64}\,\cosh(2\,v)\,\cosh(4\,s)\,\cos(2\,r_1)\,\cos(4\,r_2)\,\cos(4\,r_3)\,\cos(2\,r_5)\\
&&-\frac{1}{4}\,\cosh(2\,v)\,\cosh(2\,s)\,\cosh(6\,z)+3\,\cosh(2\,s)\,\cos(r_5)\,\sinh(2\,z)\,\sinh(v)\\
&&-\frac{1}{4}\,\cosh(2\,v)\,\cosh(2\,s)\,\cosh(6\,z)\,\cos(2\,r_5)\\
&&-9\,\cosh(2\,z)\,\cos(2\,r_3)\,\cos(r_5)\,\sinh(2\,s)\,\sinh(v)+\frac{1}{8}\,\cosh(2\,v)\,\cosh(2\,s)\,\cosh(6\,z)\,\cos(2\,r_2)\\
&&-\frac{1}{8}\,\cosh(2\,v)\,\cosh(2\,s)\,\cosh(6\,z)\,\cos(4\,r_2)-9\,\cosh(2\,s)\,\cos(2\,r_2)\,\cos(r_5)\,\sinh(2\,z)\,\sinh(v)\\
&&+\frac{1}{8}\,\cosh(2\,v)\,\cosh(2\,s)\,\cosh(6\,z)\,\cos(2\,r_2)\,\cos(2\,r_5)\\
&&-\frac{1}{8}\,\cosh(2\,v)\,\cosh(2\,s)\,\cosh(6\,z)\,\cos(4\,r_2)\,\cos(2\,r_5)\\
&&-\frac{1}{8}\,\cosh(2\,v)\,\cos(2\,r_2)\,\cos(2\,r_3)\,\sinh(6\,z)\,\sinh(2\,s)\\
&&-\frac{1}{8}\,\cosh(2\,v)\,\cos(4\,r_2)\,\cos(2\,r_3)\,\sinh(6\,z)\,\sinh(2\,s)\\
&&-9\,\cosh(2\,z)\,\cos(2\,r_2)\,\cos(2\,r_3)\,\cos(r_5)\,\sinh(2\,s)\,\sinh(v)\\
&&-\frac{1}{8}\,\cosh(2\,v)\,\cos(2\,r_2)\,\cos(2\,r_3)\,\cos(2\,r_5)\,\sinh(6\,z)\,\sinh(2\,s)\\
&&-\frac{1}{8}\,\cosh(2\,v)\,\cos(4\,r_2)\,\cos(2\,r_3)\,\cos(2\,r_5)\,\sinh(6\,z)\,\sinh(2\,s)\\
&&-3\,\cosh(2\,s)\,\cos(2\,r_1)\,\cos(r_5)\,\sinh(2\,z)\,\sinh(v)\\
&&-\frac{1}{4}\,\cosh(2\,v)\,\cos(2\,r_1)\,\cos(2\,r_3)\,\sinh(6\,z)\,\sinh(2\,s)\\
&&+9\,\cosh(2\,z)\,\cos(2\,r_1)\,\cos(2\,r_3)\,\cos(r_5)\,\sinh(2\,s)\,\sinh(v)\\
&&-\frac{1}{4}\,\cosh(2\,v)\,\cos(2\,r_1)\,\cos(2\,r_3)\,\cos(2\,r_5)\,\sinh(6\,z)\,\sinh(2\,s)\\
&&-\frac{1}{8}\,\cosh(2\,v)\,\cosh(2\,s)\,\cosh(6\,z)\,\cos(2\,r_1)\,\cos(2\,r_2)\\
&&-\frac{1}{8}\,\cosh(2\,v)\,\cosh(2\,s)\,\cosh(6\,z)\,\cos(2\,r_1)\,\cos(4\,r_2)\\
&&-3\,\cosh(2\,s)\,\cos(2\,r_1)\,\cos(2\,r_2)\,\cos(r_5)\,\sinh(2\,z)\,\sinh(v)\\
&&-\frac{1}{8}\,\cosh(2\,v)\,\cosh(2\,s)\,\cosh(6\,z)\,\cos(2\,r_1)\,\cos(2\,r_2)\,\cos(2\,r_5)\\
&&-\frac{1}{8}\,\cosh(2\,v)\,\cosh(2\,s)\,\cosh(6\,z)\,\cos(2\,r_1)\,\cos(4\,r_2)\,\cos(2\,r_5)\\
&&-\frac{1}{4}\,\cosh(2\,v)\,\sin(2\,r_3)\,\sin(r_2)\,\sin(2\,r_1)\,\sinh(6\,z)\,\sinh(2\,s)\\
&&+\frac{1}{4}\,\cosh(2\,v)\,\sin(2\,r_3)\,\sin(3\,r_2)\,\sin(2\,r_1)\,\sinh(6\,z)\,\sinh(2\,s)\\
&&+\frac{1}{8}\,\cosh(2\,v)\,\cos(2\,r_1)\,\cos(2\,r_2)\,\cos(2\,r_3)\,\sinh(6\,z)\,\sinh(2\,s)\\
&&-\frac{1}{8}\,\cosh(2\,v)\,\cos(2\,r_1)\,\cos(4\,r_2)\,\cos(2\,r_3)\,\sinh(6\,z)\,\sinh(2\,s)\\
&&+12\,\cosh(2\,z)\,\cos(r_5)\,\sin(2\,r_3)\,\sin(r_2)\,\sin(2\,r_1)\,\sinh(2\,s)\,\sinh(v)\\
&&-3\,\cosh(2\,z)\,\cos(2\,r_1)\,\cos(2\,r_2)\,\cos(2\,r_3)\,\cos(r_5)\,\sinh(2\,s)\,\sinh(v)\\
&&-\frac{1}{4}\,\cosh(2\,v)\,\cos(2\,r_5)\,\sin(2\,r_3)\,\sin(r_2)\,\sin(2\,r_1)\,\sinh(6\,z)\,\sinh(2\,s)\\
&&+\frac{1}{4}\,\cosh(2\,v)\,\cos(2\,r_5)\,\sin(2\,r_3)\,\sin(3\,r_2)\,\sin(2\,r_1)\,\sinh(6\,z)\,\sinh(2\,s)\\
&&+\frac{1}{8}\,\cosh(2\,v)\,\cos(2\,r_1)\,\cos(2\,r_2)\,\cos(2\,r_3)\,\cos(2\,r_5)\,\sinh(6\,z)\,\sinh(2\,s)\\
&&-\frac{1}{8}\,\cosh(2\,v)\,\cos(2\,r_1)\,\cos(4\,r_2)\,\cos(2\,r_3)\,\cos(2\,r_5)\,\sinh(6\,z)\,\sinh(2\,s)
\end{array}
\label{so71sl3}
\ee
}

$\bullet\;\;$ $G_0=SO(6,2)\times SO(6,2)$, potential restricted to
singlets under $SO(6)_{\rm diag}$: 
{\small
\be\label{so62phi}
\begin{array}{lcl}
-8g^{-2}V&=&27+3\,\cosh(4\,z)+3\,\cosh(4\,z)\,\cos(2\,r_2)-3\,\cosh(4\,z)\,\cos(2\,r_1)\\
&&-3\,\cosh(4\,z)\,\cos(2\,r_1)\,\cos(2\,r_2)+\frac{1}{4}\cosh(4\,s)\\
&&+\frac{1}{4}\cosh(4\,s)\,\cos(2\,r_2)-\frac{1}{4}\cosh(4\,s)\,\cos(2\,r_1)\\
&&-\frac{1}{4}\cosh(4\,s)\,\cos(2\,r_1)\,\cos(2\,r_2)+9\,\cosh(2\,s)\,\cosh(2\,z)\\
&&+\frac{3}{4}\cosh(2\,s)\,\cosh(6\,z)-3\,\cos(2\,r_3)\,\sinh(2\,z)\,\sinh(2\,s)\\
&&+\frac{1}{4}\cos(2\,r_3)\,\sinh(6\,z)\,\sinh(2\,s)-3\,\cosh(2\,s)\,\cosh(2\,z)\,\cos(2\,r_2)\\
&&-\frac{1}{4}\cosh(2\,s)\,\cosh(6\,z)\,\cos(2\,r_2)-3\,\cos(2\,r_2)\,\cos(2\,r_3)\,\sinh(2\,z)\,\sinh(2\,s)\\
&&+\frac{1}{4}\cos(2\,r_2)\,\cos(2\,r_3)\,\sinh(6\,z)\,\sinh(2\,s)+3\,\cosh(2\,s)\,\cosh(2\,z)\,\cos(2\,r_1)\\
&&+\frac{1}{4}\cosh(2\,s)\,\cosh(6\,z)\,\cos(2\,r_1)-9\,\cos(2\,r_1)\,\cos(2\,r_3)\,\sinh(2\,z)\,\sinh(2\,s)\\
&&+\frac{3}{4}\cos(2\,r_1)\,\cos(2\,r_3)\,\sinh(6\,z)\,\sinh(2\,s)\\
&&+3\,\cosh(2\,s)\,\cosh(2\,z)\,\cos(2\,r_1)\,\cos(2\,r_2)\\
&&+\frac{1}{4}\cosh(2\,s)\,\cosh(6\,z)\,\cos(2\,r_1)\,\cos(2\,r_2)\\
&&-12\,\sin(2\,r_3)\,\sin(r_2)\,\sin(2\,r_1)\,\sinh(2\,z)\,\sinh(2\,s)\\
&&+\sin(2\,r_3)\,\sin(r_2)\,\sin(2\,r_1)\,\sinh(6\,z)\,\sinh(2\,s)\\
&&+3\,\cos(2\,r_1)\,\cos(2\,r_2)\,\cos(2\,r_3)\,\sinh(2\,z)\,\sinh(2\,s)\\
&&-\frac{1}{4}\cos(2\,r_1)\,\cos(2\,r_2)\,\cos(2\,r_3)\,\sinh(6\,z)\,\sinh(2\,s)\\
&&+\,\cosh(2\,v)-9\,\cosh(v)\,\cosh(4\,z)+3\,\cosh(v)\,\cosh(4\,z)\,\cos(2\,r_2)\\&&
-3\,\cosh(v)\,\cosh(4\,z)\,\cos(2\,r_1)-3\,\cosh(v)\,\cosh(4\,z)\,\cos(2\,r_1)\,\cos(2\,r_2)\\
&&-\frac{1}{4}\cosh(2\,v)\,\cosh(4\,s)-\frac{1}{4}\cosh(2\,v)\,\cosh(4\,s)\,\cos(2\,r_2)\\
&&+\frac{1}{4}\cosh(2\,v)\,\cosh(4\,s)\,\cos(2\,r_1)\\
&&+\frac{1}{4}\cosh(2\,v)\,\cosh(4\,s)\,\cos(2\,r_1)\,\cos(2\,r_2)-15\,\cosh(v)\,\cosh(2\,s)\,\cosh(2\,z)\\
&&-\frac{3}{4}\cosh(2\,v)\,\cosh(2\,s)\,\cosh(6\,z)-3\,\cosh(v)\,\cos(2\,r_3)\,\sinh(2\,z)\,\sinh(2\,s)\\
&&-\frac{1}{4}\cosh(2\,v)\,\cos(2\,r_3)\,\sinh(6\,z)\,\sinh(2\,s)\\
&&-3\,\cosh(v)\,\cosh(2\,s)\,\cosh(2\,z)\,\cos(2\,r_2)\\
&&+\frac{1}{4}\cosh(2\,v)\,\cosh(2\,s)\,\cosh(6\,z)\,\cos(2\,r_2)\\
&&-3\,\cosh(v)\,\cos(2\,r_2)\,\cos(2\,r_3)\,\sinh(2\,z)\,\sinh(2\,s)\\
&&-\frac{1}{4}\cosh(2\,v)\,\cos(2\,r_2)\,\cos(2\,r_3)\,\sinh(6\,z)\,\sinh(2\,s)\\
&&+3\,\cosh(v)\,\cosh(2\,s)\,\cosh(2\,z)\,\cos(2\,r_1)\\
&&-\frac{1}{4}\cosh(2\,v)\,\cosh(2\,s)\,\cosh(6\,z)\,\cos(2\,r_1)\\
&&-9\,\cosh(v)\,\cos(2\,r_1)\,\cos(2\,r_3)\,\sinh(2\,z)\,\sinh(2\,s)\\
&&-\frac{3}{4}\cosh(2\,v)\,\cos(2\,r_1)\,\cos(2\,r_3)\,\sinh(6\,z)\,\sinh(2\,s)\\
&&+3\,\cosh(v)\,\cosh(2\,s)\,\cosh(2\,z)\,\cos(2\,r_1)\,\cos(2\,r_2)\\
&&-\frac{1}{4}\cosh(2\,v)\,\cosh(2\,s)\,\cosh(6\,z)\,\cos(2\,r_1)\,\cos(2\,r_2)\\
&&-12\,\cosh(v)\,\sin(2\,r_3)\,\sin(r_2)\,\sin(2\,r_1)\,\sinh(2\,z)\,\sinh(2\,s)\\
&&-\,\cosh(2\,v)\,\sin(2\,r_3)\,\sin(r_2)\,\sin(2\,r_1)\,\sinh(6\,z)\,\sinh(2\,s)\\
&&+3\,\cosh(v)\,\cos(2\,r_1)\,\cos(2\,r_2)\,\cos(2\,r_3)\,\sinh(2\,z)\,\sinh(2\,s)\\
&&+\frac{1}{4}\cosh(2\,v)\,\cos(2\,r_1)\,\cos(2\,r_2)\,\cos(2\,r_3)\,\sinh(6\,z)\,\sinh(2\,s)
\end{array}
\ee}

$\bullet\;\;$ $G_0=G_2\times F_{4(-20)}$, potential restricted to
singlets under $SU(3)\cro SU(3)$:
{\small
\ben
\begin{array}{lcl}
-8g^{-2}V&=&\frac{24125}{2048}+\frac{9}{2048}\,\cos(8\,r_5)+\frac{9}{2048}\,\cos(8\,r_2)\\
&&+\frac{115}{128}\,\sin(4\,r_5)\,\sin(4\,r_2)-\frac{27}{2048}\,\cos(8\,r_2)\,\cos(8\,r_5)\\
&&-\frac{1}{64}\,\cos(4\,r_2)\,\cos(4\,r_3-4\,r_6)\,\cos(4\,r_5)-\frac{9}{2048}\,\cos(8\,r_1-8\,r_4)\\
&&-\frac{9}{2048}\,\cos(8\,r_1-8\,r_4)\,\cos(8\,r_5)-\frac{1}{64}\,\cos(4\,r_1-4\,r_4)\,\cos(4\,r_3-4\,r_6)\\
&&+\frac{1}{64}\,\sin(4\,r_5)\,\sin(4\,r_3-4\,r_6)\,\sin(4\,r_1-4\,r_4)-\frac{9}{2048}\,\cos(8\,r_1-8\,r_4)\,\cos(8\,r_2)\\
&&-\frac{9}{512}\,\cos(4\,r_1-4\,r_4)\,\sin(8\,r_5)\,\sin(8\,r_2)\\
&&+\frac{115}{128}\,\cos(4\,r_1-4\,r_4)\,\cos(4\,r_2)\,\cos(4\,r_5)\\
&&-\frac{9}{2048}\,\cos(8\,r_1-8\,r_4)\,\cos(8\,r_2)\,\cos(8\,r_5)\\
&&+\frac{1}{64}\,\sin(4\,r_3-4\,r_6)\,\sin(4\,r_2)\,\sin(4\,r_1-4\,r_4)\\
&&-\frac{1}{64}\,\cos(4\,r_1-4\,r_4)\,\cos(4\,r_3-4\,r_6)\,\sin(4\,r_5)\,\sin(4\,r_2)+\frac{449}{512}\,\cosh(z)\\
&&-\frac{65}{2048}\,\cosh(2\,z)-\frac{3}{512}\,\cosh(z)\,\cos(8\,r_5)+\frac{3}{2048}\,\cosh(2\,z)\,\cos(8\,r_5)\\
&&-\frac{3}{512}\,\cosh(z)\,\cos(8\,r_2)+\frac{3}{2048}\,\cosh(2\,z)\,\cos(8\,r_2)\\
&&-\frac{29}{32}\,\cosh(z)\,\sin(4\,r_5)\,\sin(4\,r_2)+\frac{1}{128}\,\cosh(2\,z)\,\sin(4\,r_5)\,\sin(4\,r_2)\\
&&+\frac{9}{512}\,\cosh(z)\,\cos(8\,r_2)\,\cos(8\,r_5)-\frac{9}{2048}\,\cosh(2\,z)\,\cos(8\,r_2)\,\cos(8\,r_5)\\
&&+\frac{1}{64}\,\cosh(2\,z)\,\cos(4\,r_2)\,\cos(4\,r_3-4\,r_6)\,\cos(4\,r_5)+\frac{3}{512}\,\cosh(z)\,\cos(8\,r_1-8\,r_4)\\
&&-\frac{3}{2048}\,\cosh(2\,z)\,\cos(8\,r_1-8\,r_4)+\frac{3}{512}\,\cosh(z)\,\cos(8\,r_1-8\,r_4)\,\cos(8\,r_5)\\
&&-\frac{3}{2048}\,\cosh(2\,z)\,\cos(8\,r_1-8\,r_4)\,\cos(8\,r_5)\\
&&+\frac{1}{64}\,\cosh(2\,z)\,\cos(4\,r_1-4\,r_4)\,\cos(4\,r_3-4\,r_6)\\
&&-\frac{1}{64}\,\cosh(2\,z)\,\sin(4\,r_5)\,\sin(4\,r_3-4\,r_6)\,\sin(4\,r_1-4\,r_4)\\
&&+\frac{3}{512}\,\cosh(z)\,\cos(8\,r_1-8\,r_4)\,\cos(8\,r_2)\\
&&-\frac{3}{2048}\,\cosh(2\,z)\,\cos(8\,r_1-8\,r_4)\,\cos(8\,r_2)\\
&&+\frac{3}{128}\,\cosh(z)\,\cos(4\,r_1-4\,r_4)\,\sin(8\,r_5)\,\sin(8\,r_2)\\
&&-\frac{3}{512}\,\cosh(2\,z)\,\cos(4\,r_1-4\,r_4)\,\sin(8\,r_5)\,\sin(8\,r_2)\\
&&-\frac{29}{32}\,\cosh(z)\,\cos(4\,r_1-4\,r_4)\,\cos(4\,r_2)\,\cos(4\,r_5)\\
&&+\frac{3}{512}\,\cosh(z)\,\cos(8\,r_1-8\,r_4)\,\cos(8\,r_2)\,\cos(8\,r_5)\\
&&+\frac{1}{128}\,\cosh(2\,z)\,\cos(4\,r_1-4\,r_4)\,\cos(4\,r_2)\,\cos(4\,r_5)\\
&&-\frac{3}{2048}\,\cosh(2\,z)\,\cos(8\,r_1-8\,r_4)\,\cos(8\,r_2)\,\cos(8\,r_5)\\
&&-\frac{1}{64}\,\cosh(2\,z)\,\sin(4\,r_3-4\,r_6)\,\sin(4\,r_2)\,\sin(4\,r_1-4\,r_4)\\
&&+\frac{1}{64}\,\cosh(2\,z)\,\cos(4\,r_1-4\,r_4)\,\cos(4\,r_3-4\,r_6)\,\sin(4\,r_5)\,\sin(4\,r_2)\\
&&+\frac{449}{512}\,\cosh(s)-\frac{65}{2048}\,\cosh(2\,s)-\frac{3}{512}\,\cosh(s)\,\cos(8\,r_5)\\
&&+\frac{3}{2048}\,\cosh(2\,s)\,\cos(8\,r_5)-\frac{3}{512}\,\cosh(s)\,\cos(8\,r_2)\\
&&+\frac{3}{2048}\,\cosh(2\,s)\,\cos(8\,r_2)-\frac{29}{32}\,\cosh(s)\,\sin(4\,r_5)\,\sin(4\,r_2)\\
&&+\frac{1}{128}\,\cosh(2\,s)\,\sin(4\,r_5)\,\sin(4\,r_2)+\frac{9}{512}\,\cosh(s)\,\cos(8\,r_2)\,\cos(8\,r_5)\\
&&-\frac{9}{2048}\,\cosh(2\,s)\,\cos(8\,r_2)\,\cos(8\,r_5)\\
&&+\frac{1}{64}\,\cosh(2\,s)\,\cos(4\,r_2)\,\cos(4\,r_3-4\,r_6)\,\cos(4\,r_5)\\
&&+\frac{3}{512}\,\cosh(s)\,\cos(8\,r_1-8\,r_4)-\frac{3}{2048}\,\cosh(2\,s)\,\cos(8\,r_1-8\,r_4)\\
&&+\frac{3}{512}\,\cosh(s)\,\cos(8\,r_1-8\,r_4)\,\cos(8\,r_5)\\
&&-\frac{3}{2048}\,\cosh(2\,s)\,\cos(8\,r_1-8\,r_4)\,\cos(8\,r_5)\\
\end{array}
\een
\ben
\begin{array}{lcl}
&&+\frac{1}{64}\,\cosh(2\,s)\,\cos(4\,r_1-4\,r_4)\,\cos(4\,r_3-4\,r_6)\\
&&-\frac{1}{64}\,\cosh(2\,s)\,\sin(4\,r_5)\,\sin(4\,r_3-4\,r_6)\,\sin(4\,r_1-4\,r_4)\\
&&+\frac{3}{512}\,\cosh(s)\,\cos(8\,r_1-8\,r_4)\,\cos(8\,r_2)\\
&&-\frac{3}{2048}\,\cosh(2\,s)\,\cos(8\,r_1-8\,r_4)\,\cos(8\,r_2)\\
&&+\frac{3}{128}\,\cosh(s)\,\cos(4\,r_1-4\,r_4)\,\sin(8\,r_5)\,\sin(8\,r_2)\\
&&-\frac{3}{512}\,\cosh(2\,s)\,\cos(4\,r_1-4\,r_4)\,\sin(8\,r_5)\,\sin(8\,r_2)\\
&&-\frac{29}{32}\,\cosh(s)\,\cos(4\,r_1-4\,r_4)\,\cos(4\,r_2)\,\cos(4\,r_5)\\
&&+\frac{3}{512}\,\cosh(s)\,\cos(8\,r_1-8\,r_4)\,\cos(8\,r_2)\,\cos(8\,r_5)\\
&&+\frac{1}{128}\,\cosh(2\,s)\,\cos(4\,r_1-4\,r_4)\,\cos(4\,r_2)\,\cos(4\,r_5)\\
&&-\frac{3}{2048}\,\cosh(2\,s)\,\cos(8\,r_1-8\,r_4)\,\cos(8\,r_2)\,\cos(8\,r_5)\\
&&-\frac{1}{64}\,\cosh(2\,s)\,\sin(4\,r_3-4\,r_6)\,\sin(4\,r_2)\,\sin(4\,r_1-4\,r_4)\\
&&+\frac{1}{64}\,\cosh(2\,s)\,\cos(4\,r_1-4\,r_4)\,\cos(4\,r_3-4\,r_6)\,\sin(4\,r_5)\,\sin(4\,r_2)\\
&&+\frac{341}{128}\,\cosh(s)\,\cosh(z)-\frac{21}{512}\,\cosh(s)\,\cosh(2\,z)\\
&&-\frac{21}{512}\,\cosh(2\,s)\,\cosh(z)-\frac{107}{2048}\,\cosh(2\,s)\,\cosh(2\,z)\\
&&+\frac{1}{128}\,\cosh(s)\,\cosh(z)\,\cos(8\,r_5)-\frac{1}{512}\,\cosh(s)\,\cosh(2\,z)\,\cos(8\,r_5)\\
&&-\frac{1}{512}\,\cosh(2\,s)\,\cosh(z)\,\cos(8\,r_5)+\frac{1}{2048}\,\cosh(2\,s)\,\cosh(2\,z)\,\cos(8\,r_5)\\
&&+\frac{1}{128}\,\cosh(s)\,\cosh(z)\,\cos(8\,r_2)-\frac{1}{512}\,\cosh(s)\,\cosh(2\,z)\,\cos(8\,r_2)\\
&&-\frac{1}{512}\,\cosh(2\,s)\,\cosh(z)\,\cos(8\,r_2)+\frac{1}{2048}\,\cosh(2\,s)\,\cosh(2\,z)\,\cos(8\,r_2)\\
&&+\frac{7}{8}\,\cosh(s)\,\cosh(z)\,\sin(4\,r_5)\,\sin(4\,r_2)\\
&&+\frac{1}{32}\,\cosh(s)\,\cosh(2\,z)\,\sin(4\,r_5)\,\sin(4\,r_2)\\
&&+\frac{1}{32}\,\cosh(2\,s)\,\cosh(z)\,\sin(4\,r_5)\,\sin(4\,r_2)\\
&&-\frac{5}{128}\,\cosh(2\,s)\,\cosh(2\,z)\,\sin(4\,r_5)\,\sin(4\,r_2)\\
&&-\frac{3}{128}\,\cosh(s)\,\cosh(z)\,\cos(8\,r_2)\,\cos(8\,r_5)\\
&&+\frac{3}{512}\,\cosh(s)\,\cosh(2\,z)\,\cos(8\,r_2)\,\cos(8\,r_5)\\
&&+\frac{3}{512}\,\cosh(2\,s)\,\cosh(z)\,\cos(8\,r_2)\,\cos(8\,r_5)\\
&&-\frac{3}{2048}\,\cosh(2\,s)\,\cosh(2\,z)\,\cos(8\,r_2)\,\cos(8\,r_5)\\
&&-\frac{1}{64}\,\cosh(2\,s)\,\cosh(2\,z)\,\cos(4\,r_2)\,\cos(4\,r_3-4\,r_6)\,\cos(4\,r_5)\\
&&-\frac{1}{128}\,\cosh(s)\,\cosh(z)\,\cos(8\,r_1-8\,r_4)\\
&&+\frac{1}{512}\,\cosh(s)\,\cosh(2\,z)\,\cos(8\,r_1-8\,r_4)\\
&&+\frac{1}{512}\,\cosh(2\,s)\,\cosh(z)\,\cos(8\,r_1-8\,r_4)\\
&&-\frac{1}{2048}\,\cosh(2\,s)\,\cosh(2\,z)\,\cos(8\,r_1-8\,r_4)\\
&&-\frac{1}{128}\,\cosh(s)\,\cosh(z)\,\cos(8\,r_1-8\,r_4)\,\cos(8\,r_5)\\
&&+\frac{1}{512}\,\cosh(s)\,\cosh(2\,z)\,\cos(8\,r_1-8\,r_4)\,\cos(8\,r_5)\\
&&+\frac{1}{512}\,\cosh(2\,s)\,\cosh(z)\,\cos(8\,r_1-8\,r_4)\,\cos(8\,r_5)\\
&&-\frac{1}{2048}\,\cosh(2\,s)\,\cosh(2\,z)\,\cos(8\,r_1-8\,r_4)\,\cos(8\,r_5)\\
&&-\frac{1}{64}\,\cosh(2\,s)\,\cosh(2\,z)\,\cos(4\,r_1-4\,r_4)\,\cos(4\,r_3-4\,r_6)\\
&&+\frac{1}{64}\,\cosh(2\,s)\,\cosh(2\,z)\,\sin(4\,r_5)\,\sin(4\,r_3-4\,r_6)\,\sin(4\,r_1-4\,r_4)\\
&&-\frac{1}{128}\,\cosh(s)\,\cosh(z)\,\cos(8\,r_1-8\,r_4)\,\cos(8\,r_2)\\
&&+\frac{1}{512}\,\cosh(s)\,\cosh(2\,z)\,\cos(8\,r_1-8\,r_4)\,\cos(8\,r_2)\\
&&+\frac{1}{512}\,\cosh(2\,s)\,\cosh(z)\,\cos(8\,r_1-8\,r_4)\,\cos(8\,r_2)\\
&&-\frac{1}{2048}\,\cosh(2\,s)\,\cosh(2\,z)\,\cos(8\,r_1-8\,r_4)\,\cos(8\,r_2)\\
\end{array}
\een
\ben
\begin{array}{lcl}
&&-\frac{1}{32}\,\cosh(s)\,\cosh(z)\,\cos(4\,r_1-4\,r_4)\,\sin(8\,r_5)\,\sin(8\,r_2)\\
&&+\frac{1}{128}\,\cosh(s)\,\cosh(2\,z)\,\cos(4\,r_1-4\,r_4)\,\sin(8\,r_5)\,\sin(8\,r_2)\\
&&+\frac{1}{128}\,\cosh(2\,s)\,\cosh(z)\,\cos(4\,r_1-4\,r_4)\,\sin(8\,r_5)\,\sin(8\,r_2)\\
&&-\frac{1}{512}\,\cosh(2\,s)\,\cosh(2\,z)\,\cos(4\,r_1-4\,r_4)\,\sin(8\,r_5)\,\sin(8\,r_2)\\
&&+\frac{7}{8}\,\cosh(s)\,\cosh(z)\,\cos(4\,r_1-4\,r_4)\,\cos(4\,r_2)\,\cos(4\,r_5)\\
&&-\frac{1}{128}\,\cosh(s)\,\cosh(z)\,\cos(8\,r_1-8\,r_4)\,\cos(8\,r_2)\,\cos(8\,r_5)\\
&&+\frac{1}{32}\,\cosh(s)\,\cosh(2\,z)\,\cos(4\,r_1-4\,r_4)\,\cos(4\,r_2)\,\cos(4\,r_5)\\
&&+\frac{1}{512}\,\cosh(s)\,\cosh(2\,z)\,\cos(8\,r_1-8\,r_4)\,\cos(8\,r_2)\,\cos(8\,r_5)\\
&&+\frac{1}{32}\,\cosh(2\,s)\,\cosh(z)\,\cos(4\,r_1-4\,r_4)\,\cos(4\,r_2)\,\cos(4\,r_5)\\
&&+\frac{1}{512}\,\cosh(2\,s)\,\cosh(z)\,\cos(8\,r_1-8\,r_4)\,\cos(8\,r_2)\,\cos(8\,r_5)\\
&&-\frac{5}{128}\,\cosh(2\,s)\,\cosh(2\,z)\,\cos(4\,r_1-4\,r_4)\,\cos(4\,r_2)\,\cos(4\,r_5)\\
&&-\frac{1}{2048}\,\cosh(2\,s)\,\cosh(2\,z)\,\cos(8\,r_1-8\,r_4)\,\cos(8\,r_2)\,\cos(8\,r_5)\\
&&+\frac{1}{64}\,\cosh(2\,s)\,\cosh(2\,z)\,\sin(4\,r_3-4\,r_6)\,\sin(4\,r_2)\,\sin(4\,r_1-4\,r_4)\\
&&+\frac{1}{64}\,\sin(2\,r_3-2\,r_6)\,\sin(2\,r_2-6\,r_5)\,\sin(2\,r_1-2\,r_4)\,\sinh(z)\,\sinh(s)\\
&&+\frac{229}{64}\,\sin(2\,r_3-2\,r_6)\,\sin(2\,r_2+2\,r_5)\,\sin(2\,r_1-2\,r_4)\,\sinh(z)\,\sinh(s)\\
&&-\frac{1}{64}\,\sin(2\,r_3-2\,r_6)\,\sin(6\,r_2-2\,r_5)\,\sin(2\,r_1-2\,r_4)\,\sinh(z)\,\sinh(s)\\
&&+\frac{3}{64}\,\sin(2\,r_3-2\,r_6)\,\sin(6\,r_2+6\,r_5)\,\sin(2\,r_1-2\,r_4)\,\sinh(z)\,\sinh(s)\\
&&+\frac{1}{64}\,\sin(2\,r_3-2\,r_6)\,\sin(2\,r_2-6\,r_5)\,\sin(6\,r_1-6\,r_4)\,\sinh(z)\,\sinh(s)\\
&&+\frac{1}{64}\,\sin(2\,r_3-2\,r_6)\,\sin(2\,r_2+2\,r_5)\,\sin(6\,r_1-6\,r_4)\,\sinh(z)\,\sinh(s)\\
&&-\frac{1}{64}\,\sin(2\,r_3-2\,r_6)\,\sin(6\,r_2-2\,r_5)\,\sin(6\,r_1-6\,r_4)\,\sinh(z)\,\sinh(s)\\
&&-\frac{1}{64}\,\sin(2\,r_3-2\,r_6)\,\sin(6\,r_2+6\,r_5)\,\sin(6\,r_1-6\,r_4)\,\sinh(z)\,\sinh(s)\\
&&-\frac{1}{128}\,\sin(2\,r_3-2\,r_6)\,\sin(2\,r_2-6\,r_5)\,\sin(2\,r_1-2\,r_4)\,\sinh(2\,z)\,\sinh(s)\\
&&-\frac{5}{128}\,\sin(2\,r_3-2\,r_6)\,\sin(2\,r_2+2\,r_5)\,\sin(2\,r_1-2\,r_4)\,\sinh(2\,z)\,\sinh(s)\\
&&+\frac{1}{128}\,\sin(2\,r_3-2\,r_6)\,\sin(6\,r_2-2\,r_5)\,\sin(2\,r_1-2\,r_4)\,\sinh(2\,z)\,\sinh(s)\\
&&-\frac{3}{128}\,\sin(2\,r_3-2\,r_6)\,\sin(6\,r_2+6\,r_5)\,\sin(2\,r_1-2\,r_4)\,\sinh(2\,z)\,\sinh(s)\\
&&-\frac{1}{128}\,\sin(2\,r_3-2\,r_6)\,\sin(2\,r_2-6\,r_5)\,\sin(6\,r_1-6\,r_4)\,\sinh(2\,z)\,\sinh(s)\\
&&-\frac{1}{128}\,\sin(2\,r_3-2\,r_6)\,\sin(2\,r_2+2\,r_5)\,\sin(6\,r_1-6\,r_4)\,\sinh(2\,z)\,\sinh(s)\\
&&+\frac{1}{128}\,\sin(2\,r_3-2\,r_6)\,\sin(6\,r_2-2\,r_5)\,\sin(6\,r_1-6\,r_4)\,\sinh(2\,z)\,\sinh(s)\\
&&+\frac{1}{128}\,\sin(2\,r_3-2\,r_6)\,\sin(6\,r_2+6\,r_5)\,\sin(6\,r_1-6\,r_4)\,\sinh(2\,z)\,\sinh(s)\\
&&-\frac{1}{128}\,\sin(2\,r_3-2\,r_6)\,\sin(2\,r_2-6\,r_5)\,\sin(2\,r_1-2\,r_4)\,\sinh(z)\,\sinh(2\,s)\\
&&-\frac{5}{128}\,\sin(2\,r_3-2\,r_6)\,\sin(2\,r_2+2\,r_5)\,\sin(2\,r_1-2\,r_4)\,\sinh(z)\,\sinh(2\,s)\\
&&+\frac{1}{128}\,\sin(2\,r_3-2\,r_6)\,\sin(6\,r_2-2\,r_5)\,\sin(2\,r_1-2\,r_4)\,\sinh(z)\,\sinh(2\,s)\\
&&-\frac{3}{128}\,\sin(2\,r_3-2\,r_6)\,\sin(6\,r_2+6\,r_5)\,\sin(2\,r_1-2\,r_4)\,\sinh(z)\,\sinh(2\,s)\\
&&-\frac{1}{128}\,\sin(2\,r_3-2\,r_6)\,\sin(2\,r_2-6\,r_5)\,\sin(6\,r_1-6\,r_4)\,\sinh(z)\,\sinh(2\,s)\\
&&-\frac{1}{128}\,\sin(2\,r_3-2\,r_6)\,\sin(2\,r_2+2\,r_5)\,\sin(6\,r_1-6\,r_4)\,\sinh(z)\,\sinh(2\,s)\\
&&+\frac{1}{128}\,\sin(2\,r_3-2\,r_6)\,\sin(6\,r_2-2\,r_5)\,\sin(6\,r_1-6\,r_4)\,\sinh(z)\,\sinh(2\,s)\\
&&+\frac{1}{128}\,\sin(2\,r_3-2\,r_6)\,\sin(6\,r_2+6\,r_5)\,\sin(6\,r_1-6\,r_4)\,\sinh(z)\,\sinh(2\,s)\\
&&+\frac{1}{256}\,\sin(2\,r_3-2\,r_6)\,\sin(2\,r_2-6\,r_5)\,\sin(2\,r_1-2\,r_4)\,\sinh(2\,z)\,\sinh(2\,s)\\
&&-\frac{27}{256}\,\sin(2\,r_3-2\,r_6)\,\sin(2\,r_2+2\,r_5)\,\sin(2\,r_1-2\,r_4)\,\sinh(2\,z)\,\sinh(2\,s)\\
&&-\frac{1}{256}\,\sin(2\,r_3-2\,r_6)\,\sin(6\,r_2-2\,r_5)\,\sin(2\,r_1-2\,r_4)\,\sinh(2\,z)\,\sinh(2\,s)\\
&&+\frac{3}{256}\,\sin(2\,r_3-2\,r_6)\,\sin(6\,r_2+6\,r_5)\,\sin(2\,r_1-2\,r_4)\,\sinh(2\,z)\,\sinh(2\,s)\\
&&+\frac{1}{256}\,\sin(2\,r_3-2\,r_6)\,\sin(2\,r_2-6\,r_5)\,\sin(6\,r_1-6\,r_4)\,\sinh(2\,z)\,\sinh(2\,s)\\
\end{array}
\een
\begin{equation}
\begin{array}{lcl}
&&+\frac{1}{256}\,\sin(2\,r_3-2\,r_6)\,\sin(2\,r_2+2\,r_5)\,\sin(6\,r_1-6\,r_4)\,\sinh(2\,z)\,\sinh(2\,s)\\
&&-\frac{1}{256}\,\sin(2\,r_3-2\,r_6)\,\sin(6\,r_2-2\,r_5)\,\sin(6\,r_1-6\,r_4)\,\sinh(2\,z)\,\sinh(2\,s)\\
&&-\frac{1}{256}\,\sin(2\,r_3-2\,r_6)\,\sin(6\,r_2+6\,r_5)\,\sin(6\,r_1-6\,r_4)\,\sinh(2\,z)\,\sinh(2\,s)\\
&&-\frac{229}{64}\,\cos(2\,r_1-2\,r_4)\,\cos(2\,r_2-2\,r_5)\,\cos(2\,r_3-2\,r_6)\,\sinh(z)\,\sinh(s)\\
&&-\frac{1}{64}\,\cos(2\,r_1-2\,r_4)\,\cos(2\,r_2+6\,r_5)\,\cos(2\,r_3-2\,r_6)\,\sinh(z)\,\sinh(s)\\
&&+\frac{3}{64}\,\cos(2\,r_1-2\,r_4)\,\cos(6\,r_2-6\,r_5)\,\cos(2\,r_3-2\,r_6)\,\sinh(z)\,\sinh(s)\\
&&-\frac{1}{64}\,\cos(2\,r_1-2\,r_4)\,\cos(6\,r_2+2\,r_5)\,\cos(2\,r_3-2\,r_6)\,\sinh(z)\,\sinh(s)\\
&&+\frac{1}{64}\,\cos(6\,r_1-6\,r_4)\,\cos(2\,r_2-2\,r_5)\,\cos(2\,r_3-2\,r_6)\,\sinh(z)\,\sinh(s)\\
&&+\frac{1}{64}\,\cos(6\,r_1-6\,r_4)\,\cos(2\,r_2+6\,r_5)\,\cos(2\,r_3-2\,r_6)\,\sinh(z)\,\sinh(s)\\
&&+\frac{1}{64}\,\cos(6\,r_1-6\,r_4)\,\cos(6\,r_2-6\,r_5)\,\cos(2\,r_3-2\,r_6)\,\sinh(z)\,\sinh(s)\\
&&+\frac{1}{64}\,\cos(6\,r_1-6\,r_4)\,\cos(6\,r_2+2\,r_5)\,\cos(2\,r_3-2\,r_6)\,\sinh(z)\,\sinh(s)\\
&&+\frac{5}{128}\,\cos(2\,r_1-2\,r_4)\,\cos(2\,r_2-2\,r_5)\,\cos(2\,r_3-2\,r_6)\,\sinh(2\,z)\,\sinh(s)\\
&&+\frac{1}{128}\,\cos(2\,r_1-2\,r_4)\,\cos(2\,r_2+6\,r_5)\,\cos(2\,r_3-2\,r_6)\,\sinh(2\,z)\,\sinh(s)\\
&&-\frac{3}{128}\,\cos(2\,r_1-2\,r_4)\,\cos(6\,r_2-6\,r_5)\,\cos(2\,r_3-2\,r_6)\,\sinh(2\,z)\,\sinh(s)\\
&&+\frac{1}{128}\,\cos(2\,r_1-2\,r_4)\,\cos(6\,r_2+2\,r_5)\,\cos(2\,r_3-2\,r_6)\,\sinh(2\,z)\,\sinh(s)\\
&&-\frac{1}{128}\,\cos(6\,r_1-6\,r_4)\,\cos(2\,r_2-2\,r_5)\,\cos(2\,r_3-2\,r_6)\,\sinh(2\,z)\,\sinh(s)\\
&&-\frac{1}{128}\,\cos(6\,r_1-6\,r_4)\,\cos(2\,r_2+6\,r_5)\,\cos(2\,r_3-2\,r_6)\,\sinh(2\,z)\,\sinh(s)\\
&&-\frac{1}{128}\,\cos(6\,r_1-6\,r_4)\,\cos(6\,r_2-6\,r_5)\,\cos(2\,r_3-2\,r_6)\,\sinh(2\,z)\,\sinh(s)\\
&&-\frac{1}{128}\,\cos(6\,r_1-6\,r_4)\,\cos(6\,r_2+2\,r_5)\,\cos(2\,r_3-2\,r_6)\,\sinh(2\,z)\,\sinh(s)\\
&&+\frac{5}{128}\,\cos(2\,r_1-2\,r_4)\,\cos(2\,r_2-2\,r_5)\,\cos(2\,r_3-2\,r_6)\,\sinh(z)\,\sinh(2\,s)\\
&&+\frac{1}{128}\,\cos(2\,r_1-2\,r_4)\,\cos(2\,r_2+6\,r_5)\,\cos(2\,r_3-2\,r_6)\,\sinh(z)\,\sinh(2\,s)\\
&&-\frac{3}{128}\,\cos(2\,r_1-2\,r_4)\,\cos(6\,r_2-6\,r_5)\,\cos(2\,r_3-2\,r_6)\,\sinh(z)\,\sinh(2\,s)\\
&&+\frac{1}{128}\,\cos(2\,r_1-2\,r_4)\,\cos(6\,r_2+2\,r_5)\,\cos(2\,r_3-2\,r_6)\,\sinh(z)\,\sinh(2\,s)\\
&&-\frac{1}{128}\,\cos(6\,r_1-6\,r_4)\,\cos(2\,r_2-2\,r_5)\,\cos(2\,r_3-2\,r_6)\,\sinh(z)\,\sinh(2\,s)\\
&&-\frac{1}{128}\,\cos(6\,r_1-6\,r_4)\,\cos(2\,r_2+6\,r_5)\,\cos(2\,r_3-2\,r_6)\,\sinh(z)\,\sinh(2\,s)\\
&&-\frac{1}{128}\,\cos(6\,r_1-6\,r_4)\,\cos(6\,r_2-6\,r_5)\,\cos(2\,r_3-2\,r_6)\,\sinh(z)\,\sinh(2\,s)\\
&&-\frac{1}{128}\,\cos(6\,r_1-6\,r_4)\,\cos(6\,r_2+2\,r_5)\,\cos(2\,r_3-2\,r_6)\,\sinh(z)\,\sinh(2\,s)\\
&&+\frac{27}{256}\,\cos(2\,r_1-2\,r_4)\,\cos(2\,r_2-2\,r_5)\,\cos(2\,r_3-2\,r_6)\,\sinh(2\,z)\,\sinh(2\,s)\\
&&-\frac{1}{256}\,\cos(2\,r_1-2\,r_4)\,\cos(2\,r_2+6\,r_5)\,\cos(2\,r_3-2\,r_6)\,\sinh(2\,z)\,\sinh(2\,s)\\
&&+\frac{3}{256}\,\cos(2\,r_1-2\,r_4)\,\cos(6\,r_2-6\,r_5)\,\cos(2\,r_3-2\,r_6)\,\sinh(2\,z)\,\sinh(2\,s)\\
&&-\frac{1}{256}\,\cos(2\,r_1-2\,r_4)\,\cos(6\,r_2+2\,r_5)\,\cos(2\,r_3-2\,r_6)\,\sinh(2\,z)\,\sinh(2\,s)\\
&&+\frac{1}{256}\,\cos(6\,r_1-6\,r_4)\,\cos(2\,r_2-2\,r_5)\,\cos(2\,r_3-2\,r_6)\,\sinh(2\,z)\,\sinh(2\,s)\\
&&+\frac{1}{256}\,\cos(6\,r_1-6\,r_4)\,\cos(2\,r_2+6\,r_5)\,\cos(2\,r_3-2\,r_6)\,\sinh(2\,z)\,\sinh(2\,s)\\
&&+\frac{1}{256}\,\cos(6\,r_1-6\,r_4)\,\cos(6\,r_2-6\,r_5)\,\cos(2\,r_3-2\,r_6)\,\sinh(2\,z)\,\sinh(2\,s)\\
&&+\frac{1}{256}\,\cos(6\,r_1-6\,r_4)\,\cos(6\,r_2+2\,r_5)\,\cos(2\,r_3-2\,r_6)\,\sinh(2\,z)\,\sinh(2\,s)\\
&&-\frac{1}{64}\,\cosh(2\,s)\,\cosh(2\,z)\,\cos(4\,r_1-4\,r_4)\,\cos(4\,r_3-4\,r_6)\,\sin(4\,r_5)\,\sin(4\,r_2)
\end{array}
\label{g2xf4}
\end{equation}
}


\begin{thebibliography}{10}

\bibitem{NicSam01}
H.~Nicolai and H.~Samtleben, { Maximal gauged supergravity in three
  dimensions},  { Phys. Rev. Lett.} { 86} (2001) 1686--1689,
[\href{http://xxx.lanl.gov/abs/http://arXiv.org/abs/hep-th/0010076}{{\tt
  http://arXiv.org/abs/hep-th/0010076}}].

\bibitem{NicSam01a}
H.~Nicolai and H.~Samtleben, { Compact and noncompact gauged maximal
  supergravities in three dimensions},  { JHEP} { 0104} (2001) 022,
[\href{http://xxx.lanl.gov/abs/http://arXiv.org/abs/hep-th/0103032}{{\tt
  http://arXiv.org/abs/hep-th/0103032}}].

\bibitem{Juli83}
B.~Julia, { Application of supergravity to gravitation theories},  in { Unified
  field theories in more than 4 dimensions} (V.~D. Sabbata and E.~Schmutzer,
  eds.), (Singapore), pp.~215--236, World Scientific, 1983.

\bibitem{GuSiTo86}
M.~G{\"u}naydin, G.~Sierra, and P.~K. Townsend, { The unitary supermultiplets
  of $d\!=\!3$ {A}nti-de {S}itter and $d\!=\!2$ conformal superalgebras},  {
  Nucl. Phys.} { B274} (1986)
429--447.

\bibitem{Warn83}
N.~P. Warner, { Some new extrema of the scalar potential of gauged ${N}\!=\!8$
  supergravity},  { Phys. Lett.} { B128} (1983)
169.

\bibitem{Fisc02}
T.~Fischbacher, { Some stationary points of gauged ${N}\!=\!16$ ${D}\!=\!3$
  supergravity},
\href{http://xxx.lanl.gov/abs/http://arXiv.org/abs/hep-th/0201030}{{\tt
  http://arXiv.org/abs/hep-th/0201030}}.

\bibitem{tf1}
T.~Fischbacher, { in preparation}.

\bibitem{BreFre82}
P.~Breitenlohner and D.~Z. Freedman, { Stability in gauged extended
  supergravity},  { Ann. Phys.} { 144} (1982)
249.

\bibitem{AGMOO00}
O.~Aharony, S.~S. Gubser, J.~Maldacena, H.~Ooguri, and Y.~Oz, { Large ${N}$
  field theories, string theory and gravity},  { Phys. Rept.} { 323} (2000)
  183--386,
[\href{http://xxx.lanl.gov/abs/http://arXiv.org/abs/hep-th/9905111}{{\tt
  http://arXiv.org/abs/hep-th/9905111}}].

\bibitem{FGPW99}
D.~Z. Freedman, S.~S. Gubser, K.~Pilch, and N.~P. Warner, { Renormalization
  group flows from holography --- supersymmetry and a $c$-theorem},  { Adv.
  Theor. Math. Phys.} { 3} (1999)
[\href{http://xxx.lanl.gov/abs/http://arXiv.org/abs/hep-th/9904017}{{\tt
  http://arXiv.org/abs/hep-th/9904017}}].

\bibitem{GPPZ00}
L.~Girardello, M.~Petrini, M.~Porrati, and A.~Zaffaroni, { The supergravity
  dual of ${N}=1$ super {Y}ang-{M}ills theory},  { Nucl. Phys.} { B569} (2000)
  451--469,
[\href{http://xxx.lanl.gov/abs/http://arXiv.org/abs/hep-th/9909047}{{\tt
  http://arXiv.org/abs/hep-th/9909047}}].

\bibitem{BiFrSk01}
M.~Bianchi, D.~Z. Freedman, and K.~Skenderis, { How to go with an {R}{G} flow},
   { JHEP} { 08} (2001) 041,
[\href{http://xxx.lanl.gov/abs/http://arXiv.org/abs/hep-th/0105276}{{\tt
  http://arXiv.org/abs/hep-th/0105276}}].

\bibitem{BerSam02}
M.~Berg and H.~Samtleben, { An exact holographic {RG} flow between 2d conformal
  fixed points},  { JHEP} { 05} (2002) 006,
[\href{http://xxx.lanl.gov/abs/http://arXiv.org/abs/hep-th/0112154}{{\tt
  http://arXiv.org/abs/hep-th/0112154}}].

\bibitem{IMSY98}
N.~Itzhaki, J.~M. Maldacena, J.~Sonnenschein, and S.~Yankielowicz, {
  Supergravity and the large {N} limit of theories with sixteen supercharges},
  { Phys. Rev.} { D58} (1998) 046004,
[\href{http://xxx.lanl.gov/abs/http://arXiv.org/abs/hep-th/9802042}{{\tt
  http://arXiv.org/abs/hep-th/9802042}}].

\bibitem{MorSam02}
J.~F. Morales and H.~Samtleben, { Supergravity duals of matrix string theory},
\href{http://xxx.lanl.gov/abs/http://arXiv.org/abs/hep-th/0206247}{{\tt
  http://arXiv.org/abs/hep-th/0206247}}.

\bibitem{Verl90}
H.~Verlinde, { Conformal field theory, 2-d quantum gravity and quantization of
  {T}eichm\"uller space},  { Nucl. Phys.} { B337} (1990)
652.

\bibitem{Carl91}
S.~Carlip, { Inducing {L}iouville theory from topologically massive gravity},
  { Nucl. Phys.} { B362} (1991)
111--124.

\bibitem{CoHevD95}
O.~Coussaert, M.~Henneaux, and P.~van Driel, { The asymptotic dynamics of
  three-dimensional {E}instein gravity with a negative cosmological constant},
  { Class. Quant. Grav.} { 12} (1995) 2961--2966,
[\href{http://xxx.lanl.gov/abs/http://arXiv.org/abs/gr-qc/9506019}{{\tt
  http://arXiv.org/abs/gr-qc/9506019}}].

\bibitem{Witt89}
E.~Witten, { Quantum field theory and the {J}ones polynomial},  { Commun. Math.
  Phys.} { 121} (1989) 351--399.

\bibitem{Tura94}
V.~G. Turaev, { Quantum invariants of knots and 3-manifolds}.
\newblock Walter de Gruyter \& Co., Berlin,
1994.
\newblock

\bibitem{KasRes02}
R.~Kashaev and N.~Reshetikhin, { Invariants of tangles with flat connections in
  their complements. {I}. {I}nvariants and holonomy {R}-matrices},
  \href{http://xxx.lanl.gov/abs/http://arXiv.org/abs/math.AT/0202211}{{\tt
  http://arXiv.org/abs/math.AT/0202211}}. 

\bibitem{MarSch83}
N.~Marcus and J.~H. Schwarz, { Three-dimensional supergravity theories},  {
  Nucl. Phys.} { B228} (1983)
145--162.

\bibitem{GuRoWa86}
M.~G{\"u}naydin, L.~J. Romans, and N.~P. Warner, { Compact and noncompact
  gauged supergravity theories in five-dimensions},  { Nucl. Phys.} { B272}
  (1986)
598--646.

\bibitem{ToPivN84}
P.~K. Townsend, K.~Pilch, and P.~van Nieuwenhuizen, { Selfduality in odd
  dimensions},  { Phys. Lett.} { 136B} (1984)
38--42.

\bibitem{AchTow86}
A.~Ach{\'u}carro and P.~K. Townsend, { A {C}hern-{S}imons action for
  three-dimensional anti-de {S}itter supergravity theories},  { Phys. Lett.} {
  B180} (1986)
89--92.

\bibitem{GunSca91}
M.~G{\"u}naydin and R.~J. Scalise, { Unitary lowest weight representations of
  the noncompact supergroup ${O}{S}p(2m^*/2n)$},  { J. Math. Phys.} { 32}
  (1991)
599--606.

\bibitem{Nico87}
H.~Nicolai, { ${D}\!=\!11$ supergravity with local ${SO}(16)$ invariance},  {
  Phys. Lett.} { B187} (1987)
316--320.

\bibitem{Witt98}
E.~Witten, { Anti-de {S}itter space and holography},  { Adv. Theor. Math.
  Phys.} { 2} (1998) 253--291,
[\href{http://xxx.lanl.gov/abs/http://arXiv.org/abs/hep-th/9802150}{{\tt
  http://arXiv.org/abs/hep-th/9802150}}].

\bibitem{HenSfe98}
M.~Henningson and K.~Sfetsos, { Spinors and the {A}d{S}/{C}{F}{T}
  correspondence},  { Phys. Lett.} { B431} (1998) 63--68,
[\href{http://xxx.lanl.gov/abs/http://arXiv.org/abs/hep-th/9803251}{{\tt
  http://arXiv.org/abs/hep-th/9803251}}].

\bibitem{MezTow85}
L.~Mezincescu and P.~K. Townsend, { Stability at a local maximum in higher
  dimensional anti-de {S}itter space and applications to supergravity},  { Ann.
  Phys.} { 160} (1985)
406.

\bibitem{GunNic95}
M.~G{\"u}naydin and H.~Nicolai, { Seven-dimensional octonionic {Y}ang-{M}ills
  instanton and its extension to an heterotic string soliton},  { Phys. Lett.}
  { B351} (1995) 169--172,
[\href{http://xxx.lanl.gov/abs/http://arXiv.org/abs/hep-th/9502009}{{\tt
  http://arXiv.org/abs/hep-th/9502009}}].

\bibitem{Hull84a}
C.~M. Hull, { Noncompact gaugings of ${N}\!=\!8$ supergravity},  { Phys. Lett.}
  { B142} (1984)
39--41.

\bibitem{Hull84b}
C.~M. Hull, { More gaugings of ${N}\!=\!8$ supergravity},  { Phys. Lett.} {
  B148} (1984)
297--300.

\bibitem{deWNic82}
B.~de~Wit and H.~Nicolai, { ${N}\!=\!8$ supergravity},  { Nucl. Phys.} { B208}
  (1982)
323--364.

\bibitem{Kac77}
V.~G. Kac, { A sketch of {L}ie superalgebra theory},  { Commun. Math. Phys.} {
  53} (1977)
31--64.

\bibitem{Nahm78}
W.~Nahm, { Supersymmetries and their representations},  { Nucl. Phys.} { B135}
  (1978)
149--166.

\bibitem{HulWar85b}
C.~M. Hull and N.~P. Warner, { The potentials of the gauged ${N}\!=\!8$
  supergravity theories},  { Nucl. Phys.} { B253} (1985)
675.

\bibitem{Hull85}
C.~M. Hull, { The minimal couplings and scalar potentials of the gauged
  ${N}\!=\!8$ supergravities},  { Class. Quant. Grav.} { 2} (1985)
343.

\bibitem{AhnWoo01}
C.~Ahn and K.~Woo, { Domain wall and membrane flow from other gauged $d = 4$,
  ${N} = 8$ supergravity. {I}},
\href{http://xxx.lanl.gov/abs/http://arXiv.org/abs/hep-th/0109010}{{\tt
  http://arXiv.org/abs/hep-th/0109010}}.

\bibitem{KLPS02}
R.~Kallosh, A.~D. Linde, S.~Prokushkin, and M.~Shmakova, { Gauged
  supergravities, de {S}itter space and cosmology},  { Phys. Rev.} { D65}
  (2002) 105016,
[\href{http://xxx.lanl.gov/abs/http://arXiv.org/abs/hep-th/0110089}{{\tt
  http://arXiv.org/abs/hep-th/0110089}}].

\bibitem{BroHen86}
J.~D. Brown and M.~Henneaux, { Central charges in the canonical realization of
  asymptotic symmetries: An example from three-dimensional gravity},  { Commun.
  Math. Phys.} { 104} (1986)
207--226.

\bibitem{HenSke98}
M.~Henningson and K.~Skenderis, { The holographic {W}eyl anomaly},  { JHEP} {
  07} (1998) 023,
[\href{http://xxx.lanl.gov/abs/http://arXiv.org/abs/hep-th/9806087}{{\tt
  http://arXiv.org/abs/hep-th/9806087}}].

\bibitem{GrScWi87}
M.~Green, J.~Schwarz, and E.~Witten, { Superstring theory}.
\newblock Cambridge University Press, Cambridge, 1987.

\end{thebibliography}

\newpage

\providecommand{\href}[2]{#2}\begingroup\raggedright\endgroup

\end{document}